\documentclass[aps,pre,twocolumn,epsfig,floats]{revtex4}
\usepackage{mathrsfs}

\usepackage{amsmath}
\usepackage{color}
\usepackage{graphicx}
\usepackage{epsfig}
\usepackage{amsfonts}
\usepackage{mathrsfs}
\usepackage{mathtools}
\usepackage{float}
\usepackage{physics}
\usepackage{hyperref}
\hypersetup{
	colorlinks=true,
	citecolor=blue,
	linkcolor=blue,
	urlcolor=blue}
\usepackage{upgreek}
\usepackage{easyReview} 
\usepackage{fourier}
\usepackage{MnSymbol}
\usepackage{soul}
\usepackage{xcolor, soul}
\sethlcolor{red}
\usepackage{hyperref}
\hypersetup{
	colorlinks=true,
	citecolor=blue,
	linkcolor=blue,
	urlcolor=blue}

\begin{document}
\title{Ground-State Phase Diagram, Higher-Winding Topology, and Lifshitz Criticality in an Anisotropic Four-Spin XX Chain  
}

\author{M. Abbasi}
\email[]{mohammad.abbasi390@gmail.com}
\author{F. Valeh}
\author{S. Mahdavifar}
\affiliation{Department of Physics, University of Guilan, 41335-1914, Rasht, Iran}

\begin{abstract}
We investigate the quantum critical and topological properties of a spin-$1/2$ XX chain with anisotropic four-spin cluster interactions in a transverse magnetic field. The interplay among the exchange anisotropy, extended cluster interactions, and external field gives rise to a rich ground-state phase diagram comprising multiple gapped topological phases and distinct quantum critical boundaries. We identify several Lifshitz-type transitions driven by reconstructions of the low-energy Fermi-point structure, including an unconventional multicritical point at which distinct critical branches intersect and the momentum-space topology undergoes a singular reorganization. In the isotropic limit, we further uncover a conventional Lifshitz transition associated with the merging and annihilation of Fermi points, demonstrating that qualitatively distinct mechanisms of Fermi-point reconstruction can emerge within the same microscopic model. The gapped phases are characterized by quantized winding numbers $\nu=0,\pm1,-2,$ and $\pm3$, with the higher-winding sectors arising from the extended cluster interactions. In particular, the $\nu=\pm3$ phases exhibit a substantially richer topological structure than that of conventional short-range Kitaev-type chains and are separated from other topological sectors by field- and interaction-driven gap-closing transitions. These topological distinctions are independently manifested in the bulk entanglement spectrum through characteristic degeneracy patterns of its lowest-lying levels, providing an entanglement-based signature of the higher-winding phases. Our results demonstrate that anisotropic multispin interactions provide a versatile route to enriching both the topological and critical structures of one-dimensional quantum matter, offering a framework for realizing and controlling higher-winding topological phases and unconventional multicriticality in tunable quantum platforms.

\end{abstract}

\date{\today}
\maketitle

\section{Introduction}
\label{sec:intro}

Understanding the fundamental behavior of strongly correlated many-body systems remains one of the challenges in modern condensed matter physics \cite{fulde2012correlated,cabra2012modern}. Among their various manifestations, magnetic excitations play an important role in determining the microscopic origin of magnetic order and collective phenomena in quantum materials, providing a direct bridge between theoretical models and experimental observations \cite{cabra2012modern,giamarchi2003quantum,vasiliev2019low}. In this context, low-dimensional spin systems offer an ideal setting in which enhanced quantum fluctuations give rise to rich and often nonclassical behavior.

Quantum phase transitions (QPTs) constitute a hallmark of such systems, occurring at zero temperature where nonthermal control parameters such as interaction strength, pressure, or an external magnetic field drive qualitative changes in the ground state of a many-body system \cite{sachdev2011quantum1,sachdev2011quantum,vojta2003quantum,sondhi1997continuous}. In contrast to classical phase transitions governed by thermal fluctuations, QPTs are driven entirely by quantum fluctuations and therefore probe the intrinsic correlations of the ground state \cite{sachdev2011quantum1,sachdev2011quantum}. As a consequence, characterizing the ground state, which encodes the information about the correlations and symmetries of the system, is essential for understanding the emergence of distinct quantum phases and the mechanisms underlying their transitions \cite{dutta2010quantum,carr2010understanding}.

A useful framework for this analysis is the ground-state phase diagram, which describes the competition among different ordering tendencies and identifies the critical points that separate distinct phases \cite{sachdev2011quantum1,liu2024complexification}. The structure of this phase diagram reflects the interplay between interactions and external fields, providing insight into the nature of quantum criticality \cite{sachdev2011quantum1,auerbach2012interacting}. Beyond its importance, determining such phase diagrams has important implications for a wide range of physical systems, including magnetic materials \cite{white2007quantum,inosov2018quantum,cava2021introduction}, superconductors \cite{keimer2015quantum,li2014topological}, and topologically nontrivial phases \cite{wen1990topological,basu2023topological,moessner2021topological,grushin2020introduction}, as well as for emerging applications in quantum information processing \cite{beth2005quantum,wendin2017quantum,monroe2002quantum}, where controlling quantum correlations and entanglement is of importance. Consequently, the systematic investigation of ground-state phase diagrams in low-dimensional spin systems remains an active area of research in condensed matter physics.

Within this framework, one-dimensional spin-1/2 chains play a central role as paradigmatic models for studying collective quantum behavior \cite{giamarchi2003quantum}. In particular, the Heisenberg model and its anisotropic extensions, such as the XX model, capture the essential features of quantum magnetism while remaining analytically tractable in several regimes \cite{lieb1961two,pasnoori2023boundary,abbasi2025concurrenceless,jahangiri2023quantum,abbasi2025spin,abbasi2025hybrid}. Their exact solvability in certain regimes, notably through techniques such as Jordan–Wigner fermionization, provides a direct connection between microscopic interactions, ground-state properties, and emergent quantum correlations \cite{lieb1961two}.

Beyond nearest-neighbor exchange, the inclusion of multispin or cluster interactions has attracted considerable attention in recent years \cite{smacchia2011statistical,subrahmanyam2025quantum,verga2023entanglement,montes2012phase,arsh,kheiri2025dynamical,katibzadah2026quantum,mahdavifar2026topological,kheiri2026dynamical}. Experimentally, such interactions have been identified in a wide range of physical systems. In condensed matter, they appear in quasi-one-dimensional cuprates, influencing magnetic excitations and complex spectra, including La$_2$CuO$_4$ \cite{coldea2001spin}, the cuprate ladder La$_6$Ca$_{28}$Cu$_{24}$O$_{41}$ \cite{brehmer1999effects,matsuda2000magnetic}, and La$_4$Sr$_{10}$Cu$_{24}$O$_{41}$ \cite{notbohm2007one}.  In addition, cluster-spin or cyclic exchange terms are also relevant in two-dimensional antiferromagnets, such as well as in hydrogen-bonded ferroelectrics PbHPO$_4$ and PbDPO$_4$ \cite{chunlei1988green}, squaric acid crystals H$_2$C$_2$O$_4$ \cite{wang1990critical,wang1989first,wang1989microscopic}, and certain copolymers \cite{silva1993pseudo}. Beyond solid-state materials, cluster interactions have been applied to explain thermodynamic properties of classical fluids \cite{grimsditch1986brillouin}, solid $^3$He \cite{roger1983magnetism}, lipid bilayers \cite{scott1988anisotropic}, rare gases \cite{barker1986many}, and binary alloys \cite{styer1986multiatom}. Optical lattice experiments offer further opportunities to engineer such interactions in controlled geometries, such as square and ladder lattices, enabling direct observation of their impact on ground-state correlations \cite{greiner2002quantum,zhou2015spin,bloch2012quantum}. Collectively, these experimental realizations demonstrate that cluster interactions are broadly relevant across both natural and synthetic systems and provide strong motivation for theoretical studies of the resulting quantum phases, entanglement properties, and phase transitions.

Among low-dimensional spin models, the Heisenberg chain generalized to include four-spin cluster interactions plays an important role as a paradigmatic system \cite{valeh2026spin,katibzadah2026quantum,mahdavifar2026topological,galisova2013magnetic,zaim2009monte,javsvcur2016frustration,de2014quantum,florencio1997dynamics,de2006quantum,boechat2000dynamical,florencio2004temperature}. This extension enriches the physics of the model by introducing competing interaction channels that are absent in conventional pairwise-coupling models, leading to a rich ground-state phase diagram. Previous studies have shown that such four-spin interactions can stabilize unconventional quantum phases, including chiral and nematic orders, while simultaneously supporting topological features characterized by nonlocal invariants and boundary excitations \cite{katibzadah2026quantum,mahdavifar2026topological}. The interplay between symmetry-breaking order and topological order in these systems can be described using cluster-type local order parameters, which coexist with global topological anomalies such as winding numbers, highlighting the hybrid nature of these phases \cite{mahdavifar2026topological}. In addition, studies of quantum correlations reveal entangled ground states, making this model an ideal platform for exploring the connections between many-body entanglement, quantum criticality, and topological phenomena \cite{katibzadah2026quantum}. Consequently, the spin-1/2 Heisenberg chain with four-spin interactions not only serves as a theoretical benchmark for understanding complex quantum order but also provides a versatile framework for the study of entanglement and correlation measures in strongly interacting systems.

Despite extensive studies on four-spin cluster interactions, the impact of an external transverse magnetic field on such systems remains not yet fully understood. The introduction of a transverse magnetic field induces a competition between interaction-driven correlations and  transverse magnetic field, which can  modify the ground-state structure and potentially stabilize novel quantum phases with distinct excitation spectra. Understanding this interplay is therefore important, as it provides insight into how external control parameters can tune both conventional and topological orders in strongly correlated spin systems.

In this work, we investigate the spin-1/2 Heisenberg chain with four-spin interactions under a transverse magnetic field. We explore the ground-state phase diagram, identifying the various quantum phase transitions and characterizing the orders present in each regime.  In addition, we analyze topological phase transitions, highlighting the role of the field in shaping topological phase. This study provides a coherent picture of the interplay between cluster interactions  and magnetic fields, offering a versatile platform for understanding complex many-body phenomena in low-dimensional spin systems.

The remainder of this paper is structured as follows. In Sec.~\ref{sec:model}, we introduce the spin-1/2 XX chain with four-spin interactions and present the fermionization approach used to obtain the exact ground state and the corresponding energy spectrum. Section ~\ref{sec:results} is devoted to a detailed analysis of the system’s quantum phases, including the characterization of order parameters, magnetization, and quantum correlation measures, as well as the investigation of topological phase transitions. Finally, in Sec.~\ref{sec:conclusion}, we summarize our main findings and discuss their implications for the study of entanglement, quantum criticality, and topological phenomena in low-dimensional spin systems.

\section{Model AND Energy spectrum}
\label{sec:model}

We begin by considering the one-dimensional spin-$1/2$ XX chain in the presence of a transverse magnetic field, described by the Hamiltonian
\begin{align}\label{eq1} 
	& \mathcal{H}_{XX} = J \sum_{n = 1}^N \left( S_n^x S_{n + 1}^x + S_n^y S_{n + 1}^y \right) -  h \sum_{n=1}^{N} S^z_n,
\end{align}
where $S_n^\mu$ ($\mu = x,y,z$) denotes the spin operator at site $n$, $J>0$ is the nearest-neighbor antiferromagnetic exchange coupling, and $h$ represents the strength of the transverse magnetic field. This model captures the essential features of planar quantum magnetism while remaining analytically tractable.

\begin{figure}[!ht]
	\centerline{\includegraphics[width=0.9\linewidth,height=0.4\linewidth]{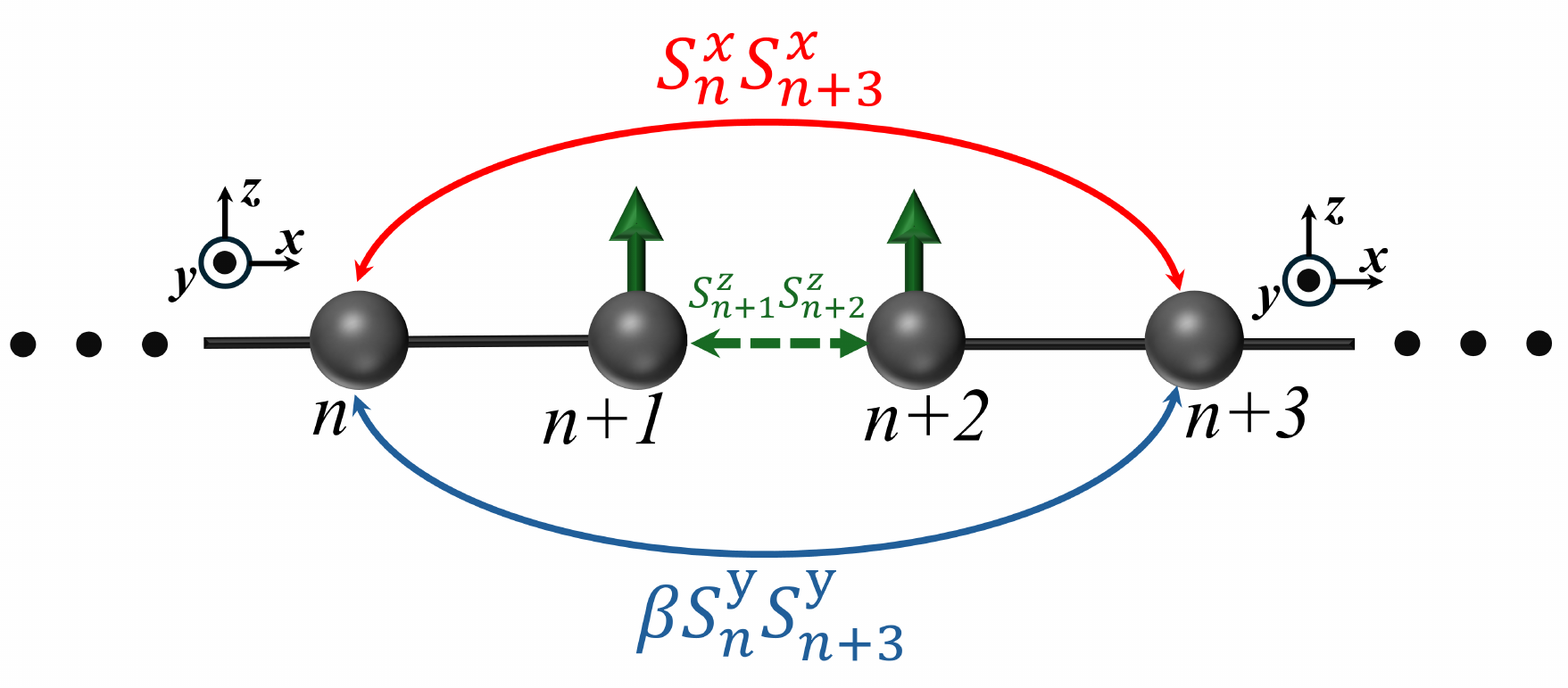}}
	\caption{Schematic representation of the four-spin cluster interaction Hamiltonian. }
	\label{Fig1}
\end{figure}

We next introduce a four-spin cluster interaction involving spins beyond nearest neighbors. The corresponding Hamiltonian is written as
\begin{align}\label{eq2} 
	& \mathcal{H}_{CI} = \alpha \sum_{n = 1}^N \left[ \left( S_n^x S_{n + 3}^x + \beta S_n^y S_{n + 3}^y \right) S_{n + 1}^z S_{n + 2}^z \right],
\end{align}
where $\alpha$ sets the strength of the cluster interaction and $\beta$ controls its anisotropy. This term introduces longer-range correlations and interaction channels not present in the conventional XX model.

Our aim is to investigate the combined effect of these contributions. The full Hamiltonian is given by
\begin{align}\label{eq3} 
	& \mathcal{H} = \mathcal{H}_{XX} + \mathcal{H}_{CI}.
\end{align}
%Throughout this work, we impose periodic boundary conditions, $S_{n+N}^\mu = S_n^\mu$.

The ground-state phase diagram in the presence of a transverse magnetic field, including the critical lines separating the distinct gapped and gapless regimes, was obtained in our previous work~\cite{mabbasi}. Here, we use this phase diagram as the basis for the analysis of the critical properties and low-energy excitation spectra presented below.
A notable multicritical point is formed by the intersection of the two critical branches $h_{c_2}$ and $h_{c_3}$, located at $\beta_m=-1-{2}/{\alpha}$, $ h_m={3J}/{4}$, and constitutes a multicritical point where distinct critical branches converge. The phase diagram further contains Lifshitz-type critical points along the symmetry-preserving line $\beta=1$, including both conventional and unconventional critical structures. In addition, the intersection of $h_{c_4}$ with $\beta=1$ gives rise to another special gapless critical point. These features highlight the rich and nontrivial organization of the critical lines generated by the interplay of the transverse field, anisotropy, and extended four-spin interactions.

To proceed, we diagonalize the Hamiltonian using the Jordan–Wigner transformation \cite{lieb1961two}, which maps the spin-$1/2$ operators onto spinless fermions:
\begin{eqnarray}\label{eq4} 
	S_n^{+} &=& a_n^\dagger \exp\left(i\pi \sum_{l<n} a_l^\dagger a_l \right), \nonumber \\ 
	S_n^{-} &=& a_n \exp\left(i\pi \sum_{l<n} a_l^\dagger a_l \right), \nonumber \\ 
	S_n^{z} &=& a_n^\dagger a_n - \frac{1}{2},
\end{eqnarray}
where $a_n^\dagger$ and $a_n$ are fermionic creation and annihilation operators, respectively. Under this mapping, the Hamiltonian can be rewritten in terms of fermionic operators, yielding a quadratic fermionic Hamiltonian,
\begin{align}\label{eq5} 
	& \mathcal{H} = \frac{J}{2} \sum_{n = 1}^{N} \left( a_{n}^\dagger a_{n+1} + a_{n+1}^\dagger a_{n} \right)\nonumber \\ 
	&+ \frac{\alpha}{16} (1+\beta) \sum_{n = 1}^{N} \left( a_{n}^\dagger a_{n+3} + a_{n+3}^\dagger a_{n} \right) \nonumber \\ 
	&+ \frac{\alpha}{16} (1-\beta) \sum_{n = 1}^{N} \left( a_{n}^\dagger a_{n+3}^\dagger + a_{n+3} a_{n} \right) \nonumber \\ 
	&- h \sum_{n = 1}^{N} \left(a_{n}^\dagger a_{n} - \frac{1}{2}\right).
\end{align}

The quadratic structure of Eq.~(\ref{eq5}) allows for exact diagonalization. We first perform a Fourier transformation, $a_n = \frac{1}{\sqrt{N}} \sum_k e^{-ikn} a_k$, which brings the Hamiltonian into momentum space. The remaining pairing terms are then diagonalized via a Bogoliubov transformation,
\begin{align}\label{eq6} 
	&a_k = \cos(\theta_k)\,\beta_k + i \sin(\theta_k)\,\beta_{-k}^\dagger,
\end{align}
by choosing
\begin{align}\label{eq7} 
	& \tan(2\theta_k) = \frac{{\cal B}_k}{{\cal A}_k}.
\end{align}
where
\begin{align}\label{eq10} 
	& \mathcal{A}_k = J \cos(k) - h + \frac{\alpha}{8}(1+\beta)\cos(3k), \nonumber \\ 
	& \mathcal{B}_k = \frac{\alpha}{8}(1-\beta)\sin(3k).
\end{align}

The Hamiltonian can thus be written in diagonal form as
\begin{align}\label{eq8} 
	& \mathcal{H} = \sum_{k=-\pi}^{\pi} \varepsilon(k) \left( \beta_k^\dagger \beta_k - \frac{1}{2} \right).
\end{align}

The corresponding single-particle energy spectrum is given by
\begin{align}\label{eq9} 
	& \varepsilon(k) = \sqrt{{\cal A}_k^2 + {\cal B}_k^2},
\end{align}

The diagonal form of the Hamiltonian in Eq.~(\ref{eq8}) implies that it commutes with the total Bogoliubov fermion number operator, $\hat{N}_B = \sum_k \beta_k^\dagger \beta_k$. The eigenvalues of $\hat{N}_B$ label the occupation of the quasiparticle modes, and the lowest-energy state corresponds to the vacuum with no Bogoliubov fermions. The ground state is therefore given by
\begin{align}\label{eqGS} 
	|GS\rangle = |0\rangle.
\end{align}

This representation serves as a useful starting point for analyzing the phase diagram of the model in the presence of the transverse magnetic field.

\section{Results}
\label{sec:results}

\subsection{Quantum phase transition in the ground state and energy gap}

Before analyzing the general anisotropic regime, it is instructive to consider the special case $\beta=1$, where the four-spin cluster interaction becomes isotropic. In this limit, the pairing contribution vanishes in the fermionic representation, the $U(1)$ symmetry remains preserved, and the excitation spectrum reduces to a purely dispersive single-particle form. Setting $J=1$, the spectrum is given by
\begin{equation}
	\varepsilon(k)=J\cos k-h+\frac{\alpha}{4}\cos(3k).
\end{equation}

The ground-state properties in this regime are determined by the Fermi points satisfying $\varepsilon(k_F)=0$. Introducing $x=\cos k$, the Fermi-point condition can be rewritten as the cubic equation
\begin{equation}
	\alpha x^{3}+\left(1-\frac{3\alpha}{4}\right)x-h=0.
\end{equation}
For a fixed magnetic field, the number of real solutions within the physical interval $x\in[-1,1]$ determines the number of Fermi points and, consequently, the structure of the gapless phase. The discriminant of the cubic equation separates regimes with one and three real roots, corresponding to different Fermi-point configurations. The transition between these regimes occurs when the discriminant vanishes, indicating the emergence of multiple roots and a change in the Fermi-point structure.

The corresponding critical condition is obtained from the extremal points of the dispersion relation. One branch is located at
\begin{equation}
	h_{c}^{(\mathrm{sp})}=1+\frac{\alpha}{4},
\end{equation}
while for sufficiently strong cluster coupling, $\alpha\geq4/3$, an additional critical branch emerges,
\begin{equation}
	h_{c_L}^{(\mathrm{sp})} = -\alpha c^{3} -\left(1-\frac{3\alpha}{4}\right)c,
\end{equation}
where 

\begin{equation}
 c = \sqrt{\frac{1}{4} - \frac{1}{3\alpha}}.
\end{equation}

These critical lines separate gapless regions with different numbers of Fermi points and show how the transverse field modifies the isotropic phase structure.

Our primary interest, however, lies in the anisotropic regime $\beta\neq1$, where pairing terms are generated in the fermionic Hamiltonian and the system generally develops a finite energy gap. In this case, quantum phase transitions are associated with the closing of the quasiparticle gap rather than simple Fermi-surface reconstruction.

We therefore begin by identifying the quantum critical points of the model, which are determined from the closing of the energy gap in the single-particle spectrum $\varepsilon(k)$. Since quantum phase transitions are associated with nonanalytic changes in the ground-state properties, they occur when the gap closes at specific momenta, signaling a qualitative change in the structure of the ground state. The resulting conditions provide direct access to the phase boundaries in terms of the control parameters $h$, $\alpha$, and $\beta$.

From the dispersion relation, the critical lines are obtained analytically. In particular, the anisotropy-driven transition occurs at
\begin{equation}\label{eqr1}
	\beta_c = 1 \qquad \text{for} \quad h \leq J + \frac{\alpha}{4},
\end{equation}

The field-induced transitions can be grouped according to the distinct regions of the anisotropy parameter $\beta$. In particular, the critical fields separate into two pairs corresponding to different regimes of the phase diagram, which we label by $(L_i, R_i)$ to distinguish the branches associated with $\beta < 0.0$ and $\beta > 0.0$, respectively. They are given by
\begin{align}
	h_c^{(L_1,R_1)} &= \mp J \mp \frac{\alpha}{8}(\beta + 1), \\
	h_c^{(L_2,R_2)} &= \mp \frac{J}{2} \pm \frac{\alpha}{8}(\beta + 1).
\end{align}
which generate four distinct critical fields,
\begin{align}\label{eqr3}
	h_{c_1} &= -J - \frac{\alpha}{8}(\beta + 1) \qquad \text{for} \quad \beta \leq -(1 + \frac{8}{\alpha}),  \nonumber \\
	h_{c_2} &= J + \frac{\alpha}{8}(\beta + 1)  \qquad \text{for} \quad \beta > -(1 + \frac{8}{\alpha}),  \nonumber \\
	h_{c_3} &= \frac{J}{2} - \frac{\alpha}{8}(\beta + 1) \qquad \text{for} \quad \beta \leq ( \frac{4}{\alpha} -1 ),  \nonumber \\
	h_{c_4} &= -\frac{J}{2} + \frac{\alpha}{8}(\beta + 1) \qquad \text{for} \quad \beta > ( \frac{4}{\alpha} -1 ).
\end{align}

The critical fields can be further understood from the momentum-space structure of the quasiparticle spectrum. At a quantum critical point, the energy gap closes at specific momenta $k_F$, which determine the corresponding Fermi points of the transition. For the present model, the allowed gap-closing momenta are constrained to the discrete set
\begin{equation}
	k_F=\frac{n\pi}{3}, \qquad n=0,\pm1,\pm2,\pm3,
\end{equation}
reflecting the momentum dependence introduced by the four-spin interaction term.

Each critical branch is associated with a characteristic set of gap-closing momenta. In particular, the critical line $h_{c_1}$ is characterized by gap closing at the Brillouin-zone boundary, $k_F=\pm\pi$, while the transition line $h_{c_2}$ corresponds to a gap closing at the zone center, $k_F=0$. The remaining two branches originate from finite-momentum critical modes: the line $h_{c_3}$ is associated with gap closing at $k_F=\pm\pi/3$, whereas $h_{c_4}$ corresponds to $k_F=\pm2\pi/3$. These distinct Fermi-point structures distinguish the different critical mechanisms realized in the phase diagram and show how the extended four-spin interaction modifies the low-energy excitation spectrum. These critical lines delineate the different quantum phases of the model. In the following, we analyze the evolution of the energy gap and the corresponding changes in the ground-state properties across the phase boundaries.

We now turn to a quantitative characterization of the phase structure by examining the ground-state energy and the energy gap. The ground-state energy density provides a direct measure of the stability of the many-body state and serves as a sensitive probe of quantum phase transitions. In the following, we fix the energy scale by setting $J=1$ and consider a sufficiently large system size, $ N=1000$, to approximate the thermodynamic limit.

Within the fermionic representation, the ground-state energy density is obtained from the diagonal form of the Hamiltonian as
\begin{eqnarray}\label{eqeg} 
	E_{G} = -\frac{1}{2N} \sum_{k} \varepsilon(k),
\end{eqnarray}
where $\varepsilon(k)$ is the single-particle spectrum. This expression follows directly from the vacuum structure of the Bogoliubov quasiparticles and defines the contribution of all momentum modes to the ground-state energy.

Quantum phase transitions manifest as nonanalytic behavior in the ground-state energy or its derivatives with respect to the control parameters. In particular, the second derivative of $E_G$ provides a useful indicator of continuous transitions, where singularities signal changes in the underlying ground-state structure.

Complementary information is obtained from the energy gap, defined by the minimum of the spectrum, $\Delta = \min_k \varepsilon(k)$. The closing of the gap identifies critical points and separates distinct quantum phases, while a finite gap characterizes gapped phases with well-defined quasiparticle energy. The combined analysis of the ground-state energy and the energy gap, therefore  allows for a consistent determination of the phase boundaries and the nature of the transitions in the model.

\begin{figure}[h]
	\centerline{\includegraphics[width=0.5\linewidth,height=0.42\linewidth]{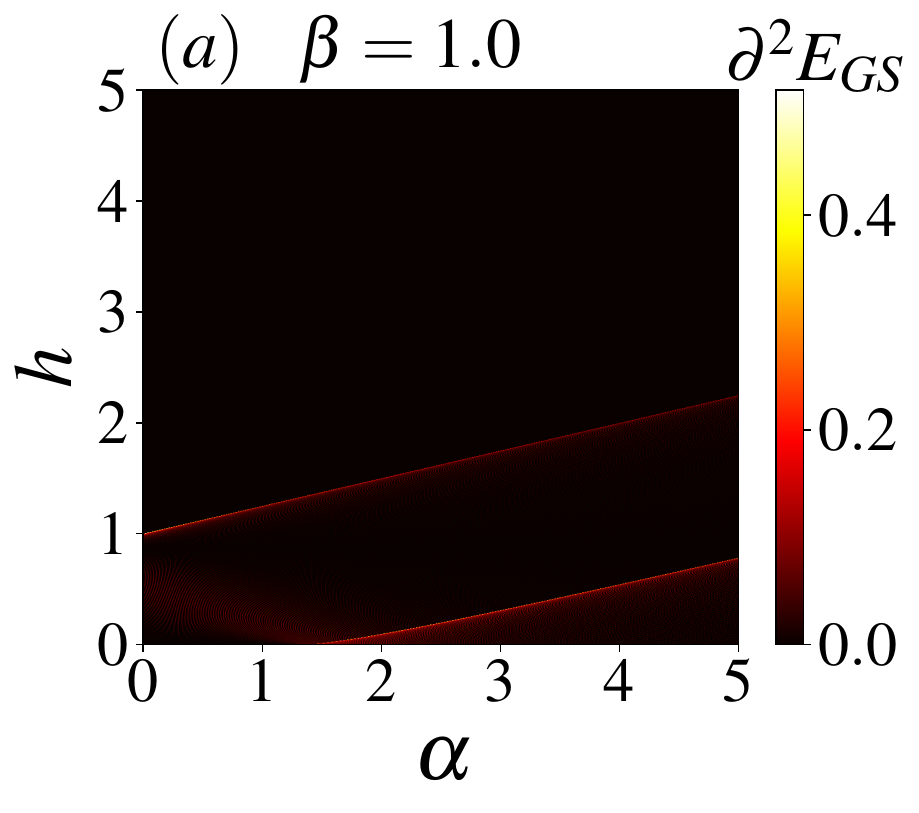} \includegraphics[width=0.5\linewidth,height=0.42\linewidth]{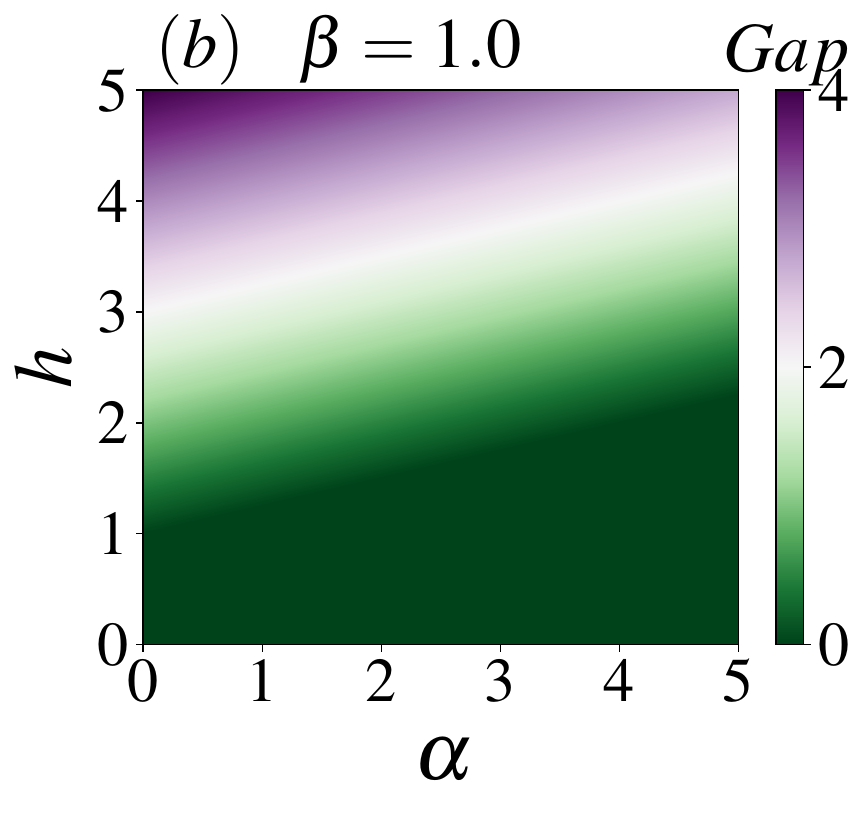} 
		
	}

	\caption{(a) Density plots of the second derivative of the ground-state energy in the $(\alpha,h)$ plane and (b) density plots of the energy gap in the $(\alpha,h)$ plane for general anisotropic regime $\beta=1.0$.
	}
	\label{Fig3}
	
\end{figure}

We first examine the special isotropic limit $\beta=1$, for which the quasiparticle spectrum reduces to the purely dispersive form discussed above. The corresponding ground-state and energy gap behavior is shown in Fig.~\ref{Fig3}. Figure~\ref{Fig3}(a) presents the density plot of the second derivative of the ground-state energy, while Fig.~\ref{Fig3}(b) shows the corresponding energy gap in the $(\alpha,h)$ plane. These results provide a direct view of the field evolution of the gapless structure in the isotropic regime.

At zero field, the system exhibits a change in the number of Fermi points at $\alpha=4/3$, where the two gapless regimes meet. Upon applying a finite transverse field, this critical condition evolves into the field-dependent branch $h_{c_L}^{(\mathrm{sp})}$. The pronounced nonanalytic structure in $\partial^2 E_G$ in Fig.~\ref{Fig3}(a) follows this branch, demonstrating that the zero-field critical point continuously develops into a line of gapless-to-gapless transitions in the $(\alpha,h)$ plane. Across this line, the number of Fermi points changes, indicating a reconstruction of the momentum-space structure of the gapless phase. The same boundary is visible in the energy gap map in Fig.~\ref{Fig3}(b), where the system remains gapless on both sides of $h_{c_L}^{(\mathrm{sp})}$.

A second critical boundary is also evident in both panels. The line $h_c^{(\mathrm{sp})}=1+\alpha/4$ marks the termination of the gapless regime as the magnetic field is increased. In Fig.~\ref{Fig3}(a), it is accompanied by a pronounced nonanalytic feature in the ground-state-energy derivative, while Fig.~\ref{Fig3}(b) shows the corresponding closing of the gap along this boundary. Beyond $h_c^{(\mathrm{sp})}$, the energy spectrum becomes finite and the system enters a gapped regime. Thus, the two boundaries visible in Fig.~\ref{Fig3} have distinct physical origins: $h_{c_L}^{(\mathrm{sp})}$ separates two gapless regimes with different Fermi-point structures, whereas $h_c^{(\mathrm{sp})}$ separates the gapless regime from a gapped phase. This distinction is a characteristic feature of the isotropic limit and provides a useful reference for the more general anisotropic phase diagram considered below.

To examine the evolution away from the special isotropic line while retaining the critical structures identified at $\beta=1$, we focus on the representative interaction strengths $\alpha=2$ and $\alpha=3$, for which the field-dependent critical branches are well separated and the influence of the four-spin interaction on the anisotropy-driven phase structure can be clearly resolved.

\begin{figure}[h]
	\centerline{\includegraphics[width=0.5\linewidth,height=0.42\linewidth]{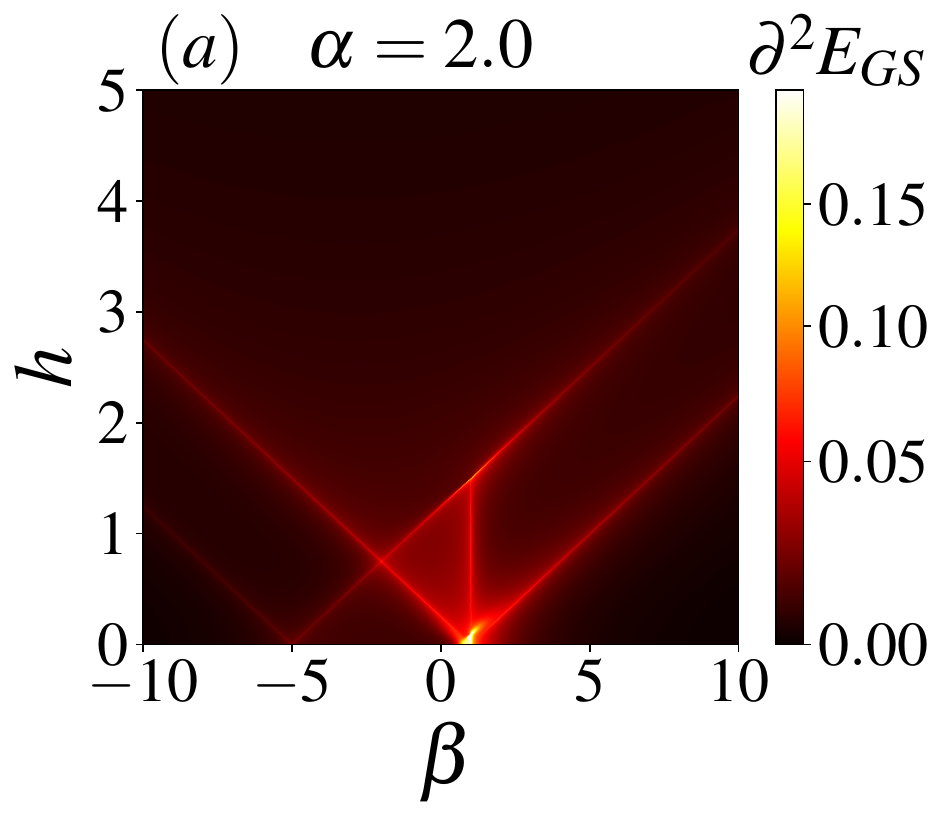} \includegraphics[width=0.5\linewidth,height=0.42\linewidth]{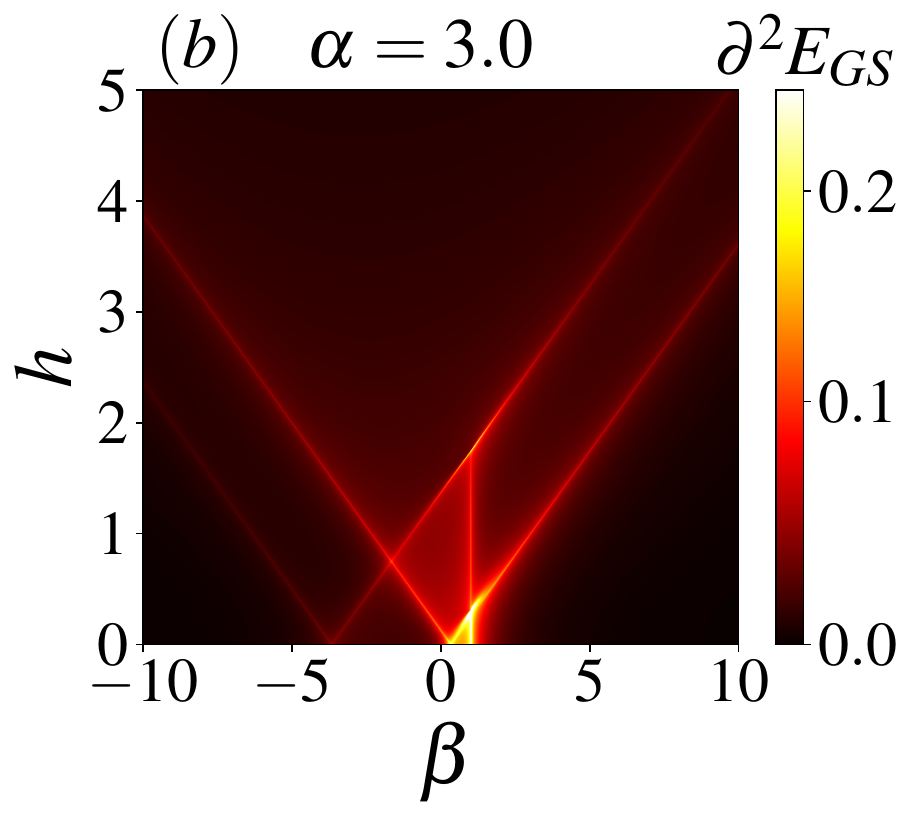} 
		
	}
	\centerline{\includegraphics[width=0.5\linewidth,height=0.42\linewidth]{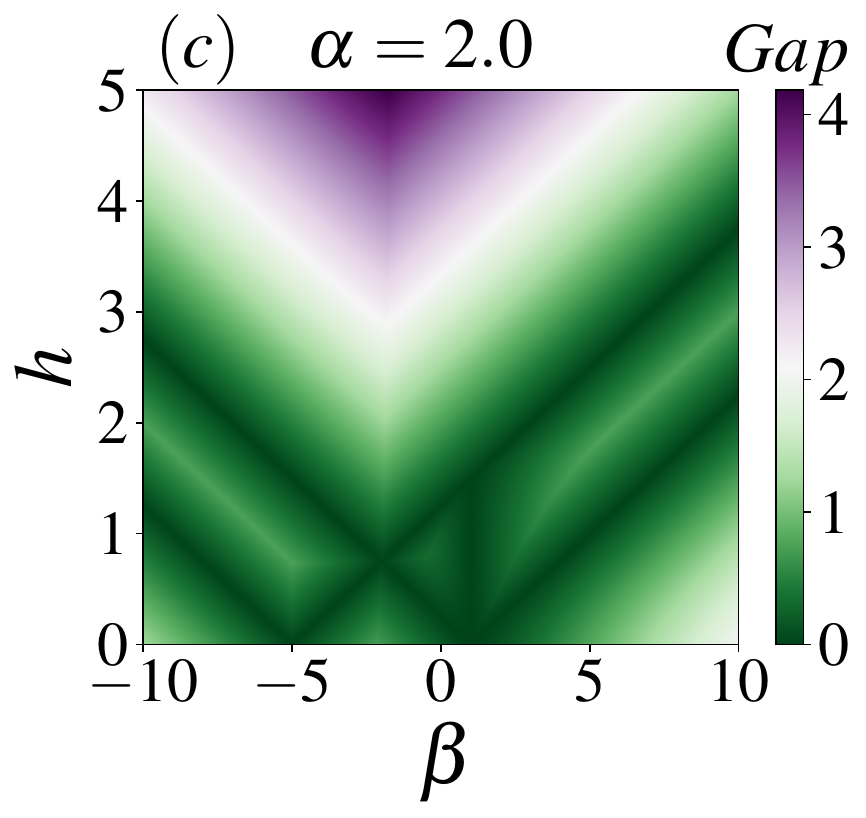} \includegraphics[width=0.5\linewidth,height=0.42\linewidth]{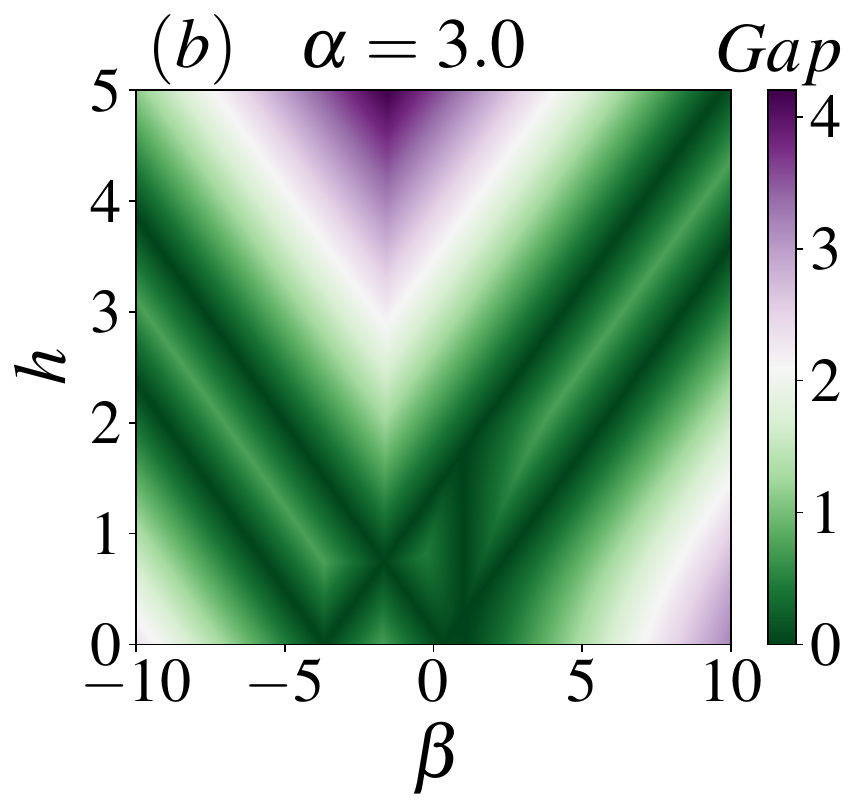} 
		
	}

	\caption{(color online).(a,b) Density plots of the second derivative of the ground-state energy in the $(\beta,h)$ plane and (c,d) density plots of the energy gap in the $(\beta,h)$ plane for $\alpha=2.0$ and $\alpha=3.0$, respectively.
	}
	\label{Fig4}
	
\end{figure}

Figure~\ref{Fig4} summarizes the behavior of the ground-state energy and energy gap for two representative values of the cluster-interaction strength, $\alpha=2.0$ and $\alpha=3.0$. Figures~\ref{Fig4}(a) and \ref{Fig4}(b) show the density plots of the second derivative of the ground-state energy, $\partial^{2}E_G$, in the $(\beta,h)$ plane, while Figs.~\ref{Fig3}(c) and \ref{Fig3}(d) display the corresponding energy-gap distributions. In all panels, the horizontal axis denotes the anisotropy parameter $\beta$ and the vertical axis the transverse magnetic field $h$.

The second-derivative plots in Figs.~\ref{Fig4}(a) and \ref{Fig4}(b) exhibit pronounced nonanalytic features along well-defined lines in parameter space. These singular structures provide a direct signature of the critical boundaries and are consistent with the critical conditions obtained analytically. Their locations evolve with the interaction strength: increasing $\alpha$ from $2$ to $3$ shifts the field-dependent critical structures toward larger values of $h$, reflecting the enhanced energy scale associated with the four-spin cluster interaction. The overall structure of the nonanalytic lines is preserved, indicating that the cluster interaction modifies the locations of the critical boundaries without changing their underlying organization.

The corresponding gap maps in Figs.~\ref{Fig4}(c) and \ref{Fig4}(d) provide an independent characterization of the same critical structure. In both cases, the energy gap vanishes along the critical lines identified in the ground-state-energy analysis, separating regions with finite gaps from gapless boundaries. The agreement between the locations of the gap closings and the nonanalytic features of $\partial^{2}E_G$ demonstrates the consistency of the two diagnostics and confirms the critical lines obtained from the analytical gap-closing conditions. As $\alpha$ increases, the critical lines shift toward larger magnetic fields, in agreement with the analytical expressions for the critical fields. Thus, the combined energy-gap and ground-state-energy analysis establishes the phase-boundary structure in the $(\beta,h)$ plane and provides the starting point for the momentum-resolved analysis of the individual critical branches presented below.

\subsection{Magnetization and Cluster order parameters}

\begin{figure*}
	\centerline{\includegraphics[width=0.32\linewidth,height=0.24\linewidth]{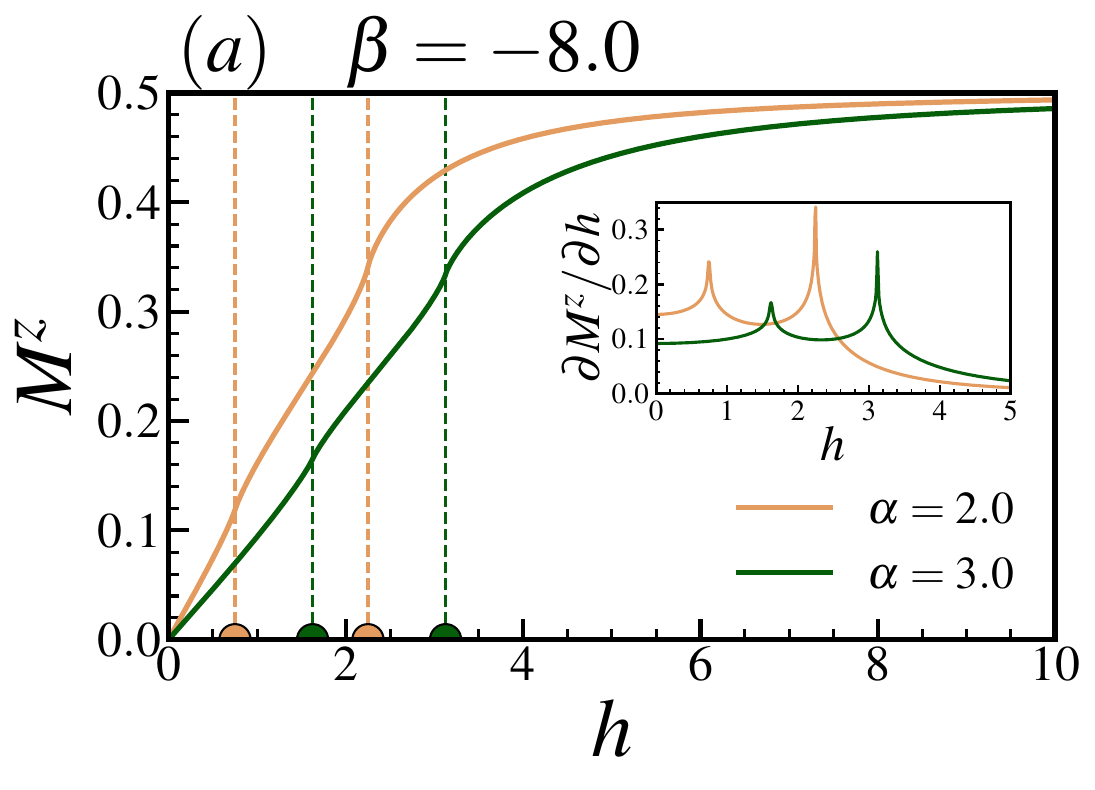} \includegraphics[width=0.32\linewidth,height=0.24\linewidth]{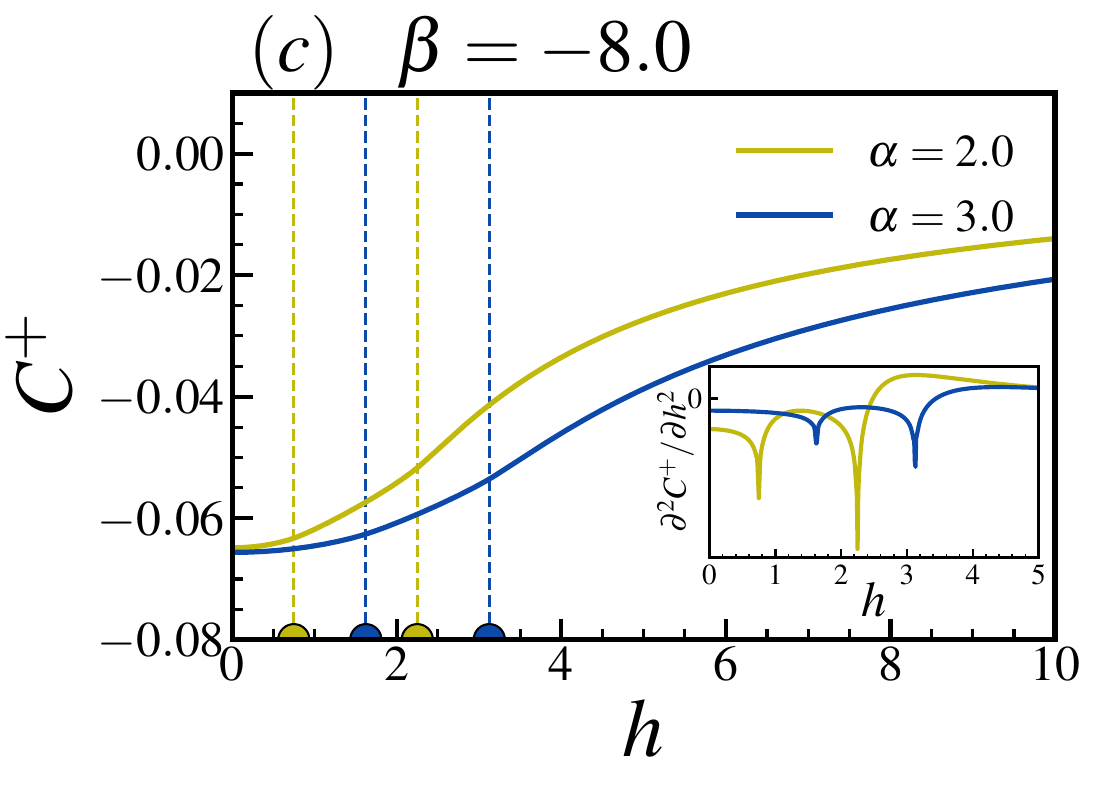} 
		\includegraphics[width=0.32\linewidth,height=0.24\linewidth]{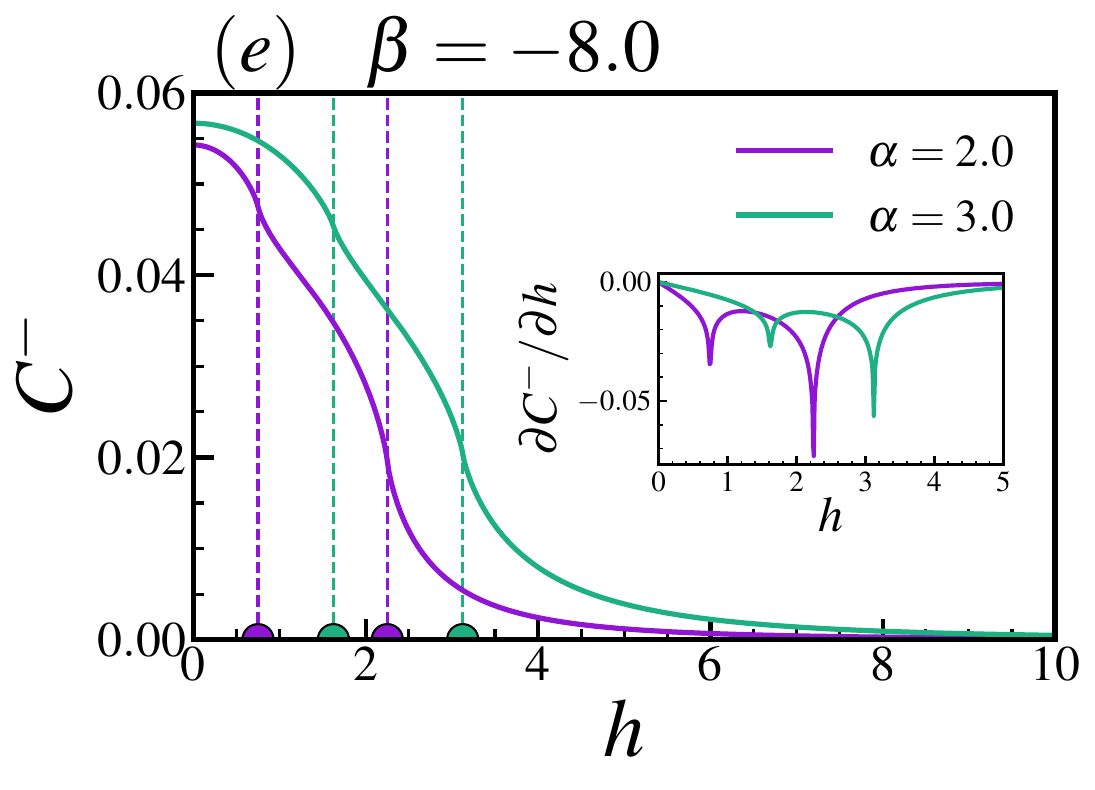}
	}
	\centerline{\includegraphics[width=0.32\linewidth,height=0.24\linewidth]{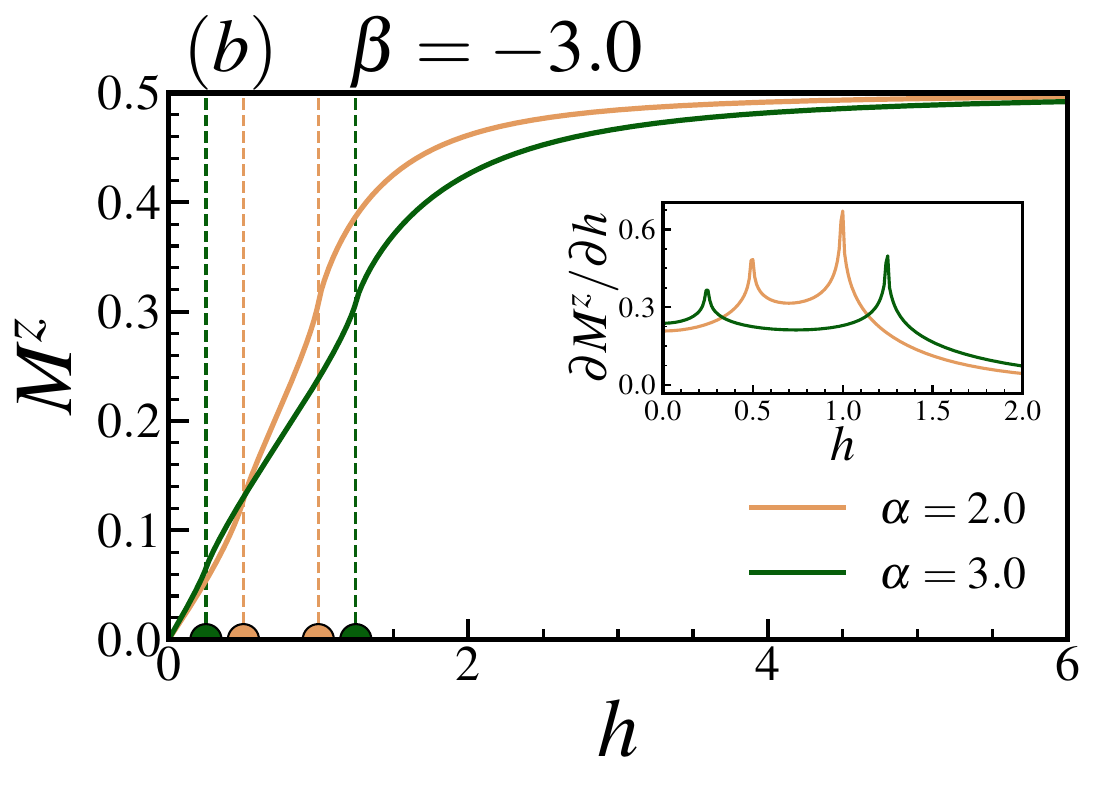} \includegraphics[width=0.32\linewidth,height=0.24\linewidth]{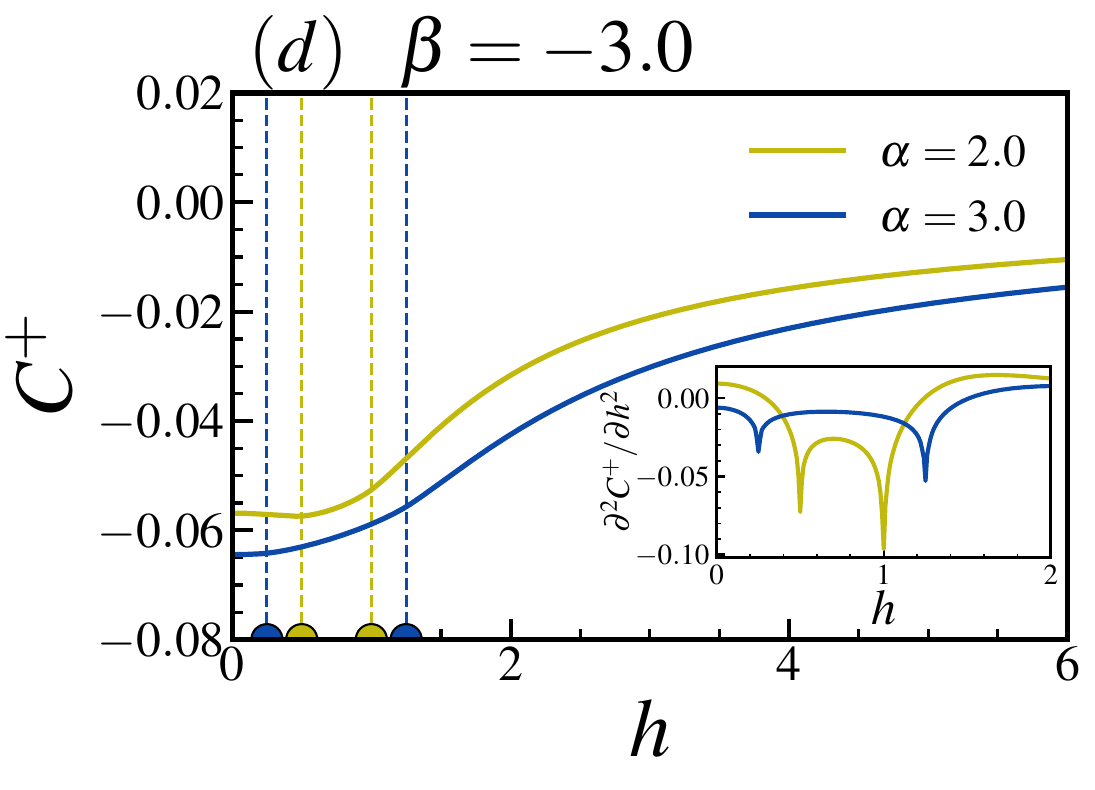} 
		\includegraphics[width=0.32\linewidth,height=0.24\linewidth]{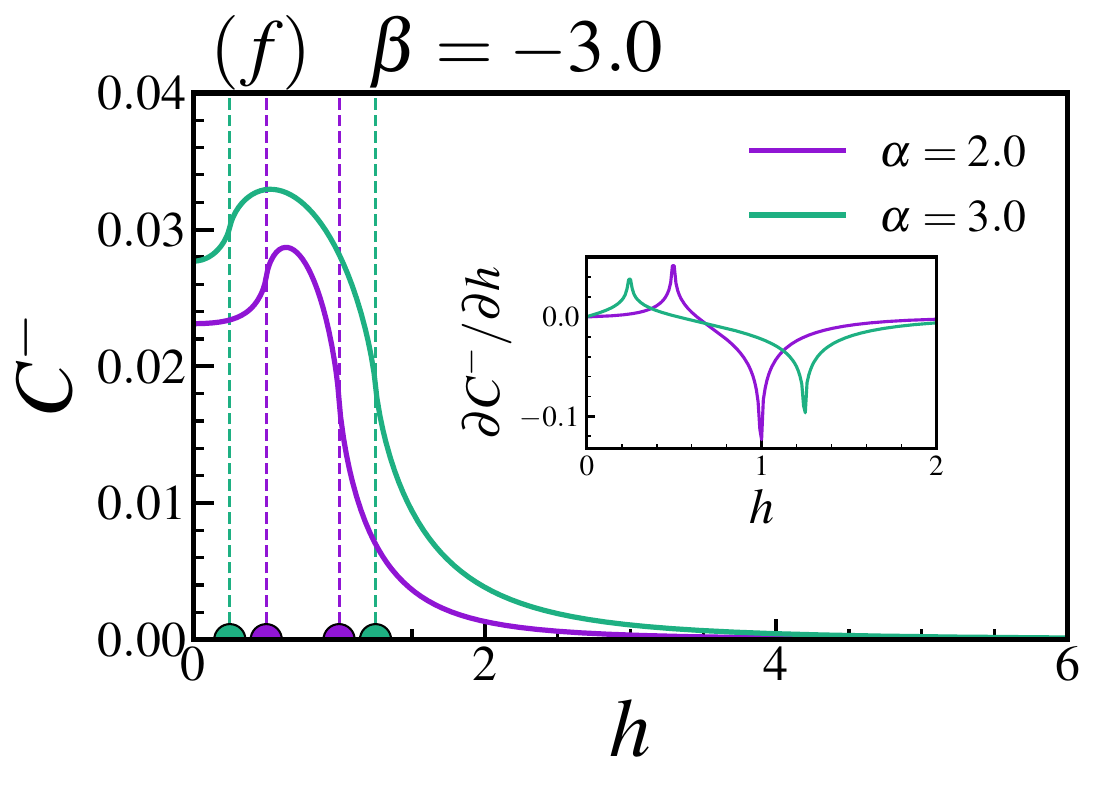}
	}

	\caption{Magnetization in panels (a) and (b), nematic order parameters in panels (c) and (d), and chiral order parameters in panels (e) and (f) as functions of the magnetic field $h$. The upper panels correspond to $\beta=-8.0$, while the lower panels are obtained for $\beta=-3.0$. All results are calculated for a chain of size $N=1000$.		
	}
	\label{Fig5}
	
\end{figure*}

To further characterize these phases and the nature of the associated transitions, it is necessary to examine observables that directly probe magnetic ordering and correlation properties. In particular, quantities such as magnetization and appropriate order parameters provide complementary information about symmetry breaking and the underlying structure of the ground state, allowing one to distinguish between different phases that may appear similar at the level of the energy spectrum alone.

We begin with the magnetization, which directly probes the response of the system to the applied transverse field and provides a simple characterization of the magnetic state. The magnetization along the field direction is defined as
\begin{align}\label{eqr4}
	M^{z} = \frac{1}{N} \sum_{n=1}^{N} \langle S^{z}_n \rangle,
\end{align}
which, in the fermionic representation, can be expressed in terms of the Bogoliubov angles as
\begin{align}
	M^{z} = -\frac{1}{2N} \sum_{k} \frac{\mathcal{A}_k}{\varepsilon(k)}
\end{align}

This quantity reflects the competition between the exchange interactions and the external field. In the field-dominated regime, the spins tend to align along the $z$ direction, leading to a finite magnetization, whereas in regions where interactions are dominant, quantum fluctuations reduce the net polarization. The evolution of $M^{z}$ across the parameter space therefore provides direct information on the reorganization of the ground state and allows one to identify distinct magnetic regimes and the associated phase transitions.

While magnetization captures the global response of the system to a transverse field, it does not fully resolve the internal structure of the gapped phases induced by the cluster interaction. To access this information, we therefore introduce order parameters that probe the cluster spin correlations  defined by the Hamiltonian. In the present model, a natural choice is cluster order parameters constructed from four-spin operators, which directly reflect the form of the interaction.

We consider the following cluster correlators
\begin{align}\label{eqr5}
	C^{+} &= \frac{1}{N} \sum_{n=1}^{N} \left\langle \left(S_n^x S_{n+3}^x + S_n^y S_{n+3}^y\right) S_{n+1}^z S_{n+2}^z \right\rangle, \\
	C^{-} &= \frac{1}{N} \sum_{n=1}^{N} \left\langle \left(S_n^x S_{n+3}^x - S_n^y S_{n+3}^y\right) S_{n+1}^z S_{n+2}^z \right\rangle,
\end{align}
which can be written in the fermionic representation as
\begin{align}\label{eqr6}
	C^{+} &= -\frac{1}{8N} \sum_k \cos(3k)\,\frac{\mathcal{A}_k}{\varepsilon(k)}, \\
	C^{-} &= -\frac{1}{8N} \sum_k \sin(3k)\,\frac{\mathcal{B}_k}{\varepsilon(k)}.
\end{align}

These quantities capture correlations extending over four consecutive sites and thus reveal ordering patterns that are not visible in local observables. In particular, $C^{-}$ is sensitive to the relative phase between spin components and provides a measure of chiral correlations, while $C^{+}$ reflects anisotropic correlations between the $x$ and $y$ spin components and is associated with nematic-like correlations. From the symmetry point of view, $C^{-}$ is odd under spatial inversion and indicates the onset of chiral ordering, whereas $C^{+}$ preserves inversion symmetry but distinguishes between different anisotropic spin configurations of the transverse spin components.

In contrast to conventional magnetic order parameters, these cluster quantities do not depend on spontaneous magnetization and can remain finite even in phases where $M^z$ varies smoothly. They provide a more sensitive characterization of the gapped phases generated by the four-spin interaction. The combined analysis of magnetization and cluster order parameters allows us to distinguish between phases that appear similar at the level of the excitation spectrum, and to identify the underlying symmetry-breaking patterns induced by the cluster interaction.

Throughout this analysis, we set $J=1.0$ and consider a system size $N=1000$ to approach the thermodynamic limit, while choosing $\alpha=2.0$ and $\alpha=3.0$ as representative values of the interaction strength. We first consider the regime $\beta<0.0$, focusing on the behavior of the magnetization and cluster order parameters. Figure~\ref{Fig5} summarizes the results for two characteristic anisotropies, $\beta=-8.0$ (upper panels) and $\beta=-3.0$ (lower panels). 

Figures~\ref{Fig5}(a) and ~\ref{Fig5}(b) display the magnetization as a function of the transverse field. In both cases, the application of the field leads to a gradual increase in $M^z$, reflecting the alignment of spins along the $z$ direction. Upon further increasing $h$, the system approaches a fully polarized state, and the magnetization saturates beyond the second critical field $h_{c_3}$. The insets show the field derivative $\partial M^z/\partial h$, which exhibits pronounced peaks at the critical points ($h_{c_1}, h_{c_3} \text{ at } \beta = -8.0;\; h_{c_2}, h_{c_3} \text{ at } \beta = -3.0$), indicating continuous phase transitions.

The behavior of the nematic order parameter $C^{+}$ is shown in Figs.~\ref{Fig5}(c) and ~\ref{Fig5}(d) for $\beta=-8.0$ and $\beta=-3.0$, respectively. In both cases, $C^{+}$ takes a finite negative value in the absence of the field, indicating nematic correlations induced by the cluster interaction. As the field increases, the magnitude of $C^{+}$ is gradually suppressed, indicating the suppression of these correlations. The insets display the second derivative $\partial^2 C^{+}/\partial h^2$, which shows clear peaks at the critical fields $h_{c_1}$,$h_{c_2}$ and $h_{c_3}$. Notably, the first derivative remains smooth across these points, indicating that the transition is not accompanied by a discontinuity at that level. Instead, the nonanalytic behavior emerges only in higher-order derivatives, similar to what is observed in related systems and reflecting the nature of quantum criticality \cite{abbasi2025hybrid}.

The chiral order parameter $C^{-}$, shown in Figs.~\ref{Fig5}(e) and ~\ref{Fig5}(f), exhibits a distinct response. For both values of $\beta$, $C^{-}$ takes a finite value at zero field, determined by the anisotropy of the interaction. With increasing $h$, the chiral order is gradually reduced and eventually vanishes in the large-field regime, where the system approaches a fully polarized state. The insets show the first derivative $\partial C^{-}/\partial h$, which shows clear peaks at $h_{c_1}$,$h_{c_2}$ and $h_{c_3}$, indicating the locations of the phase transitions.

\begin{figure*}
	\centerline{\includegraphics[width=0.32\linewidth,height=0.24\linewidth]{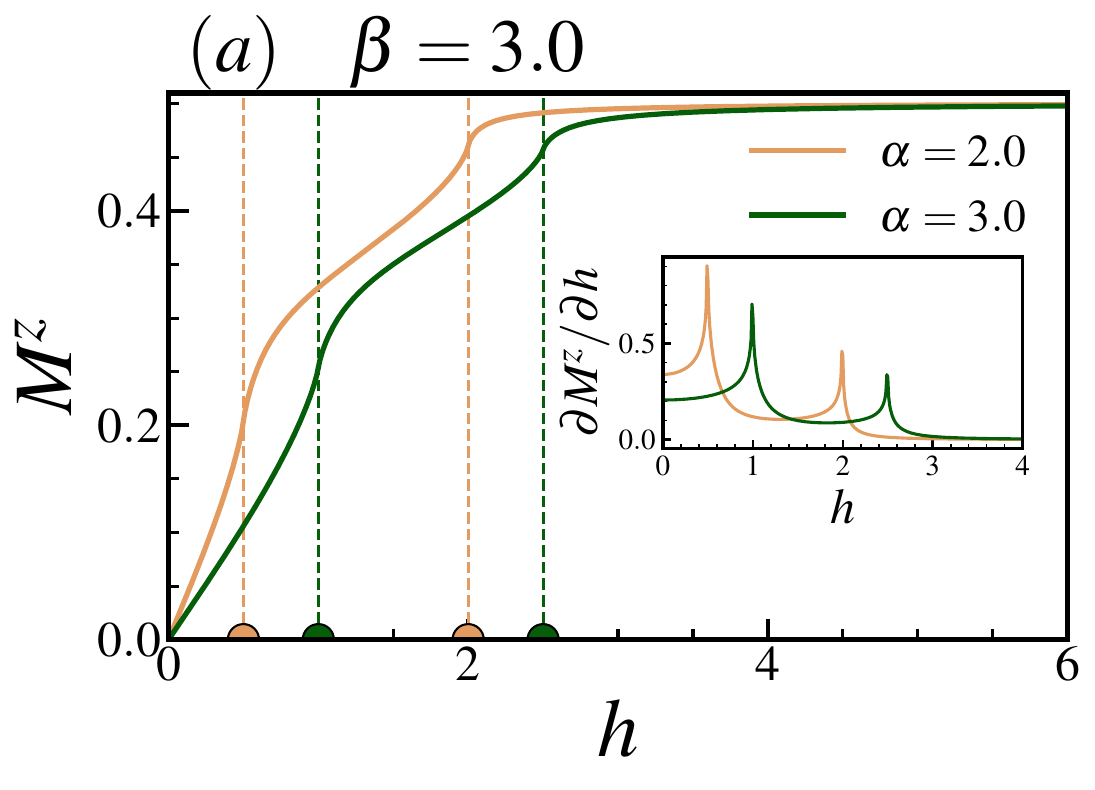} \includegraphics[width=0.32\linewidth,height=0.24\linewidth]{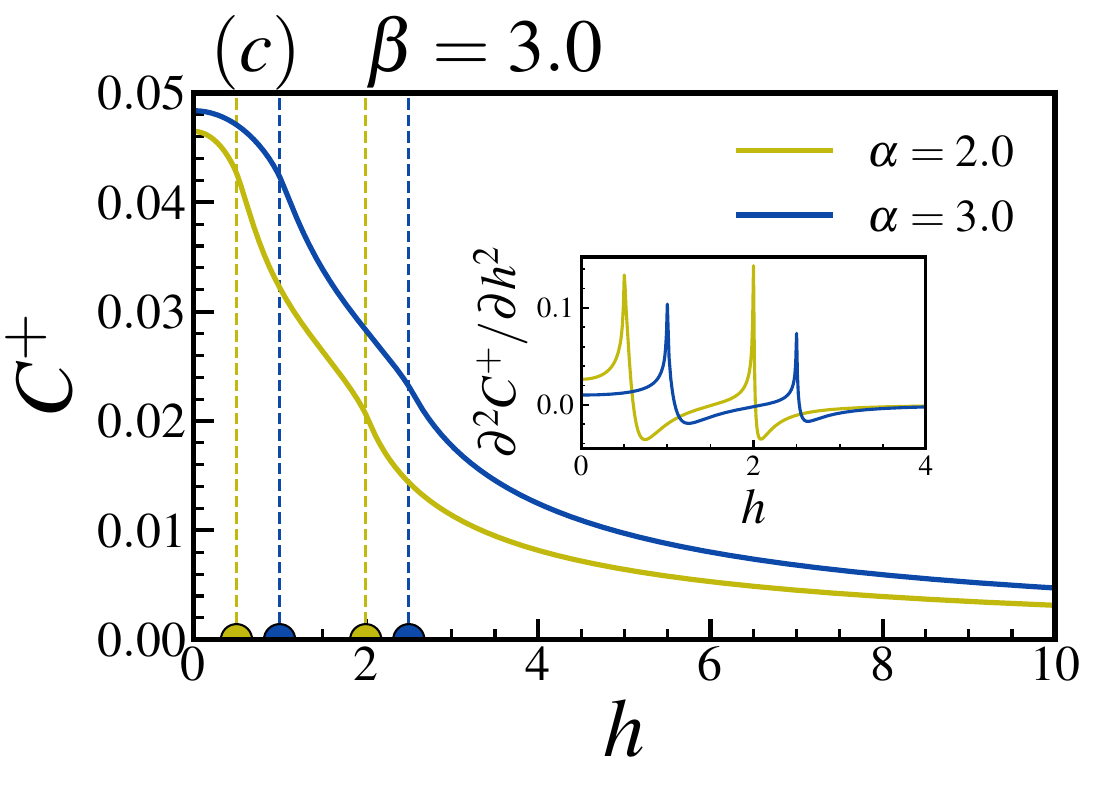} 
		\includegraphics[width=0.32\linewidth,height=0.24\linewidth]{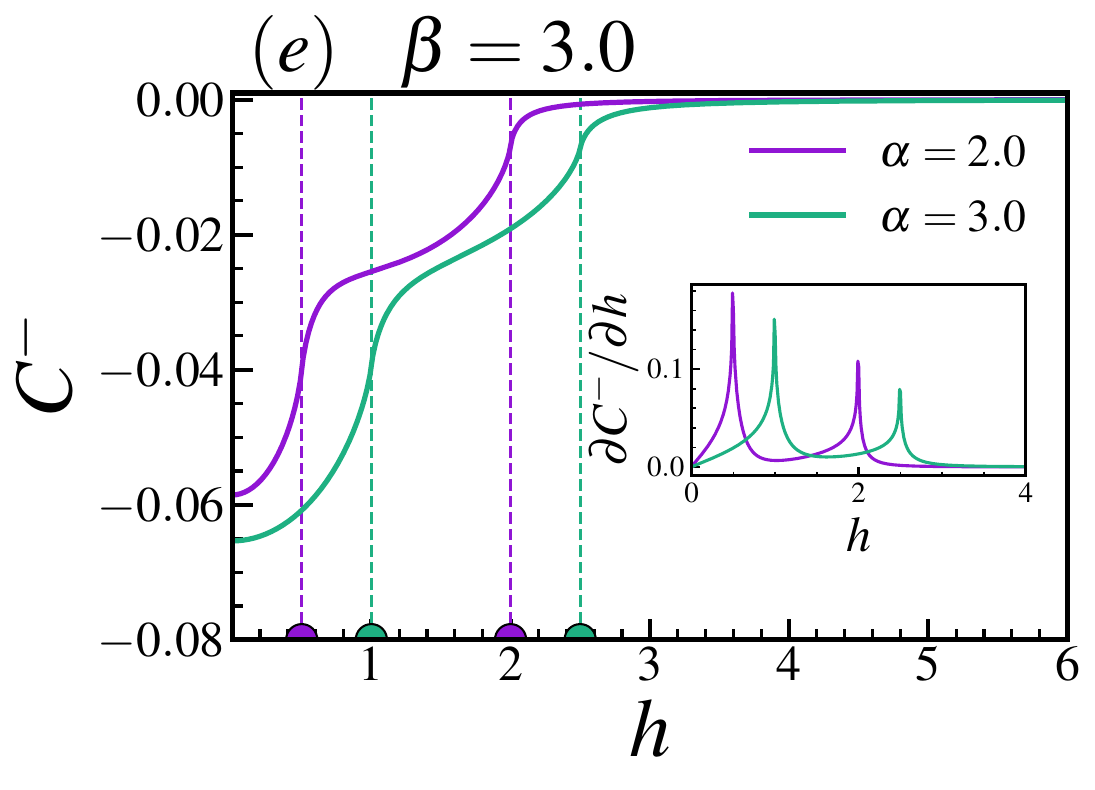}
	}
	\centerline{\includegraphics[width=0.32\linewidth,height=0.24\linewidth]{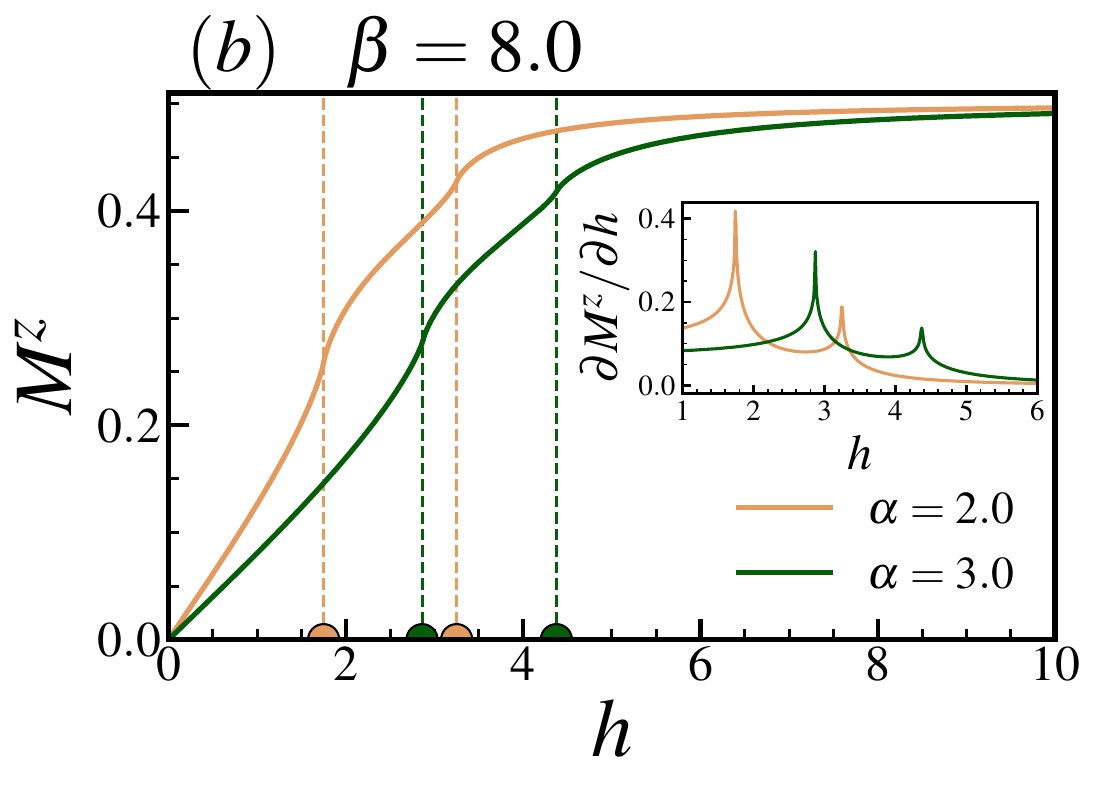} \includegraphics[width=0.32\linewidth,height=0.24\linewidth]{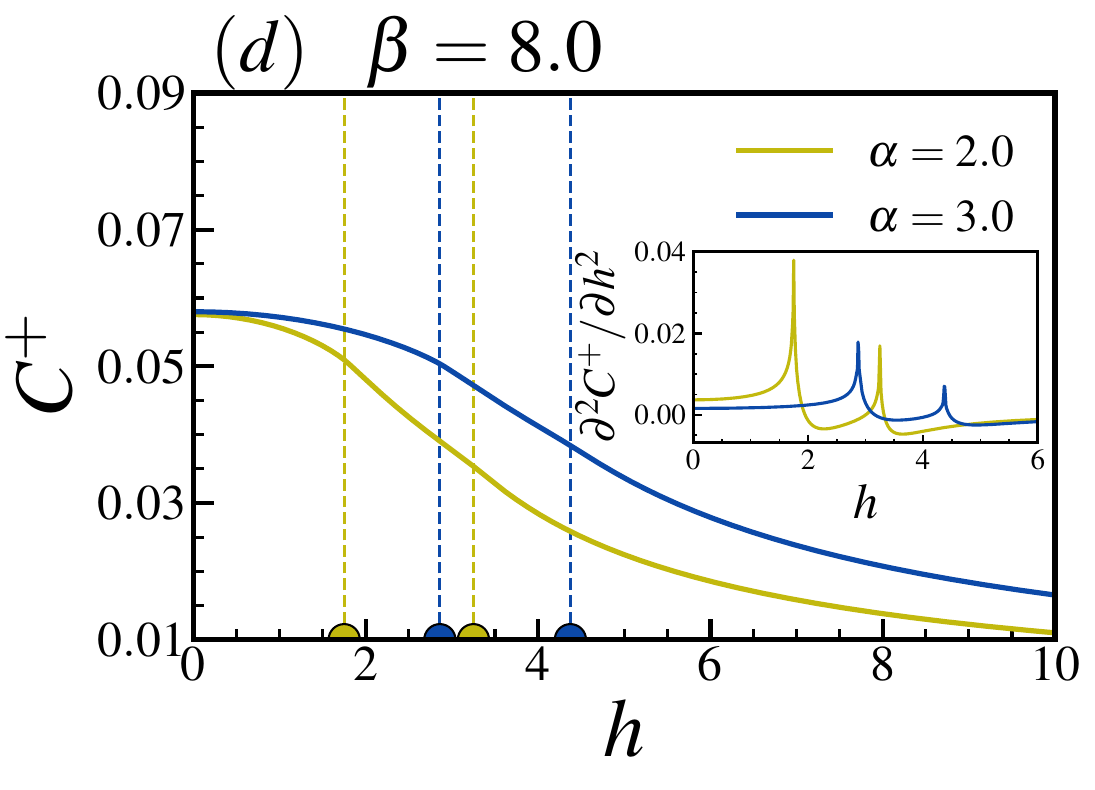} 
		\includegraphics[width=0.32\linewidth,height=0.24\linewidth]{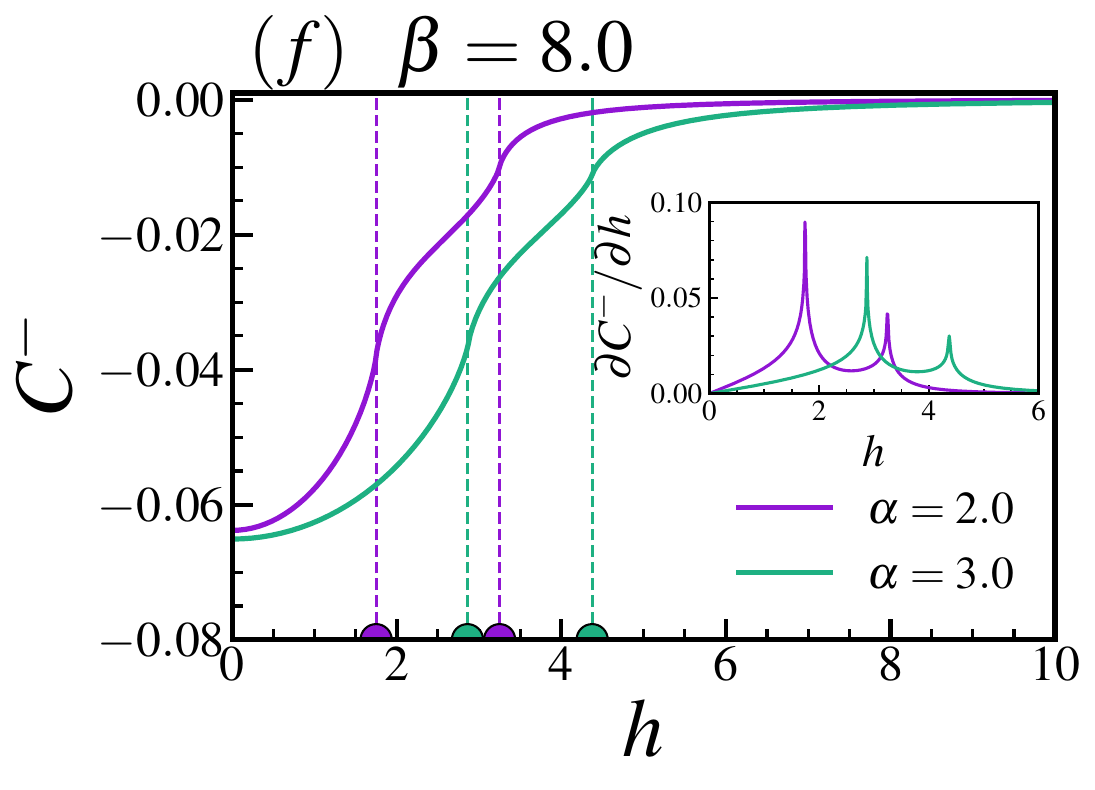}
	}

	\caption{ Magnetization in panels (a) and (b), nematic order parameters in panels (c) and (d), and chiral order parameters in panels (e) and (f) as functions of the magnetic field $h$. The upper panels correspond to $\beta=3.0$, while the lower panels are obtained for $\beta=8.0$. All results are calculated for a chain of size $N=1000$.
	}
	\label{Fig6}
	
\end{figure*}

%%%%%%%%%%%%%%%%%%%%%%%%%%%%%%%%%%%%%%%%%%%
We next turn to the regime $\beta>0.0$, where the behavior of the magnetization and cluster order parameters is shown in Fig.~\ref{Fig6} for two representative values, $\beta=3.0$ (upper panels) and $\beta=8.0$ (lower panels). 

Figures~\ref{Fig6}(a) and \ref{Fig6}(b) show the magnetization as a function of the transverse field. In both cases, $M^z$ increases monotonically with $h$, reflecting the alignment of spins along the field direction. Beyond the second critical field $h_{c_2}$, the system approaches a fully polarized state and the magnetization saturates. The corresponding field derivative, displayed in the accompanying panels, exhibits pronounced maxima at $h_{c_2}$ and $h_{c_4}$, indicating continuous phase transitions in the system.

\begin{figure*}
	\centerline{\includegraphics[width=0.32\linewidth,height=0.24\linewidth]{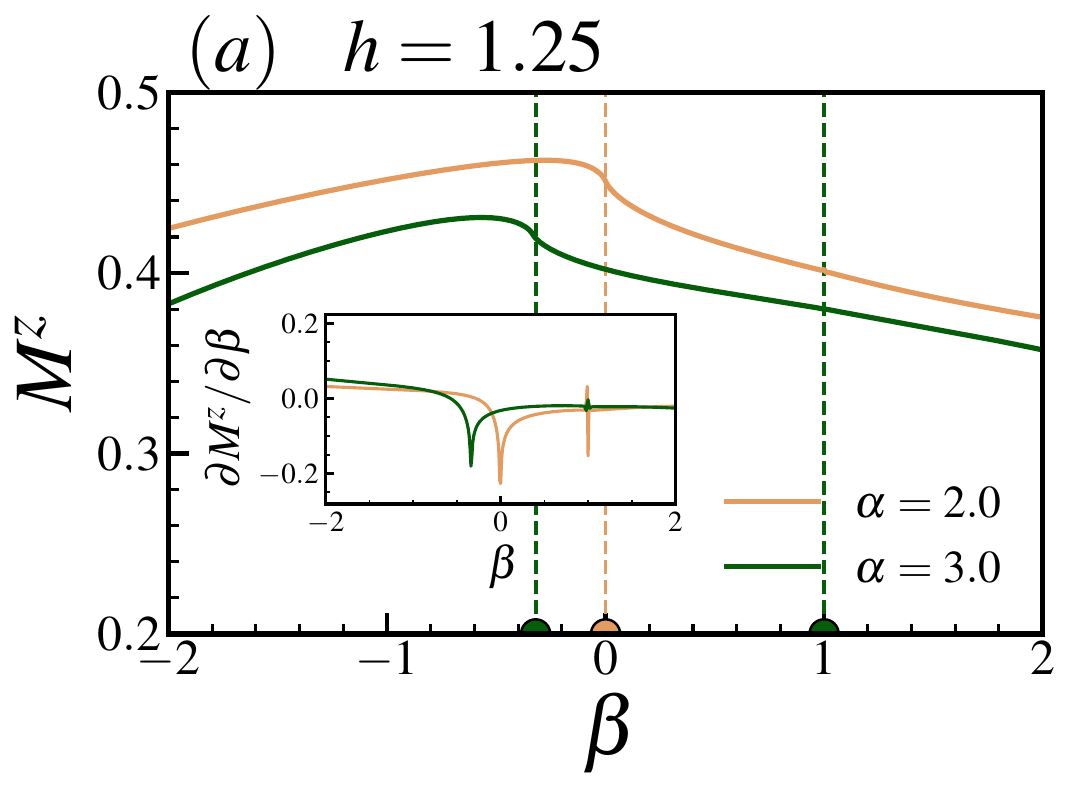} \includegraphics[width=0.32\linewidth,height=0.24\linewidth]{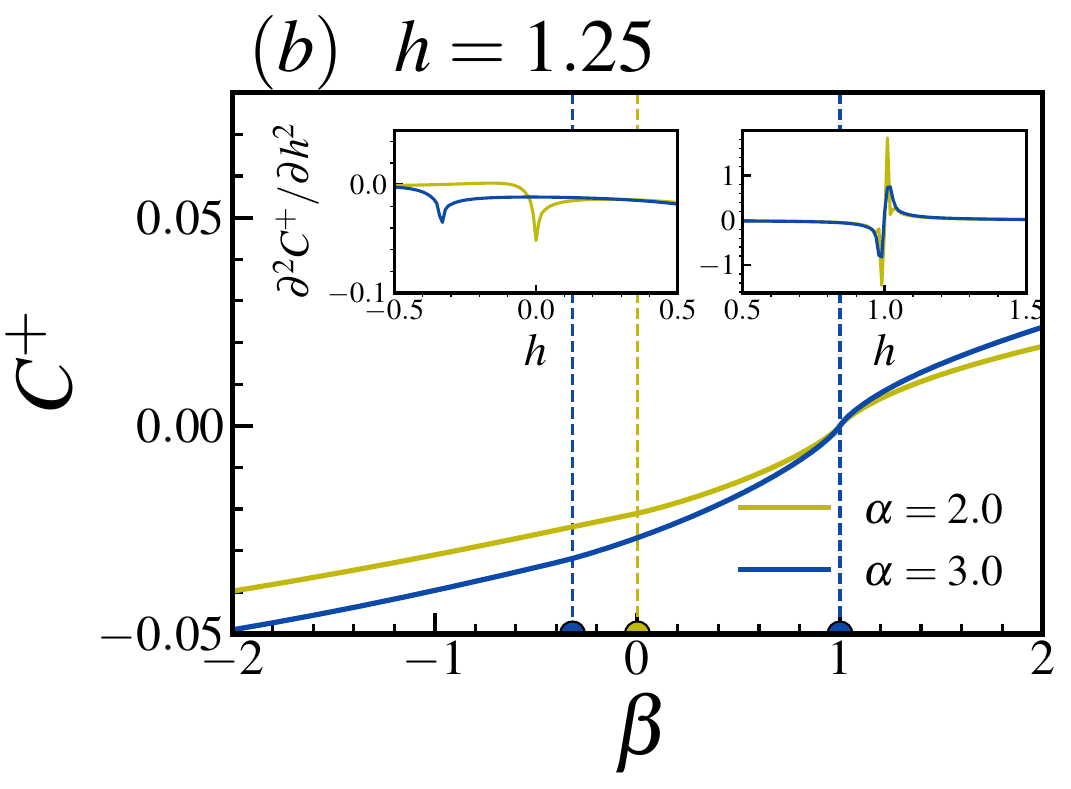} 
		\includegraphics[width=0.32\linewidth,height=0.24\linewidth]{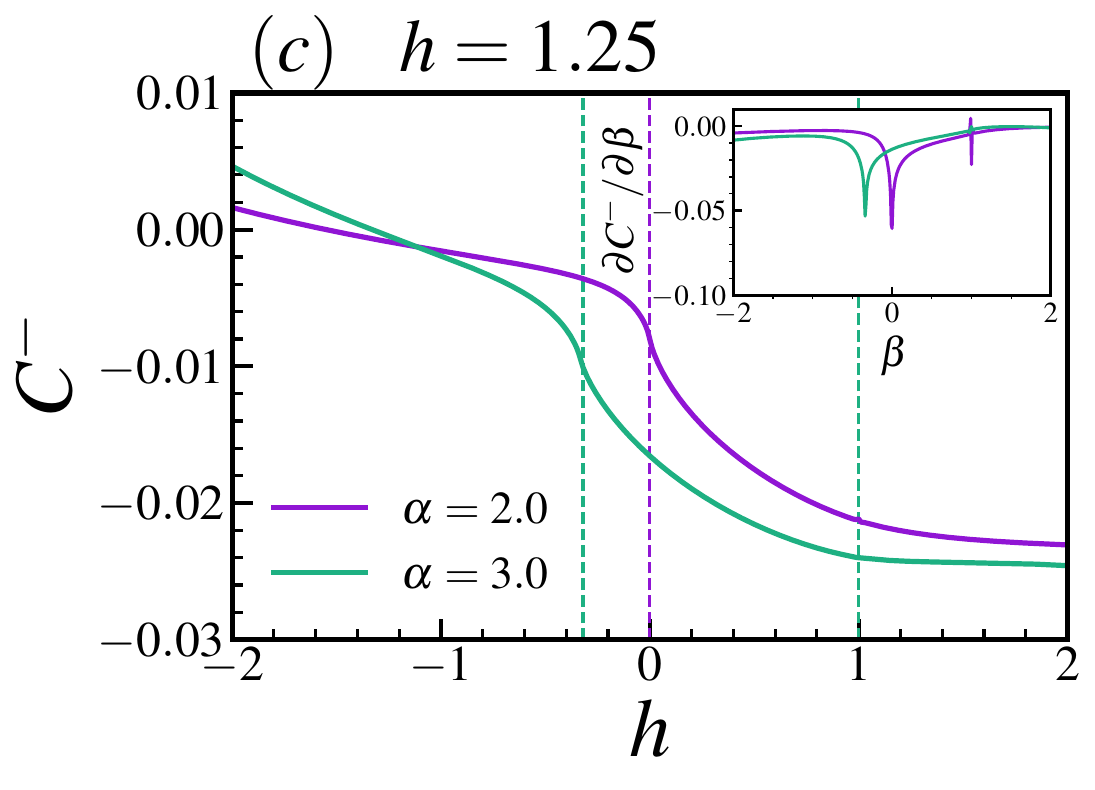}
	}

	\caption{Magnetization in panel (a), nematic order parameter in panel (b), and chiral order parameter in panel (c) as functions of  $\beta$. All results are calculated for a chain of size $N=1000$ and $h=1.25$.
	}
	\label{Fig7}
	
\end{figure*}

\begin{figure*}
	\centerline{\includegraphics[width=0.25\linewidth,height=0.2\linewidth]{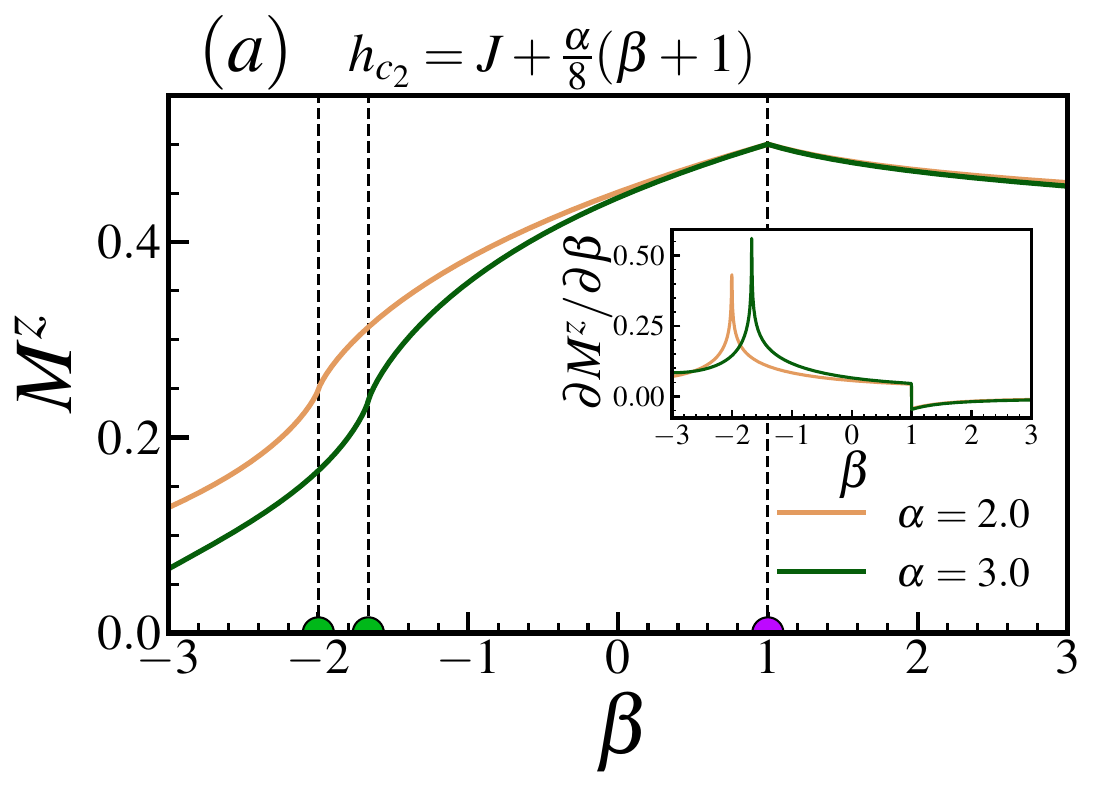} \includegraphics[width=0.25\linewidth,height=0.2\linewidth]{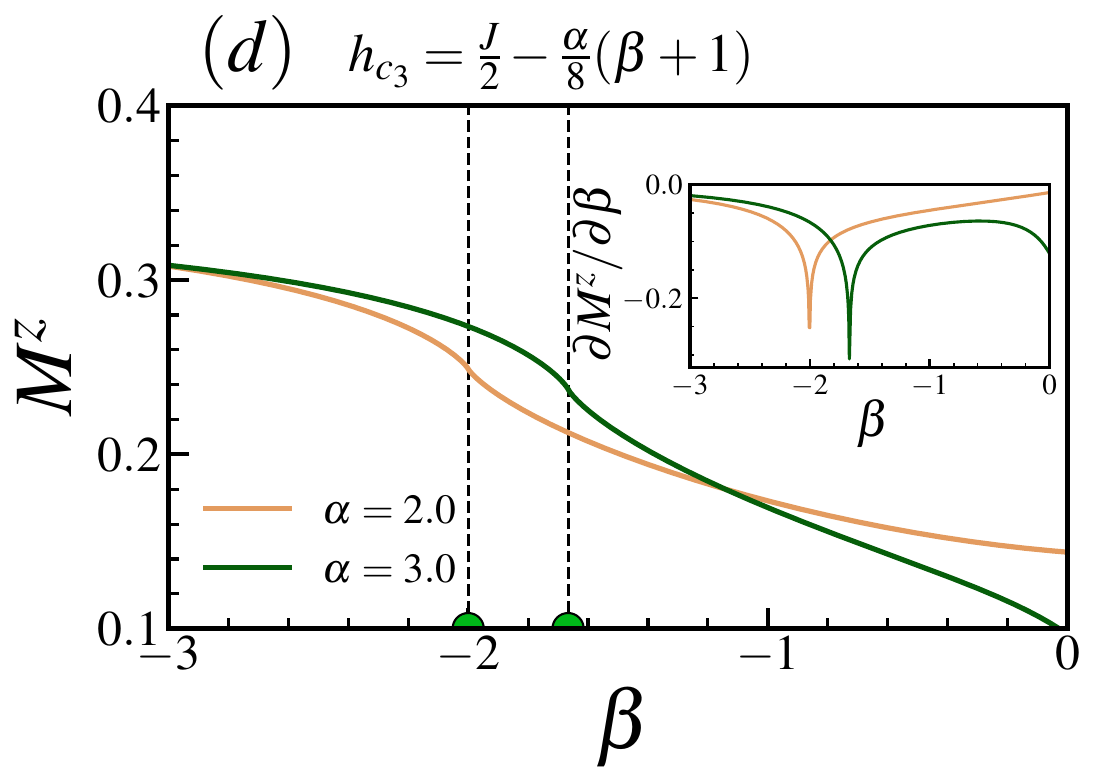} 
		\includegraphics[width=0.25\linewidth,height=0.2\linewidth]{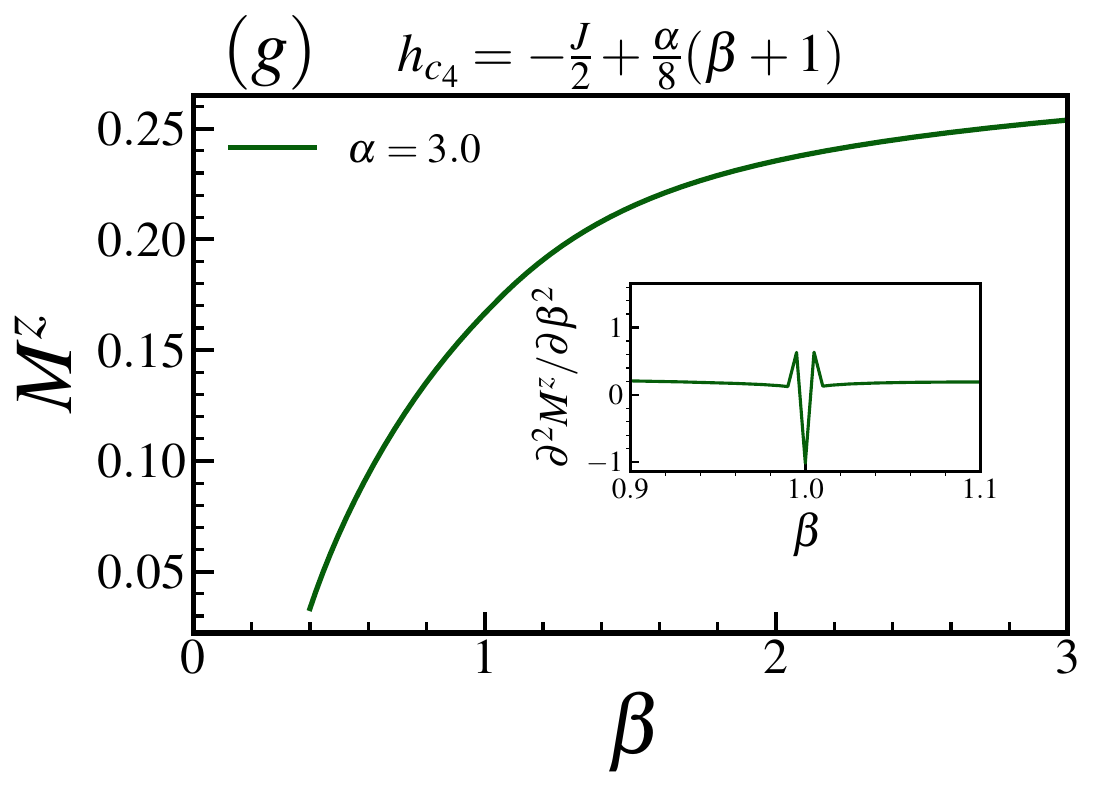}
		\includegraphics[width=0.25\linewidth,height=0.2\linewidth]{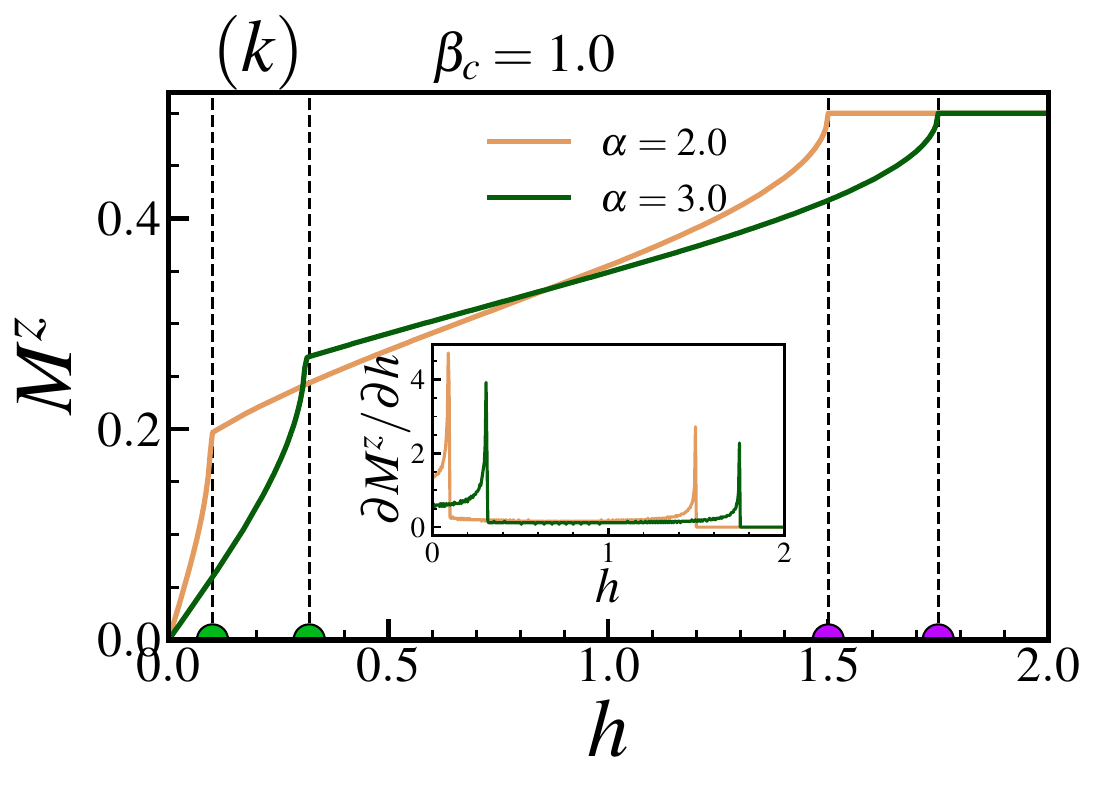}
	}
	
	\centerline{\includegraphics[width=0.25\linewidth,height=0.2\linewidth]{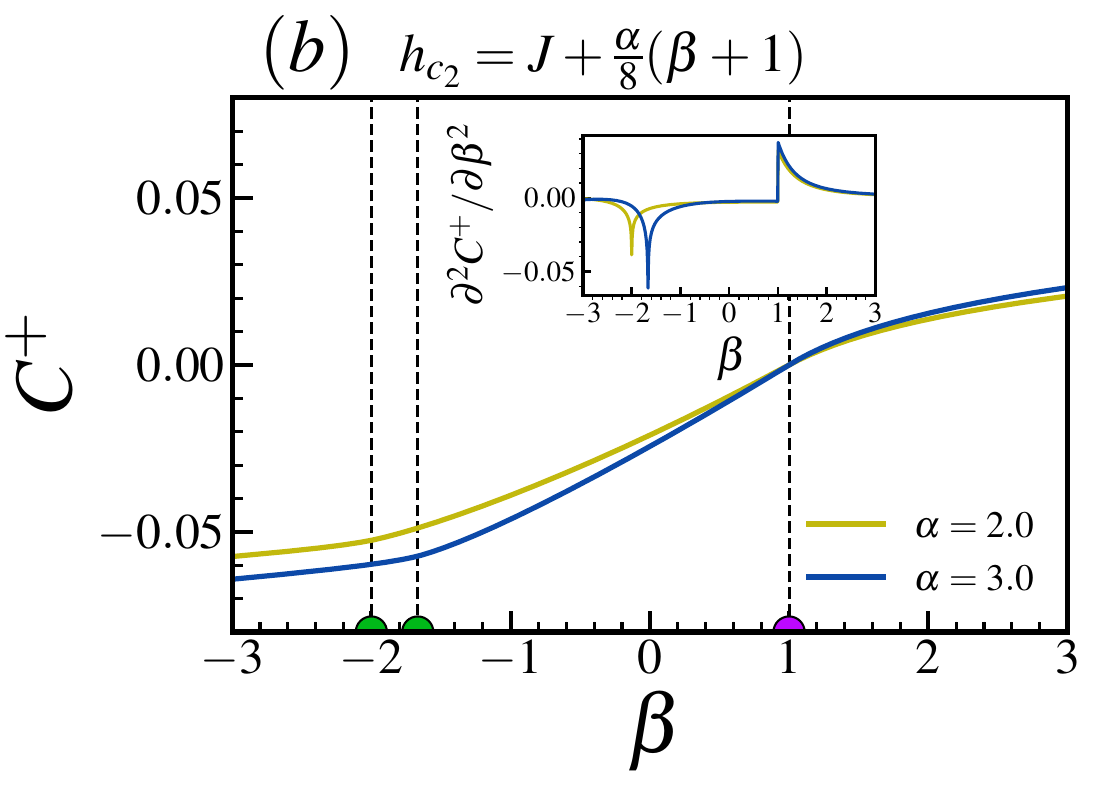} \includegraphics[width=0.25\linewidth,height=0.2\linewidth]{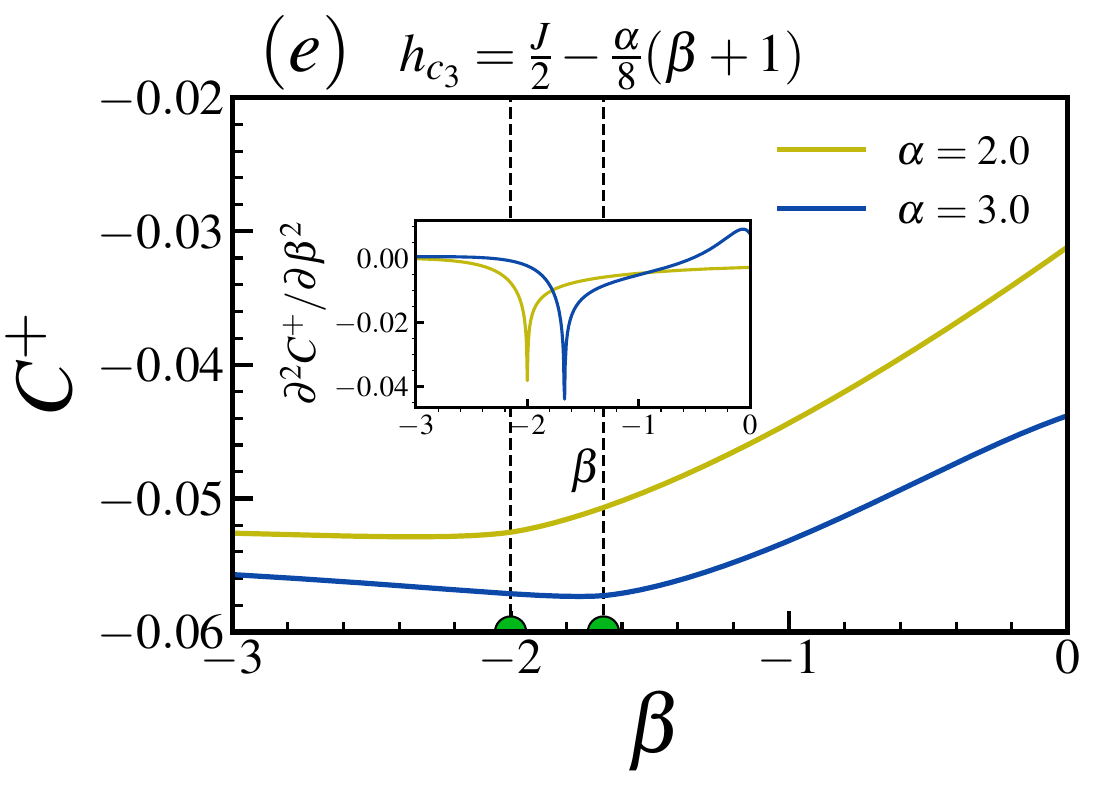} 
		\includegraphics[width=0.25\linewidth,height=0.2\linewidth]{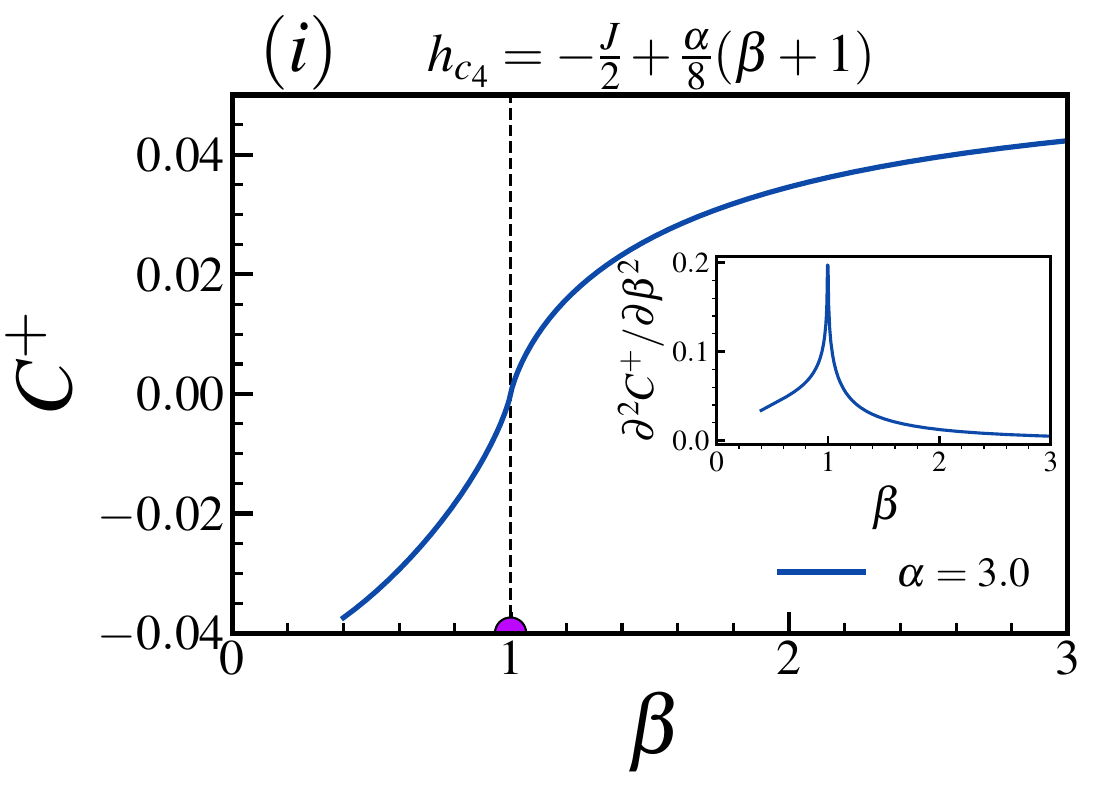}
		\includegraphics[width=0.25\linewidth,height=0.2\linewidth]{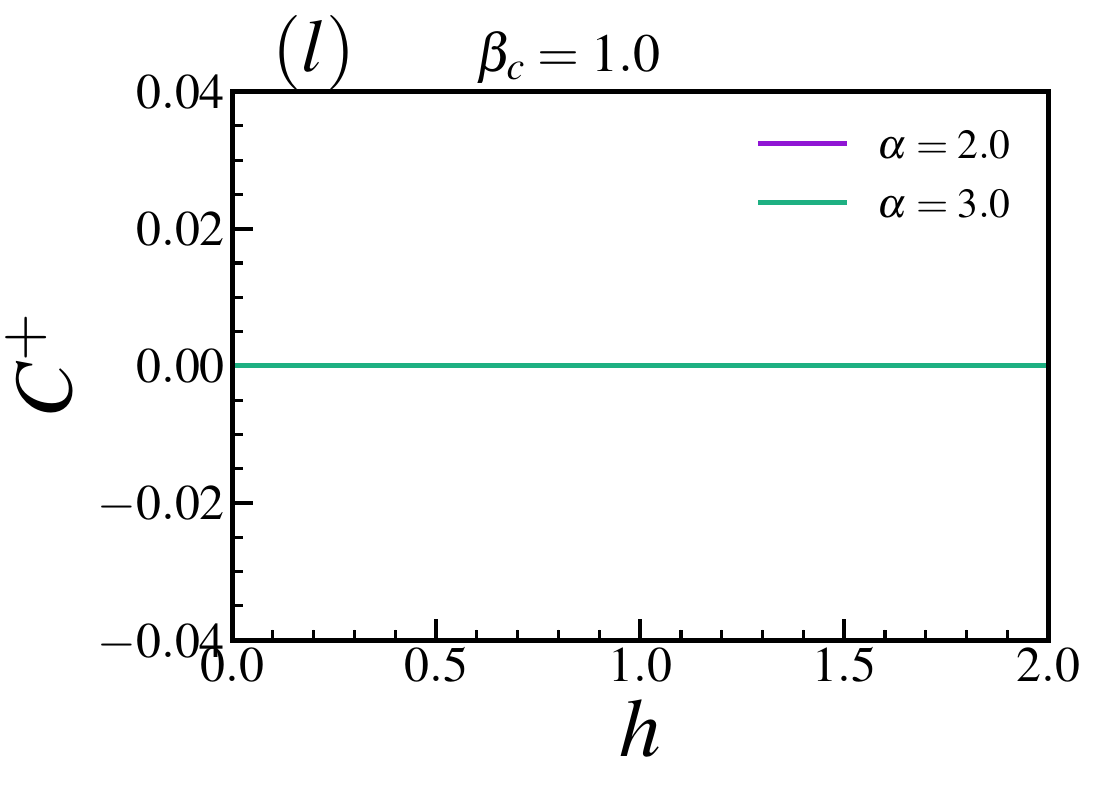}
	}
	
		\centerline{\includegraphics[width=0.25\linewidth,height=0.2\linewidth]{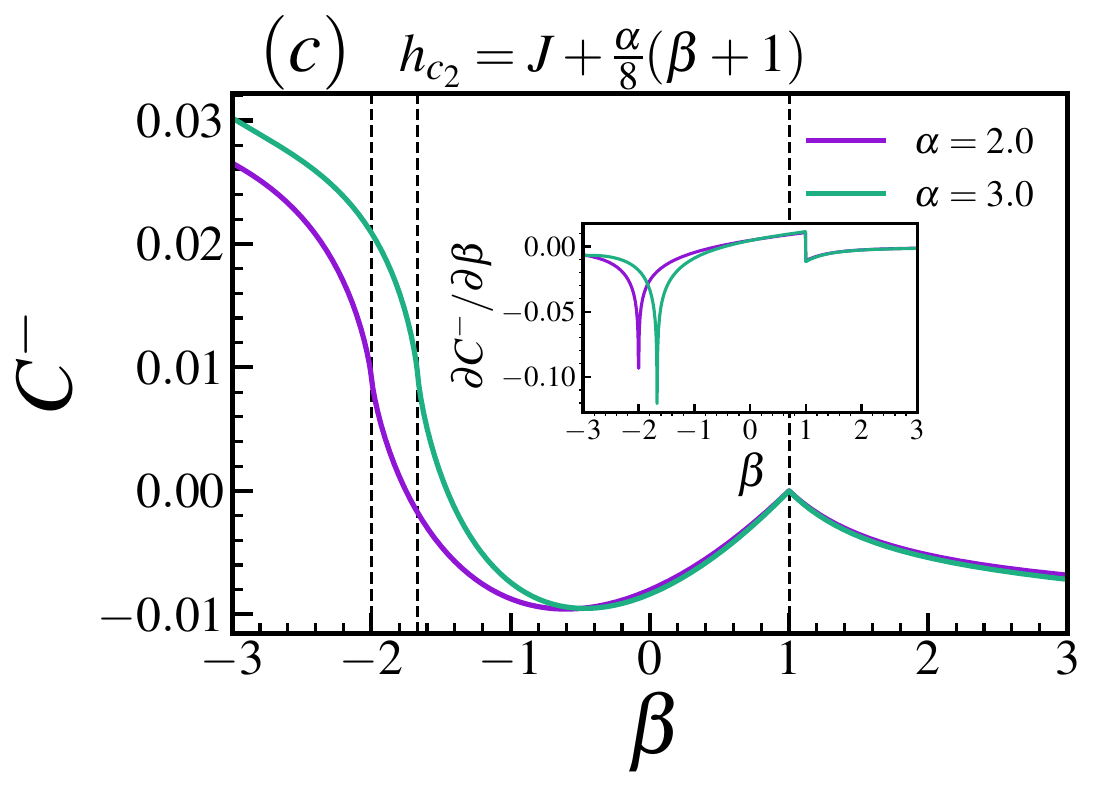} \includegraphics[width=0.25\linewidth,height=0.2\linewidth]{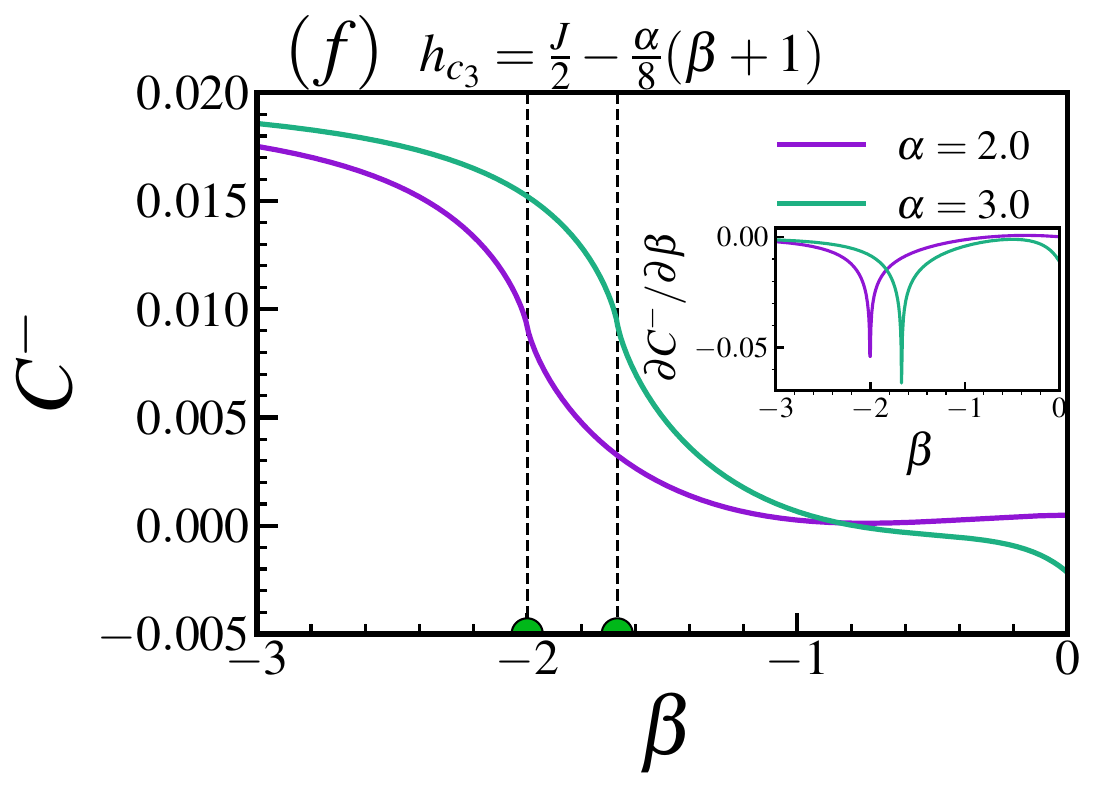} 
		\includegraphics[width=0.25\linewidth,height=0.2\linewidth]{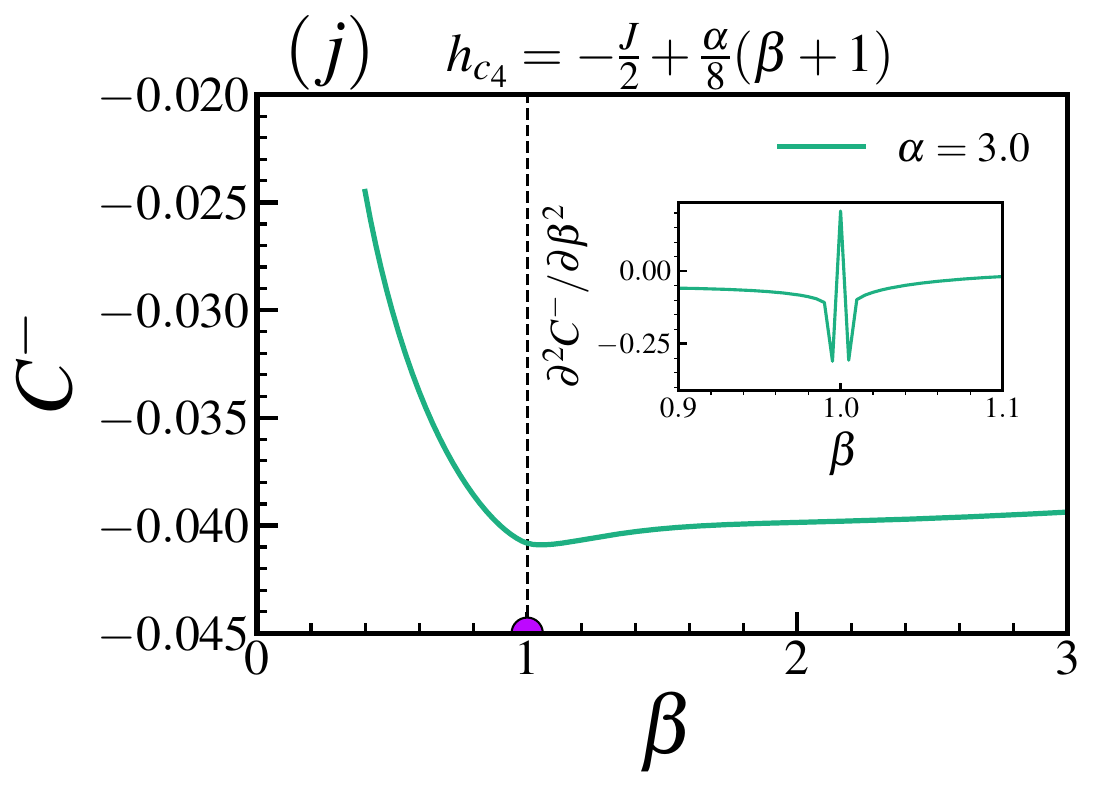}
		\includegraphics[width=0.25\linewidth,height=0.2\linewidth]{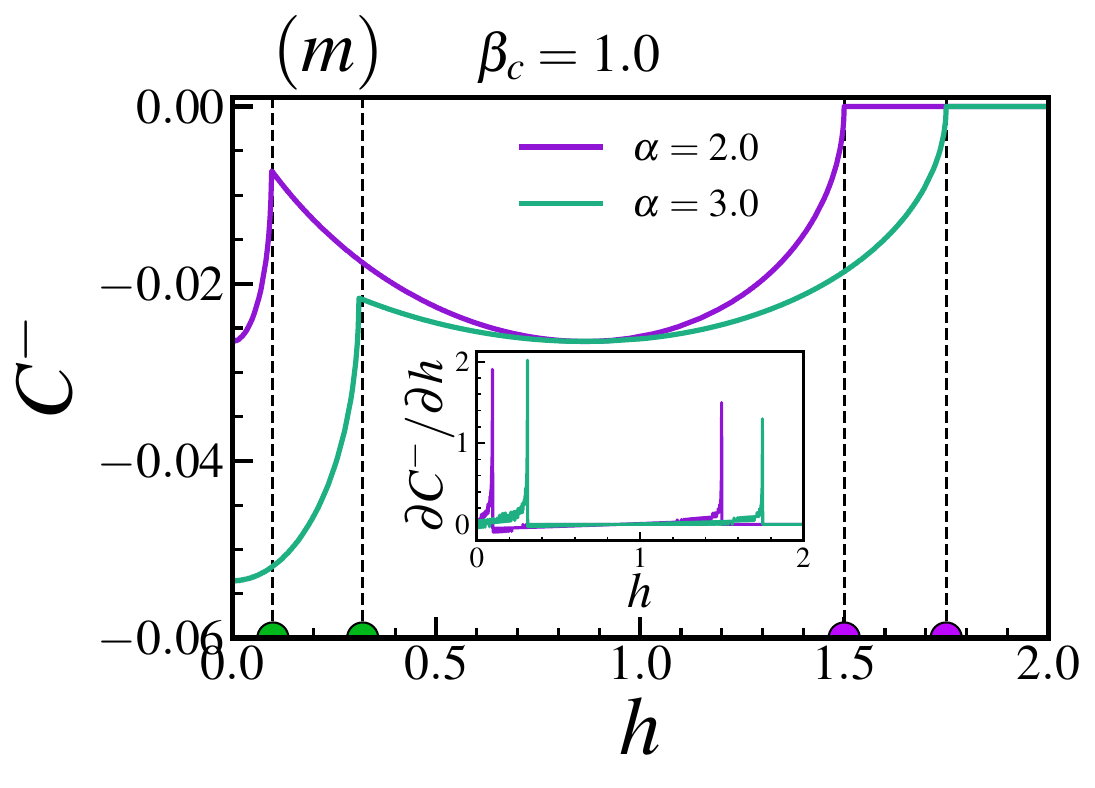}
	}

	\caption{ Magnetization (top row),  nematic order parameter  (middle row), and chiral order parameter  (bottom row) for the critical lines $h_{c_2}$, $h_{c_3}$, $h_{c_4}$, and $\beta_{c}$ shown in the first, second, third, and fourth columns, respectively.
	}
	\label{Fig8}
	
\end{figure*}

The evolution of the nematic order parameter $C^{+}$ is presented in Figs.~\ref{Fig6}(c) and \ref{Fig6}(d). For both values of $\beta$, $C^{+}$ is finite at zero field, indicating chiral correlations in the ground state. As the field is increased, these correlations are gradually suppressed, and $C^{+}$ vanishes at sufficiently large $h$, signaling the disappearance of nematic order. The corresponding second derivative with respect to the field develops sharp peaks at $h_{c_2}$ and $h_{c_4}$, identifying the critical points. As in the $\beta<0.0$ regime, the absence of singular features in lower-order derivatives and the emergence of nonanalytic behavior at higher order indicate that the transition is more clearly resolved at the level of the second derivative.

The chiral order parameter $C^{-}$, shown in Figs.~\ref{Fig6}(e) and \ref{Fig6}(f), displays a qualitatively similar trend. It takes a finite value at zero field and decreases continuously as $h$ increases, eventually vanishing in the high-field regime. The field derivative of $C^{-}$, shown alongside the main curves, exhibits distinct peaks at $h_{c_2}$ and $h_{c_4}$, again marking the transition points.

Overall, the behavior in the $\beta>0.0$ region closely parallels that observed for $\beta<0.0$, with the magnetization capturing the global polarization. At the same time, the cluster order parameters resolve the suppression of chiral and nematic correlations and consistently identify the critical fields across the regime.

%%%%%%%%%%%%%%%%%%%%%%%%%%%%%%%5

 To further examine the phase boundaries, and in particular the vertical critical line at $\beta_c=1.0$, we analyze the behavior of observables as a function of the anisotropy parameter at fixed field. For this purpose, we consider a representative field value $h=1.25$ and study the magnetization and cluster order parameters as functions of $\beta$, as shown in Fig.~\ref{Fig7}.

The magnetization, nematic, and chirality order parameters all exhibit clear changes across the phase boundaries as $\beta$ is varied. These features are more clearly seen in the corresponding derivatives with respect to $\beta$, displayed alongside the main curves, where pronounced extrema indicate the transition points. In particular, while most transition points depend on both $\alpha$ and $\beta$, a distinct behavior is observed at $\beta=1.0$. At this point, the extrema of the derivatives for different values of $\alpha$ coincide, indicating that the transition line is independent of the interaction strength. This observation is consistent with the behavior of the second derivative of the ground-state energy and provides independent confirmation of the vertical critical line at $\beta_c=1.0$.

This analysis complements the field-dependent results by directly probing the anisotropy-driven transitions and confirms the robustness of the $\beta_c=1.0$ boundary across the parameter space.
\\
To complement the spectral and ground-state-energy analysis, we examine the behavior of the magnetization, nematic, and chiral order parameters directly along the critical branches. Figure~\ref{Fig8} shows these quantities evaluated for representative values $\alpha=2$ and $\alpha=3$, together with their derivatives, which provide enhanced sensitivity to the underlying critical behavior.

Along the intersecting branches $h_{c_2}$, shown in Figs.~\ref{Fig8}(a)–\ref{Fig8}(c), and $h_{c_3}$, shown in Figs.~\ref{Fig8}(d)–\ref{Fig8}(f), the derivatives of all three observables develop pronounced extrema at the special point $(\beta_m,h_m)$, providing independent confirmation of the unconventional Lifshitz-type multicriticality identified from the momentum-space analysis. The same branches also exhibit additional signatures upon approaching the isotropic limit $\beta=1$, consistent with the special role of the symmetry-preserving line discussed previously. Although the magnitude of the chiral order parameter remains small, its derivative retains a clear critical signal, indicating that the reconstruction of the low-energy spectrum leaves detectable fingerprints even in weak local correlations. A different behavior is observed along the branch $h_{c_4}$, shown in Figs.~\ref{Fig8}(g)–\ref{Fig8}(i). In this case, the derivatives of the magnetization, nematic, and chiral order parameters exhibit clear discontinuities upon reaching $\beta=1$, reflecting the unconventional Lifshitz-type transition associated with the isotropic limit.

The critical line $\beta_c=1$, shown in Figs.~\ref{Fig8}(j)–\ref{Fig8}(l), further reinforces this picture. Along this branch, the derivatives of the magnetization display two clear signatures corresponding to the two distinct transitions identified in the isotropic regime: the conventional Lifshitz transition and the gapless-to-gapped transition. While the nematic order parameter vanishes identically at $\beta=1$, the derivative of the chiral correlation continues to exhibit pronounced anomalies at the same transition points.

The combined behavior of these observables demonstrates that local observables and their derivatives provide a robust and complementary probe of the quantum critical structure, capable of detecting not only conventional phase boundaries but also gapless-to-gapless and Lifshitz-type transitions driven by changes in the momentum-space organization of the low-energy spectrum.

\subsection{Topological phase transitions}

Topological phases of matter represent a paradigm shift beyond the conventional Landau framework \cite{landau1937theory}, as they cannot be characterized by local order parameters or spontaneous symmetry breaking \cite{landau1937theory,hasan2011three}. Although parts of the phase diagram can be understood from spectral properties and conventional quantum critical behavior, such a description is not sufficient to fully classify the ground-state phases.  In particular, distinct gapped phases may possess identical symmetries while remaining topologically inequivalent \cite{hasan2011three,wen2013topological,ryu2010topological}. This limitation motivates a topological characterization in terms of global invariants of the ground-state wavefunction rather than symmetry-breaking mechanisms \cite{hasan2011three,wen2013topological,ryu2010topological}. In one-dimensional systems, topological phases are classified by quantized invariants that remain unchanged under continuous deformations of the Hamiltonian, provided that the bulk energy gap stays finite \cite{hasan2011three,ryu2010topological}. Consequently, phases that are indistinguishable within the Landau framework may belong to fundamentally different topological sectors \cite{ryu2010topological}. The robustness of these phases originates from the quantization of the invariant, which prevents any continuous evolution between distinct topological states without a closing of the bulk gap \cite{ryu2010topological,qi2011topological}.

A central consequence of this framework is that topological phase transitions can occur only at quantum critical points where the bulk energy gap closes \cite{hasan2011three,ryu2010topological}. At these critical points, the topological invariant becomes ill-defined, allowing the system to undergo a discontinuous change between distinct topological sectors. In open geometries, such bulk transitions are accompanied by the appearance or disappearance of symmetry-protected zero-energy boundary states localized at the edges of the chain, establishing the bulk–boundary correspondence \cite{hatsugai1993chern,hasan2010colloquium}. In Kitaev-type realizations, the topological invariant is directly related to the number of Majorana zero modes supported at the boundaries \cite{kitaev2001unpaired}. More generally, extended interactions may stabilize higher winding numbers associated with multiple edge states, enriching the topological structure of the phase diagram \cite{mahdavifar2026topological}.

\begin{figure*}
	\centerline{\includegraphics[width=0.25\linewidth,height=0.2\linewidth]{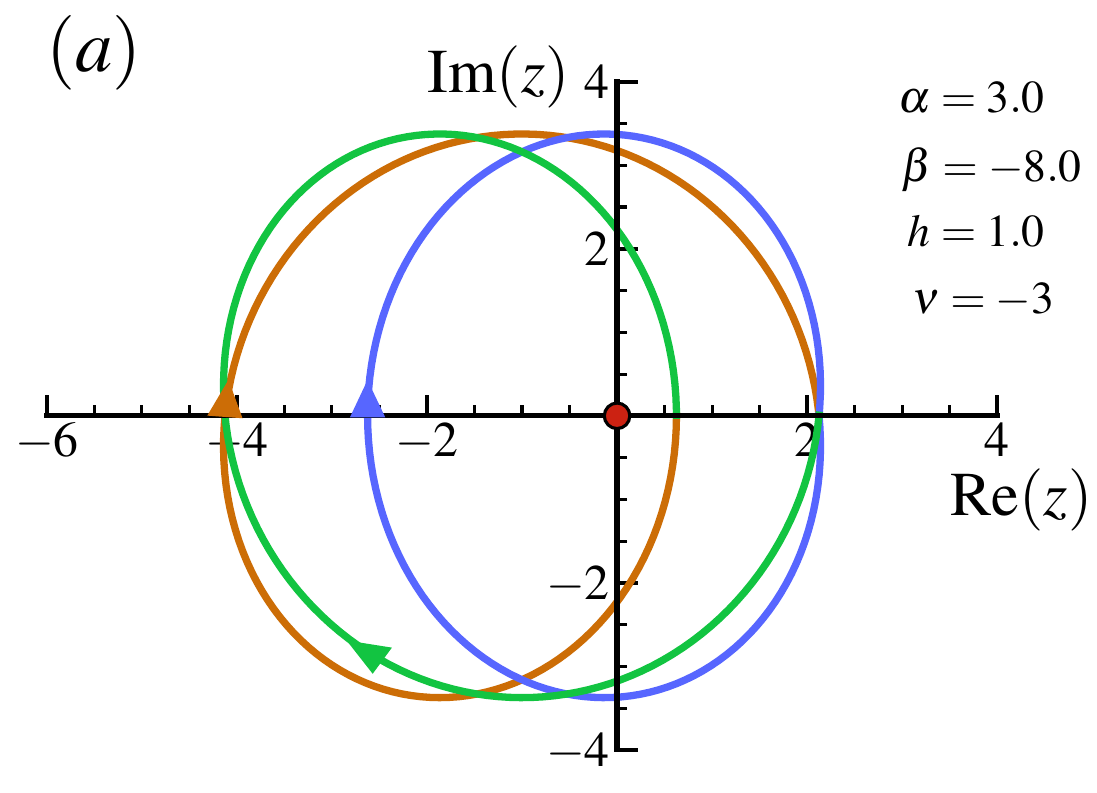} \includegraphics[width=0.25\linewidth,height=0.2\linewidth]{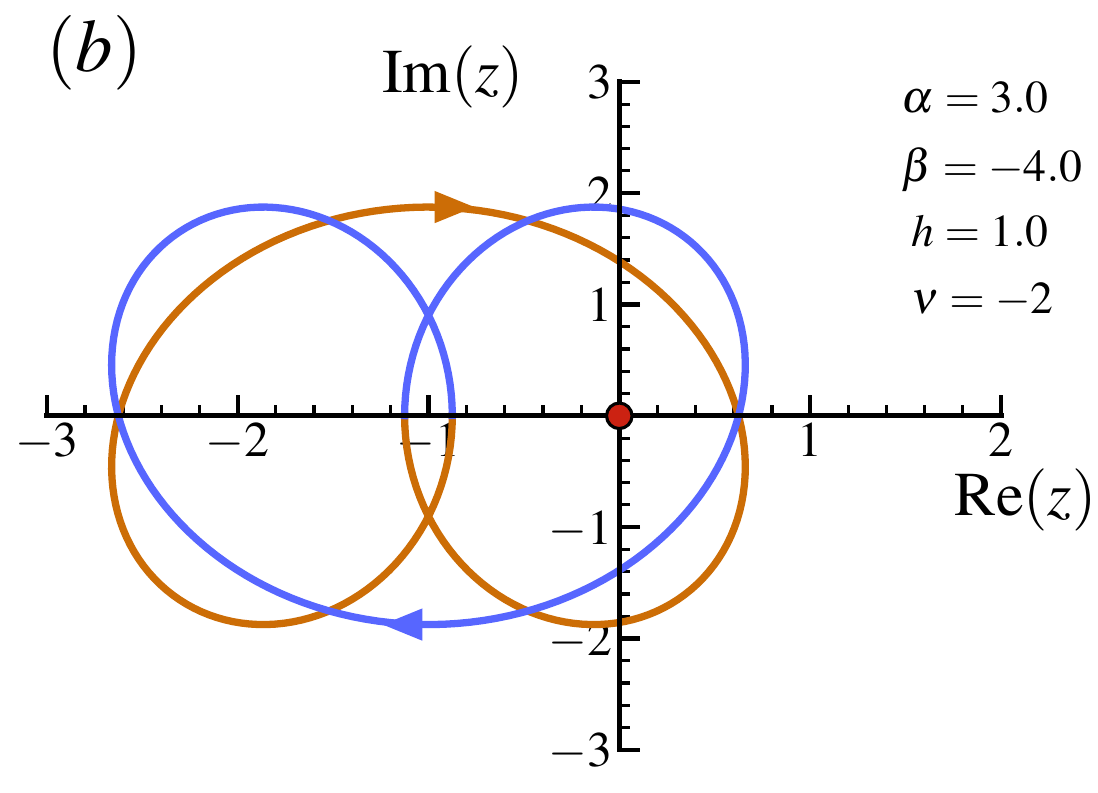} 
		\includegraphics[width=0.25\linewidth,height=0.2\linewidth]{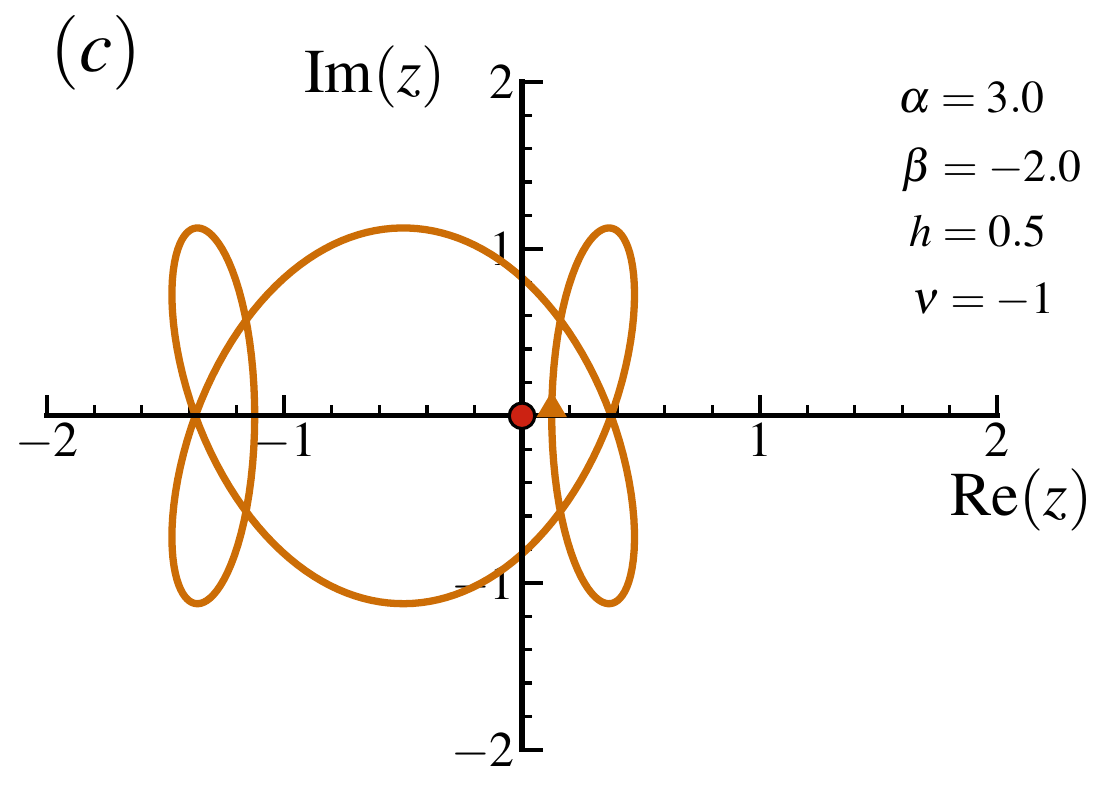}
		\includegraphics[width=0.25\linewidth,height=0.2\linewidth]{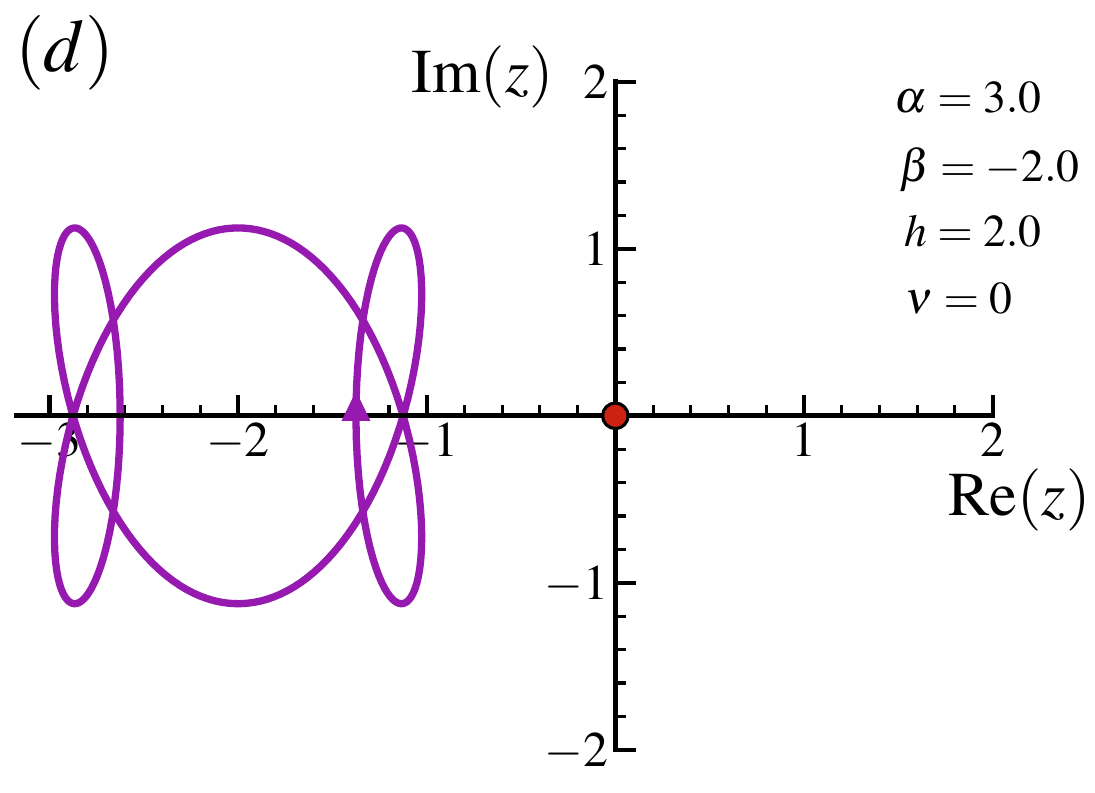}
	}
	
	\centerline{\includegraphics[width=0.25\linewidth,height=0.2\linewidth]{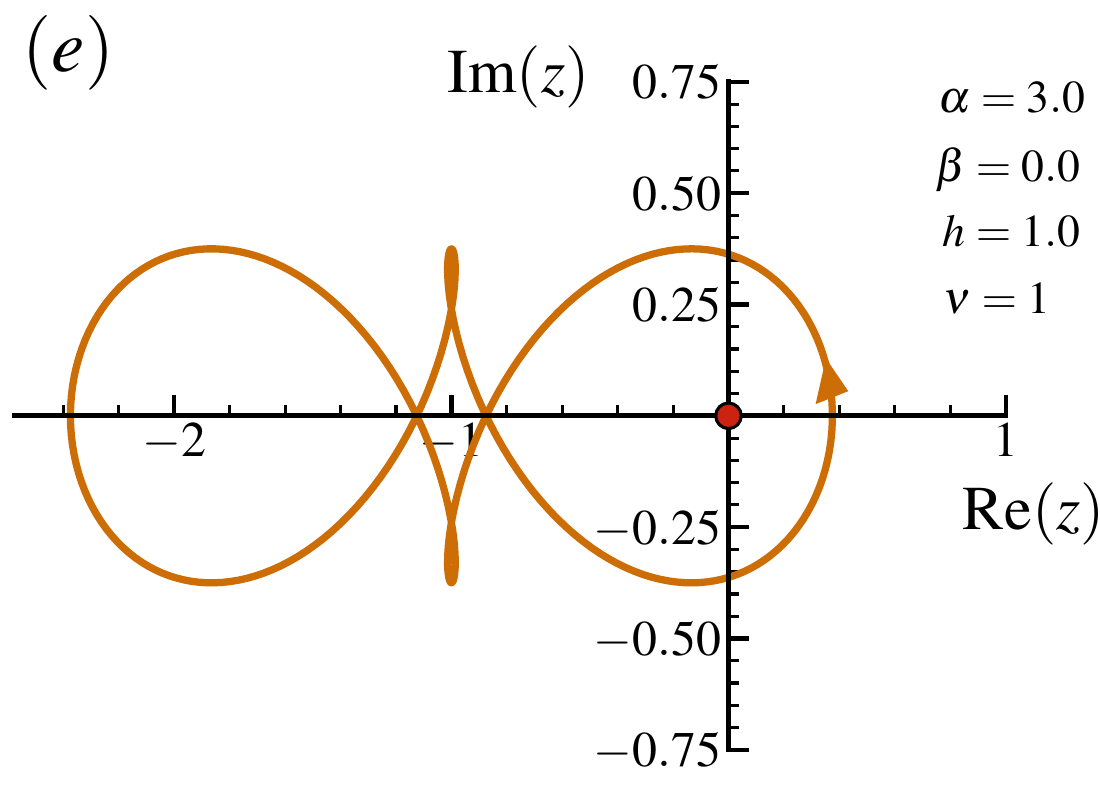} \includegraphics[width=0.25\linewidth,height=0.2\linewidth]{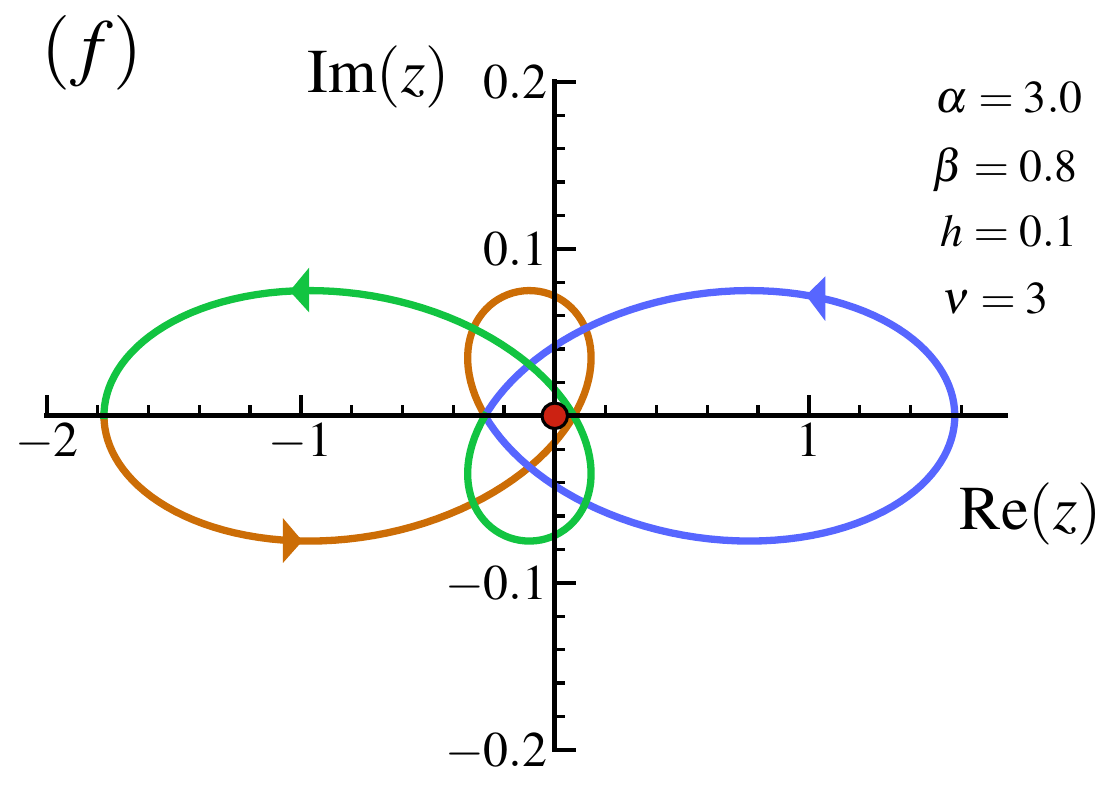} 
		\includegraphics[width=0.25\linewidth,height=0.2\linewidth]{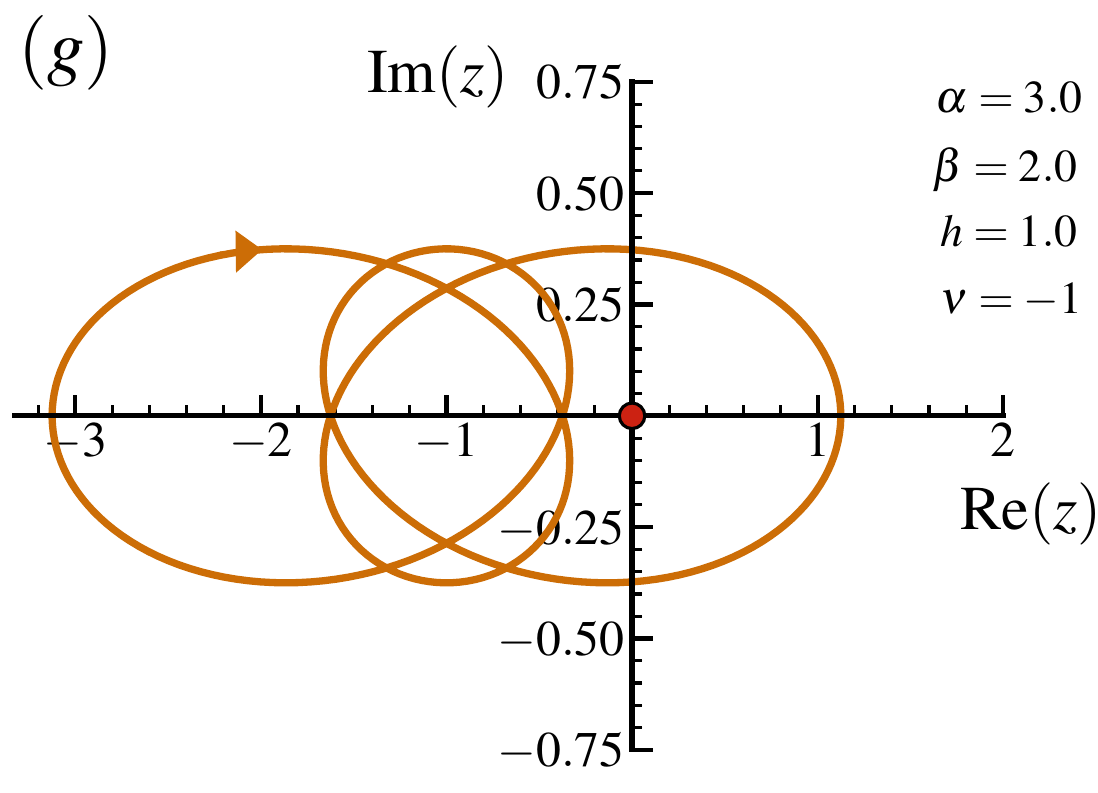}
		\includegraphics[width=0.25\linewidth,height=0.2\linewidth]{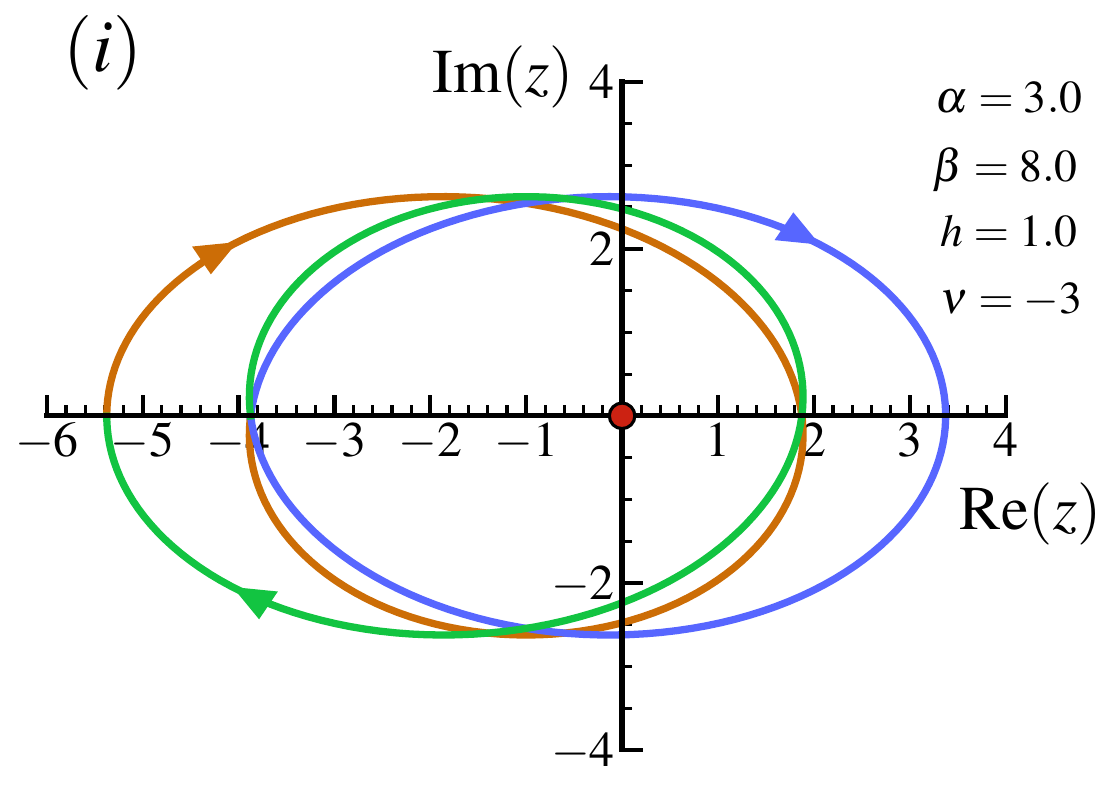}
	}

	\caption{ The complex trajectory $z(k)=A_k+iB_k$ for representative points in the phase diagram. All results are obtained for $\alpha=3$ and a chain of size $N=1000$.
	}
	\label{Fig9}
	
\end{figure*}

To make the topological characterization explicit, we now consider the momentum-space representation of the fermionic Hamiltonian. After the Jordan-Wigner and Bogoliubov transformations, the present model can be mapped onto an effective two-band Hamiltonian of the form
\begin{equation}
	\mathcal{H}(k)=\vec d(k)\cdot \vec \sigma=A_k \sigma_x + B_k \sigma_y,
\end{equation}
where $A_k$ and $B_k$ are momentum-dependent real functions determined by the microscopic couplings. This form admits a natural geometrical interpretation through the vector
\begin{equation}
	\vec{d}(k)=(A_k,B_k,0),
\end{equation}
defined over the Brillouin zone. Equivalently, one may introduce the complex function
\begin{equation}
	z(k)=A_k+iB_k,
\end{equation}
whose evolution as $k\in[-\pi,\pi]$ traces a closed trajectory in the complex plane \cite{ryu2010topological}.

Both the spectral and topological properties of the system are encoded in the geometry of this trajectory. The quasiparticle spectrum is determined by the magnitude of the vector $\vec d(k)$,
\begin{equation}
	\varepsilon(k)=\pm |\vec d(k)|=\pm\sqrt{A_k^2+B_k^2},
\end{equation}
such that the energy gap closes whenever $|\vec d(k)|=0$. Consequently, the origin of the complex plane represents a critical point where the bulk spectrum becomes gapless.

The topological character of the phase is determined by how the vector $\vec{d}(k)$ winds around the origin. Distinct phases correspond to topologically inequivalent loops that cannot be continuously deformed into one another without crossing the origin. In this way, the Hamiltonian defines a mapping from the Brillouin zone onto the parameter space traced by $\vec d(k)$, with different winding numbers corresponding to distinct topological phases.

The topological invariant associated with this mapping is the winding number $\nu$, which classifies the distinct ground-state phases of the system. Geometrically, $\nu$ measures how many times the trajectory $z(k)$ encircles the origin as the momentum space the Brillouin zone. It is formally defined as
\begin{align}
	\nu &= \frac{1}{2\pi} \int_{-\pi}^{\pi}
	\frac{\partial \theta(k)}{\partial k}\, dk, \nonumber \\
	\theta(k)&=\arg[z(k)]
	=\arctan\!\left(\frac{B_k}{A_k}\right).
\end{align}

Since the Brillouin zone forms a closed loop, the total accumulated phase is constrained to integer multiples of $2\pi$. As a result, the winding number is quantized to integer values, $\nu\in\mathbb Z$.  Its magnitude determines the number of windings around the origin, while its sign specifies the orientation of the trajectory. Consequently, $\nu \neq 0$ characterizes a topologically nontrivial phase, whereas $\nu = 0$ corresponds to a trivial phase \cite{ryu2010topological,zhang2015topological}.

In numerical calculations, it is often convenient to evaluate the winding number from the total accumulated phase,
\begin{equation}
	\nu =
	\frac{1}{2\pi}
	\left[
	\theta(\pi)-\theta(-\pi)
	\right],
\end{equation}
where $\theta(k)$ is treated as a continuous phase along the Brillouin zone. This expression is fully equivalent to the integral definition in the continuum limit and provides a direct measure of the total phase winding of the trajectory.

A topological phase transition is signaled by a discontinuous change in the winding number, which can occur only when the excitation gap closes at some critical momentum $k_c$, namely when
\begin{equation}
	A_{k_c}=B_{k_c}=0,
\end{equation}
or equivalently,
\begin{equation}
	|\vec d(k_c)|=0.
\end{equation}

At this point, the trajectory crosses the origin in the complex plane, allowing the system to change between distinct topological sectors. As the control parameters are tuned across the critical point, the gap closes and subsequently reopens with a different winding number, signaling a topological phase transition.
After establishing the winding number characterization of the model, we now examine the resulting topological phases numerically. In particular, we investigate the geometric trajectories generated by the complex function $z(k)=A_k+iB_k$ in the complex plane for different values of the control parameters. Because the winding number depends on how these trajectories wind around the origin, the evolution of the loops provides a direct picture of the topological structure of the phase diagram. Distinct topological phases are identified through different winding patterns, while topological phase transitions are signaled by trajectories that cross the origin, where the bulk excitation gap closes. The evolution of these loops connects the spectral properties of the model to its topological classification. In the following, we analyze the representative trajectories shown in Fig.~\ref{Fig9} and relate the corresponding winding number structures to different regions of the phase diagram.

Figure~\ref{Fig9} shows the complex trajectory $z(k)=A_k+iB_k$ for representative points in the phase diagram at fixed $\alpha=3$. We identify multiple topological phases characterized by integer winding numbers $\nu = 0, \pm1, -2, \pm3$. Higher winding numbers arise from the presence of longer-range interactions in the Hamiltonian, leading to a richer topological phase structure beyond the conventional single-winding number. 

Upon varying the control parameters $(\beta,h)$, the topology of the trajectory changes significantly, resulting in distinct winding cases with different values of the topological invariant $\nu$. Panels (a)-(c) correspond to phases with $\nu=-3,-2$, and $-1$, respectively, where the trajectory encircles the origin multiple times in the clockwise direction. These winding patterns correspond to topologically nontrivial phases with different winding multiplicities. In contrast, panel (d) represents the topologically trivial phase $\nu=0$, where the trajectory does not enclose the origin, indicating the absence of topological winding. Panels (e) and (f) illustrate positive winding numbers with $\nu=+1$ and $\nu=+3$, where the trajectory winds counterclockwise around the origin. Further variation of $\beta$ drives the system back into negative winding numbers, as shown in panels (g) and (i), demonstrating the reentrant nature of the topological phases in the extended parameter space. Although the geometry of the trajectory evolves continuously with the system parameters, the winding number changes only when the loop crosses the origin, consistent with gap closing at topological phase transitions.

\begin{figure}[h]
	\centerline{\includegraphics[width=0.5\linewidth,height=0.42\linewidth]{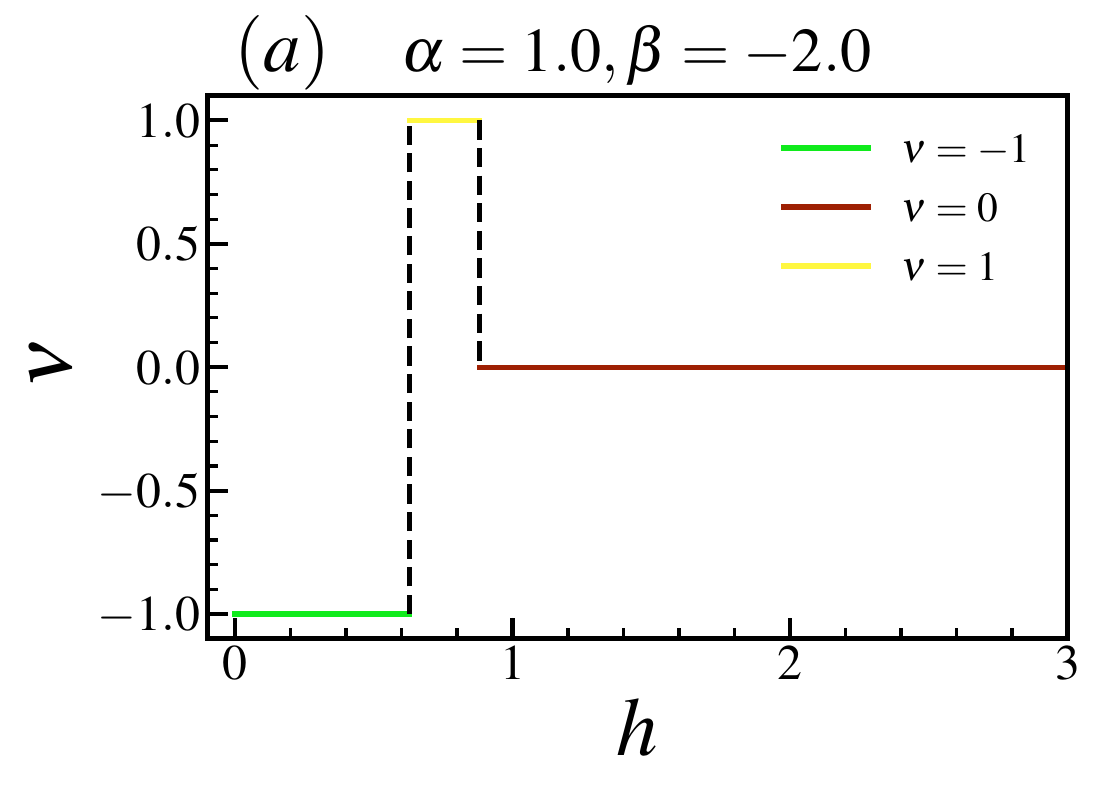} \includegraphics[width=0.5\linewidth,height=0.42\linewidth]{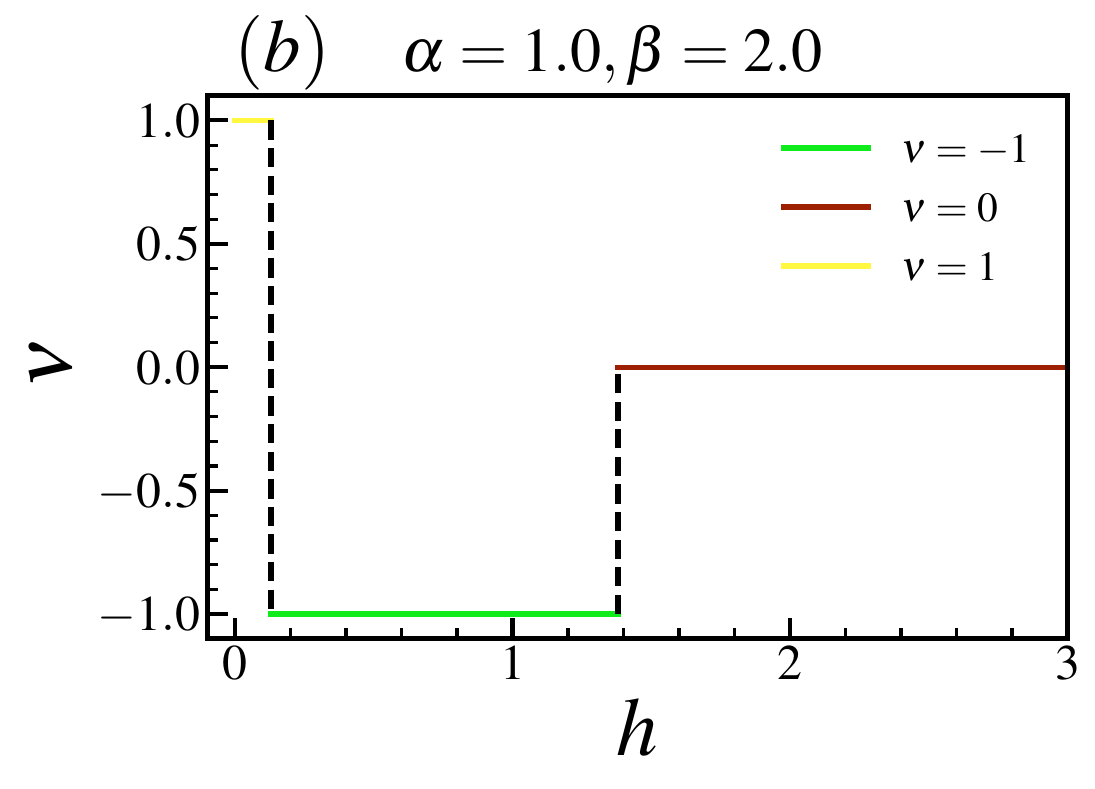} 
		
	}
	\centerline{\includegraphics[width=0.5\linewidth,height=0.42\linewidth]{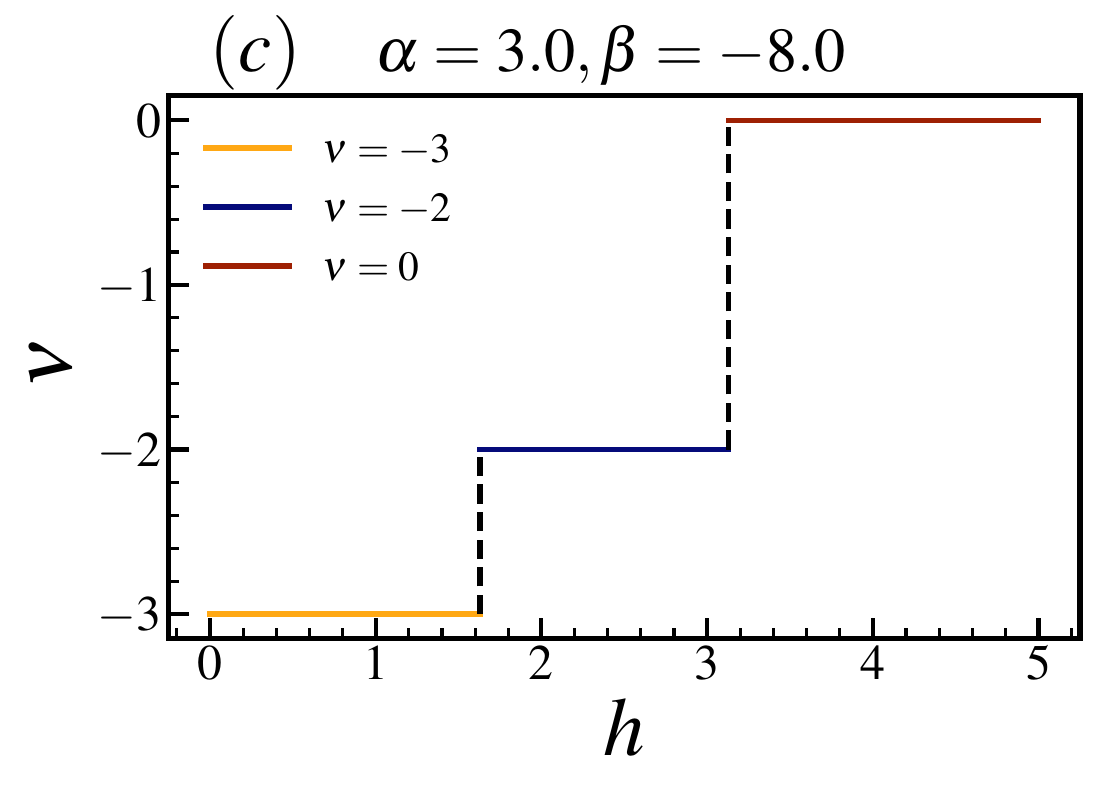} \includegraphics[width=0.5\linewidth,height=0.42\linewidth]{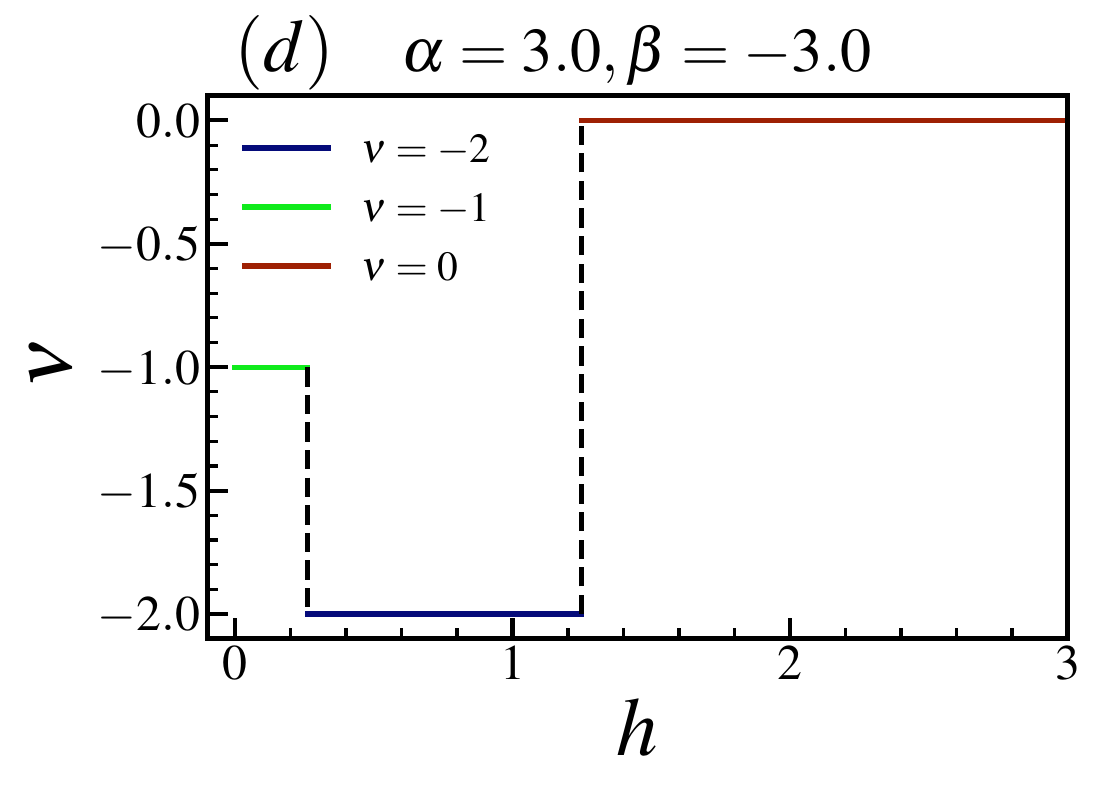} 
		
	}
	
	\centerline{\includegraphics[width=0.5\linewidth,height=0.42\linewidth]{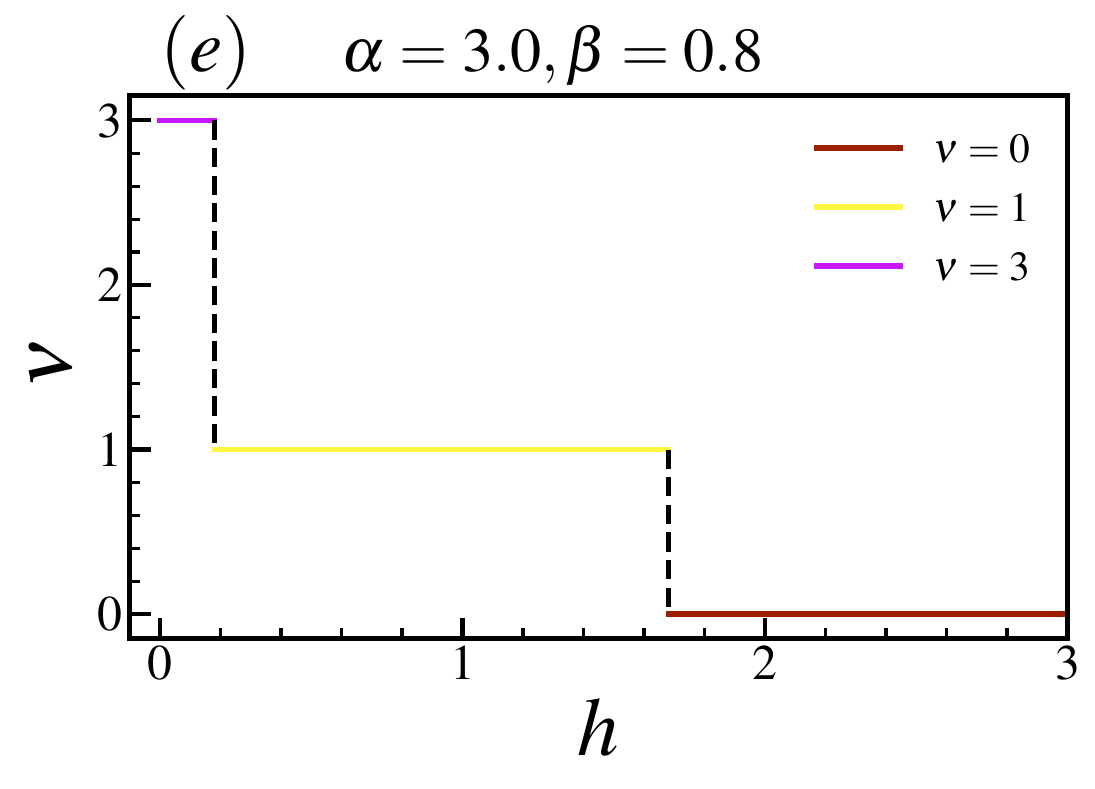} \includegraphics[width=0.5\linewidth,height=0.42\linewidth]{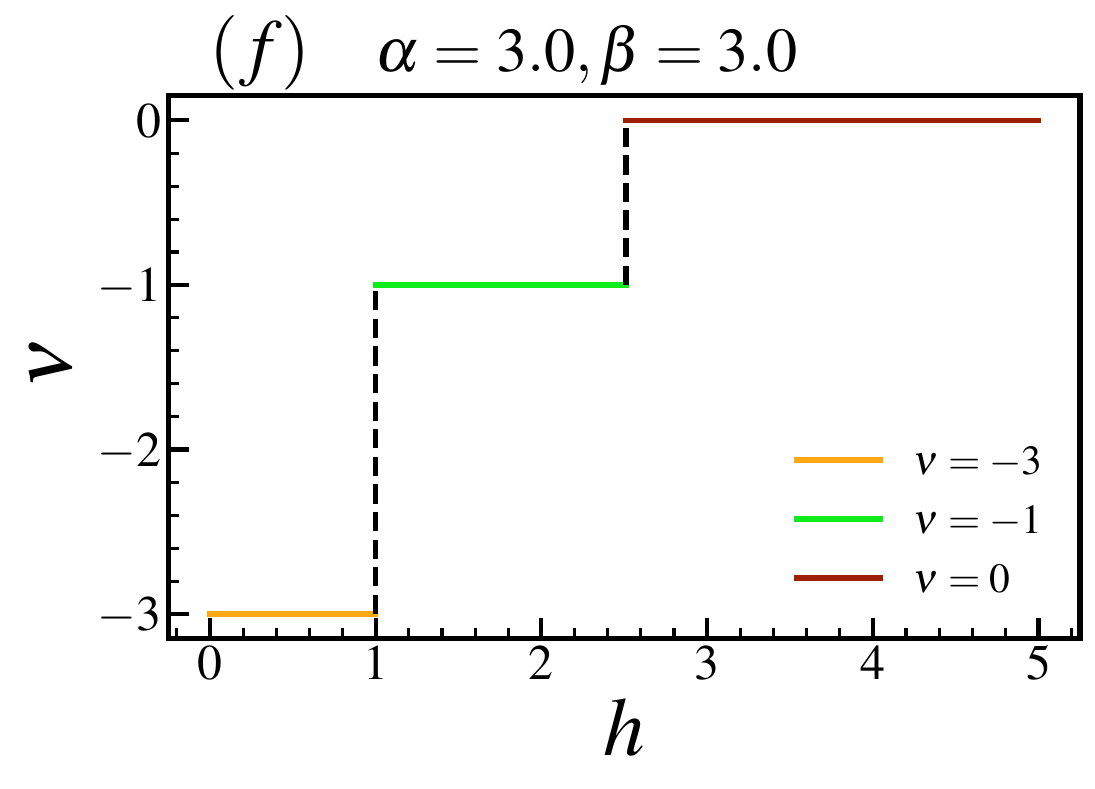} 
		
	}

	\caption{ The winding number as a function of the magnetic field $h$ for different parameter sets: (a,b) $\alpha=1$ with $\beta=-2$ and $2$, respectively; (c--f) $\alpha=3$ with $\beta=-8$, $-3$, $0.8$, and $3$, respectively. All results are obtained for a chain of size $N=1000$.
	}
	\label{Fig10}
	
\end{figure}

After identifying multiple topological cases from the trajectories shown in Fig.~\ref{Fig9}, we now examine the evolution of the winding number as a function of the transverse field to identify the corresponding phase boundaries and examine the stability of the topological phases.  In reference \cite{mahdavifar2026topological}, the Hamiltonian Eq(\ref{eq3}) shows that, in the absence of the field, the model supports several distinct topological regions depending on the values of $\alpha$ and $\beta$. Based on this phase structure, we compute the winding number for representative parameter sets corresponding to both $\beta<0$ and $\beta>0$ at fixed $\alpha=1$ and $\alpha=3$, chosen to sample all topological parts present in the zero-field phase diagram. This analysis allows us to follow the evolution of each phase under the application of the field, identify the critical points where the winding number changes, and characterize the robustness of the corresponding topological states against field-induced gap closing. In this way,  the winding number diagrams directly capture the field-driven topological transitions and their connection to the phase structure of the model.

Figure~\ref{Fig10} shows the evolution of the winding number as a function of the transverse field for representative values of $\alpha$ and $\beta$. For $\alpha=1$ and $\beta=-2$, shown in Fig.~\ref{Fig10}(a), the system is initially in a topological phase with $\nu=-1$ at zero field. Upon increasing the field, the winding number changes from $\nu=-1$ to $\nu=+1$ at the critical line $h_{c_3}$, indicating a transition between distinct nontrivial topological sectors. A further increase of the field drives the system across $h_{c_2}$ into the trivial phase with $\nu=0$. A similar sequence is observed for $\alpha=1$ and $\beta=2$, shown in Fig.~\ref{Fig10}(b), where the system evolves from the $\nu=+1$ phase to $\nu=-1$ at $h_{c_3}$ before entering the trivial sector at $h_{c_2}$.

For $\alpha=3$ and $\beta=-8$, shown in Fig.~\ref{Fig10}(c), the zero-field phase is characterized by $\nu=-3$. Increasing the field first induces a transition at $h_{c_1}$ into a phase with $\nu=-2$, the system then enters the trivial phase at $h_{c_3}$ with $\nu=0$. For $\alpha=3$ and $\beta=-3$, shown in Fig.~\ref{Fig10}(d), the system starts from the $\nu=-1$ part and undergoes a transition to $\nu=-2$ at $h_{c_2}$, before finally reaching the trivial phase at $h_{c_3}$.

The positive-$\beta$ regime for $\alpha=3$ is illustrated in Figs.~\ref{Fig10}(e) and \ref{Fig10}(f). For $\beta=0.8$, shown in Fig.~\ref{Fig10}(e), the system initially lies in the $\nu=+3$ phase and undergoes a transition to the $\nu=+1$ sector at $h_{c_4}$, followed by a transition into the trivial phase at $h_{c_2}$. For $\beta=3$, shown in Fig.~\ref{Fig10}(f), the winding number changes from $\nu=-3$ to $\nu=-1$ across $h_{c_4}$ and subsequently vanishes at $h_{c_2}$. In each case, the discontinuous changes in $\nu$ occur precisely at the critical fields obtained from the gap-closing conditions, confirming the connection between the spectral transitions and the topological structure of the phase diagram.

\begin{figure*}
	\centerline{\includegraphics[width=0.34\linewidth,height=0.28\linewidth]{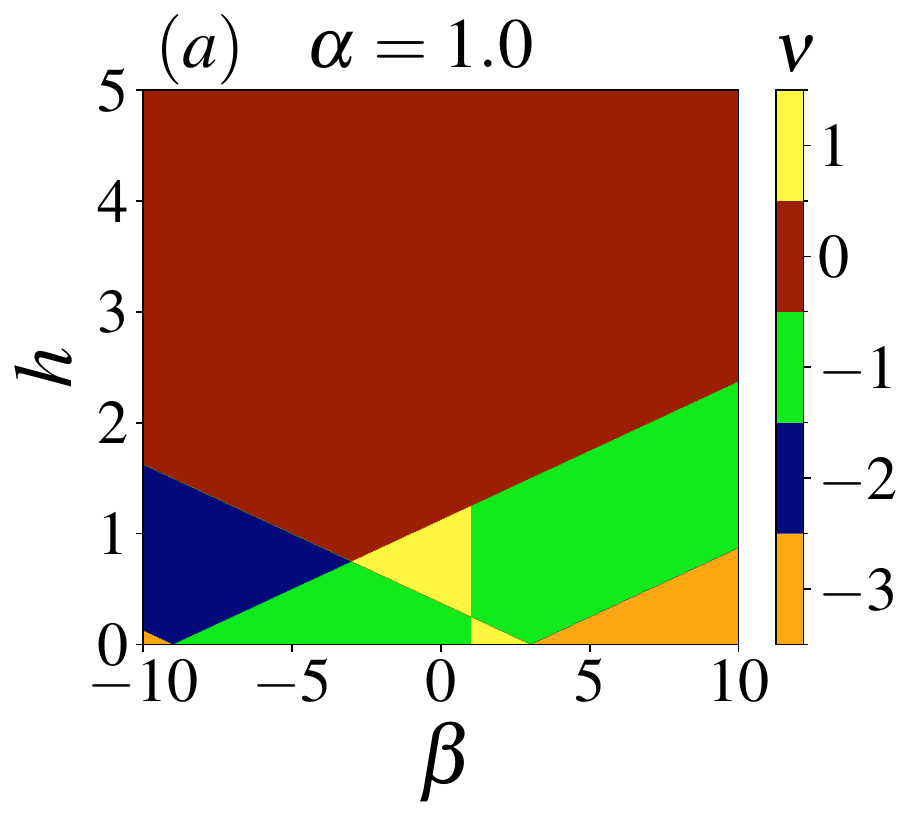} \includegraphics[width=0.34\linewidth,height=0.28\linewidth]{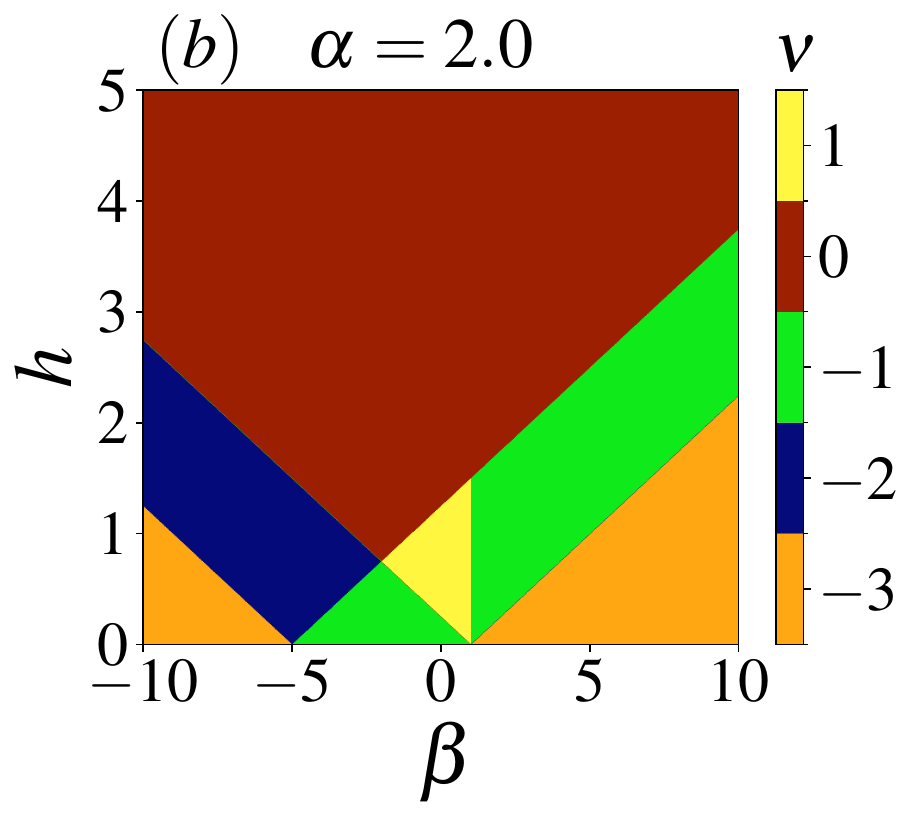} 
		\includegraphics[width=0.34\linewidth,height=0.28\linewidth]{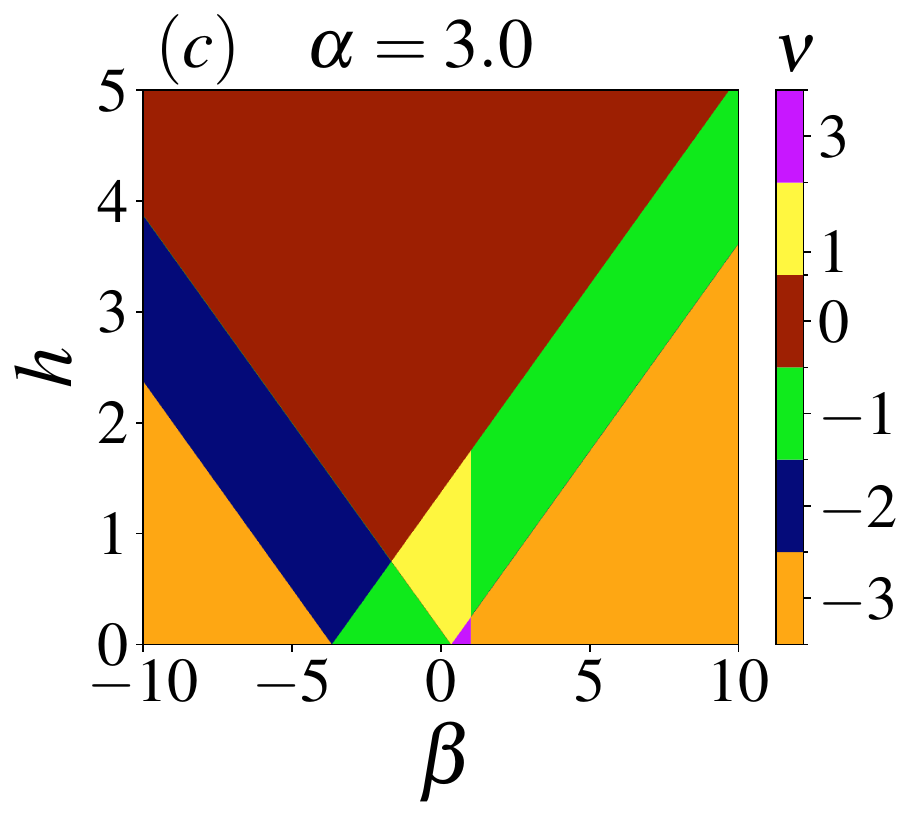}
	}
	
	\caption{ Density plot of the topological phase diagram characterized by the winding number in the $(\beta, h)$ parameter plane. All results are obtained for a chain of size $N=1000$.
	}
	\label{Fig11}
	
\end{figure*}

Figure~\ref{Fig11} shows the winding number density plots for $\alpha=1$, $2$, and $3$, illustrating the overall structure of the topological phase diagram. For $\alpha=1$ and $\alpha=2$, the system supports five distinct topological sectors characterized by $\nu=-3,-2,-1,0$, and $1$, whereas for $\alpha=3$ an additional sector with $\nu=3$ emerges, reflecting the enhanced topological richness induced by stronger extended interactions. Each integer value of $\nu$ identifies a distinct topological phase, with neighboring sectors separated by critical lines at which the bulk excitation gap closes. The sign of $\nu$ determines the orientation of the winding, whereas larger values of $|\nu|$ indicate trajectories that encircle the origin multiple times.

The phase boundaries obtained from the winding number agree with the gap-closing conditions derived from the excitation spectrum, showing that the topological transitions coincide with spectral criticality. As the transverse field increases, all nontrivial sectors eventually evolve into the trivial phase with $\nu=0$, indicating that sufficiently strong fields suppress the topological order by driving the system through successive gap-closing transitions. These results provide a global picture of how the topological sectors evolve under the combined effects of the magnetic field and cluster interactions.

\begin{figure*}
	\centerline{\includegraphics[width=0.25\linewidth,height=0.2\linewidth]{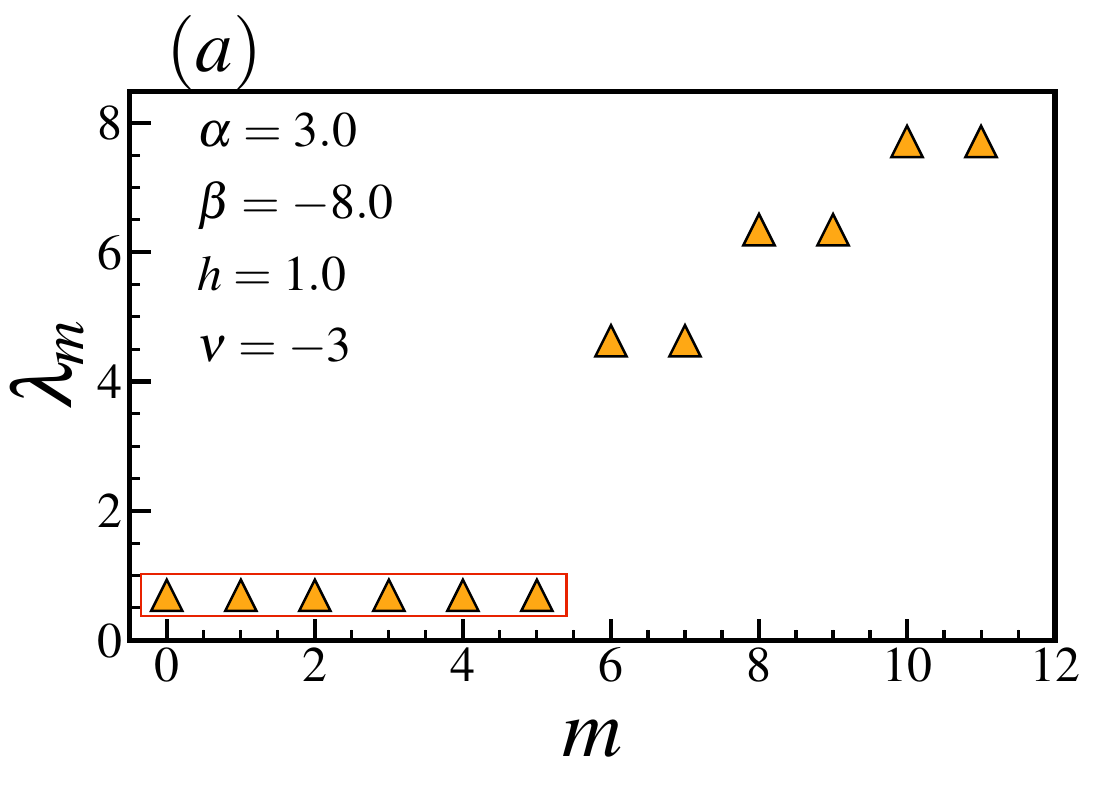} \includegraphics[width=0.25\linewidth,height=0.2\linewidth]{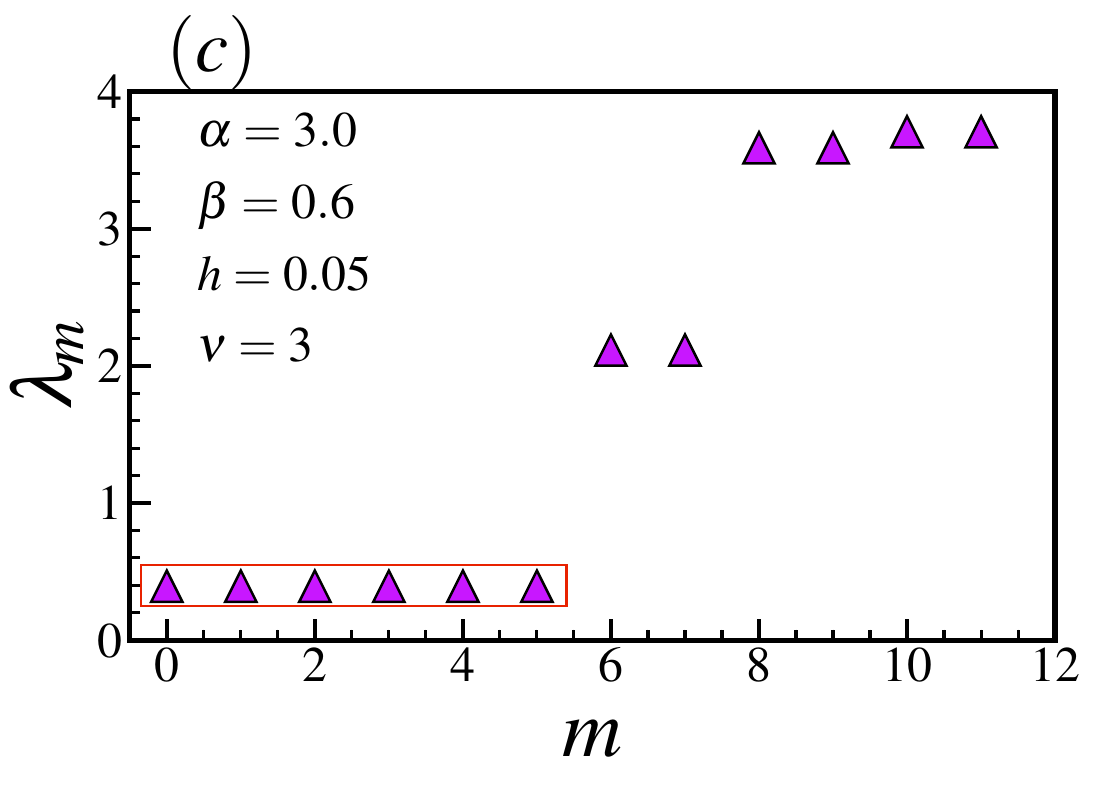} 
		\includegraphics[width=0.25\linewidth,height=0.2\linewidth]{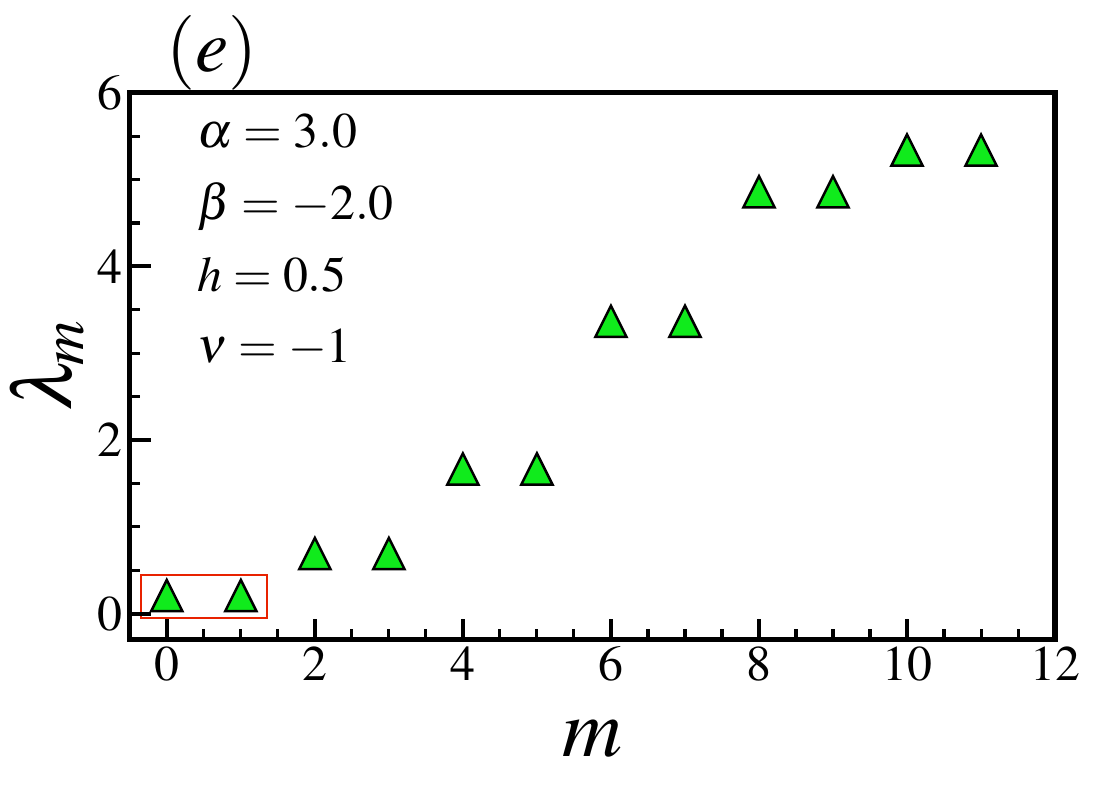}
		\includegraphics[width=0.25\linewidth,height=0.2\linewidth]{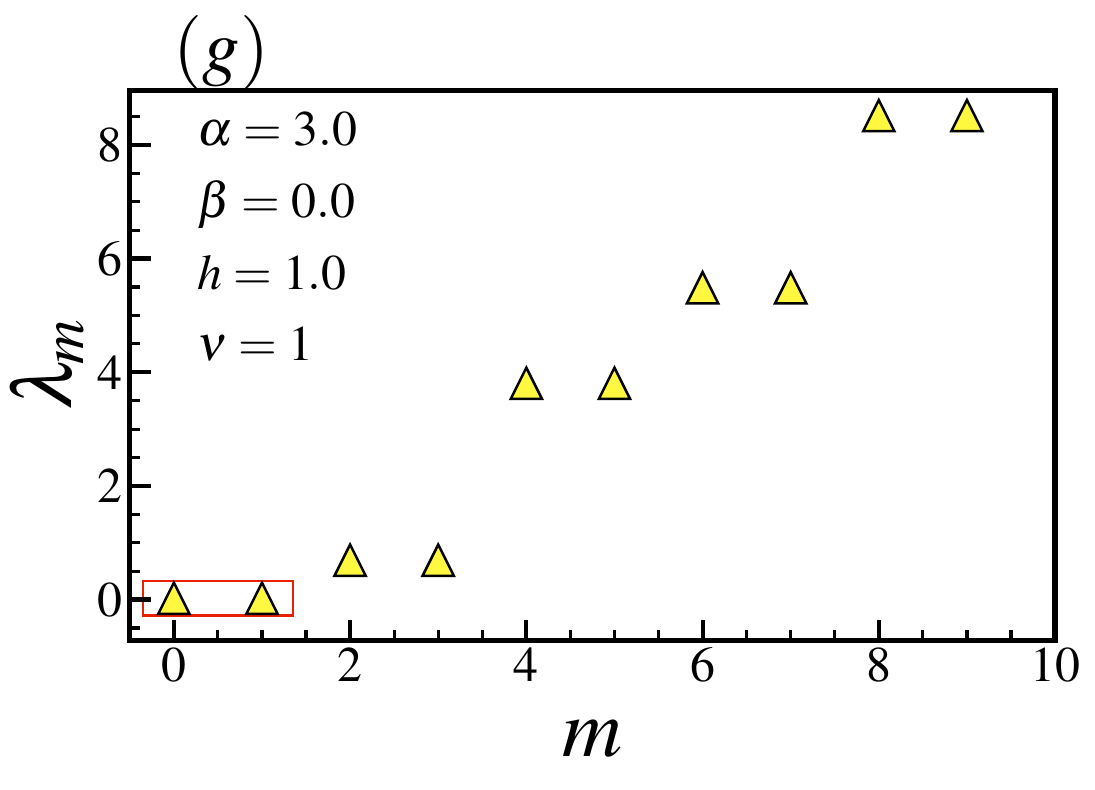}
	}
	
	\centerline{\includegraphics[width=0.25\linewidth,height=0.2\linewidth]{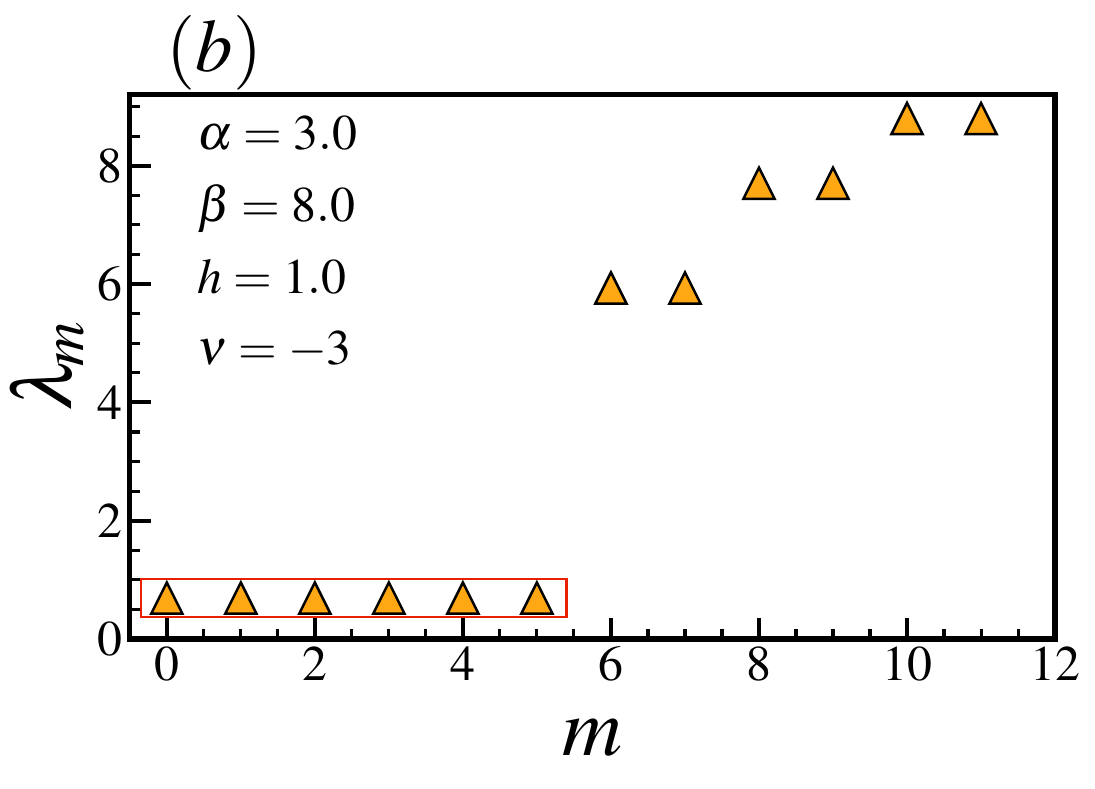} \includegraphics[width=0.25\linewidth,height=0.2\linewidth]{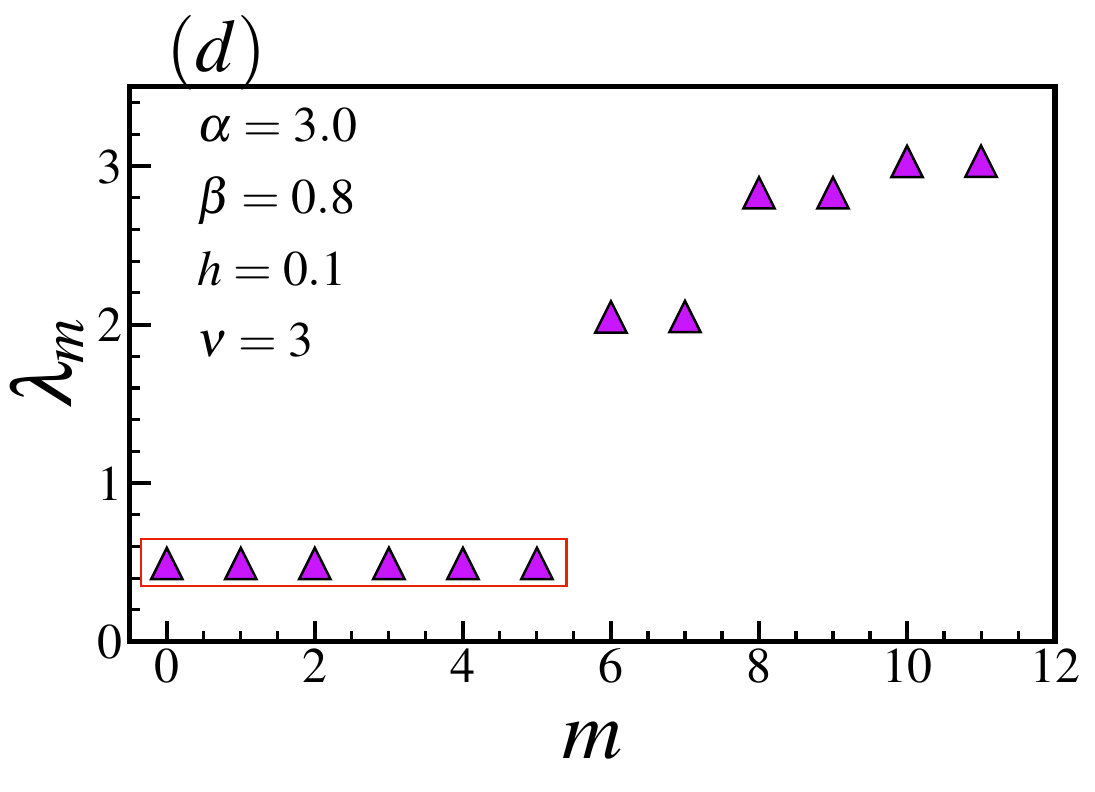} 
		\includegraphics[width=0.25\linewidth,height=0.2\linewidth]{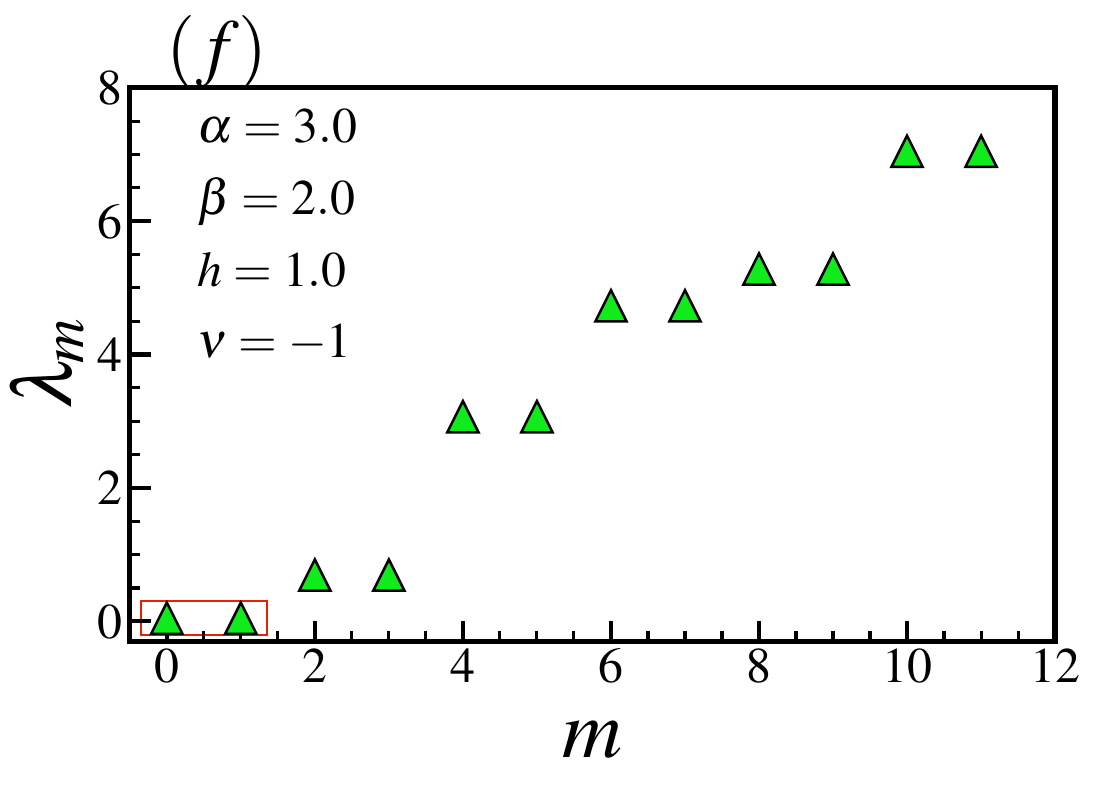}
		\includegraphics[width=0.25\linewidth,height=0.2\linewidth]{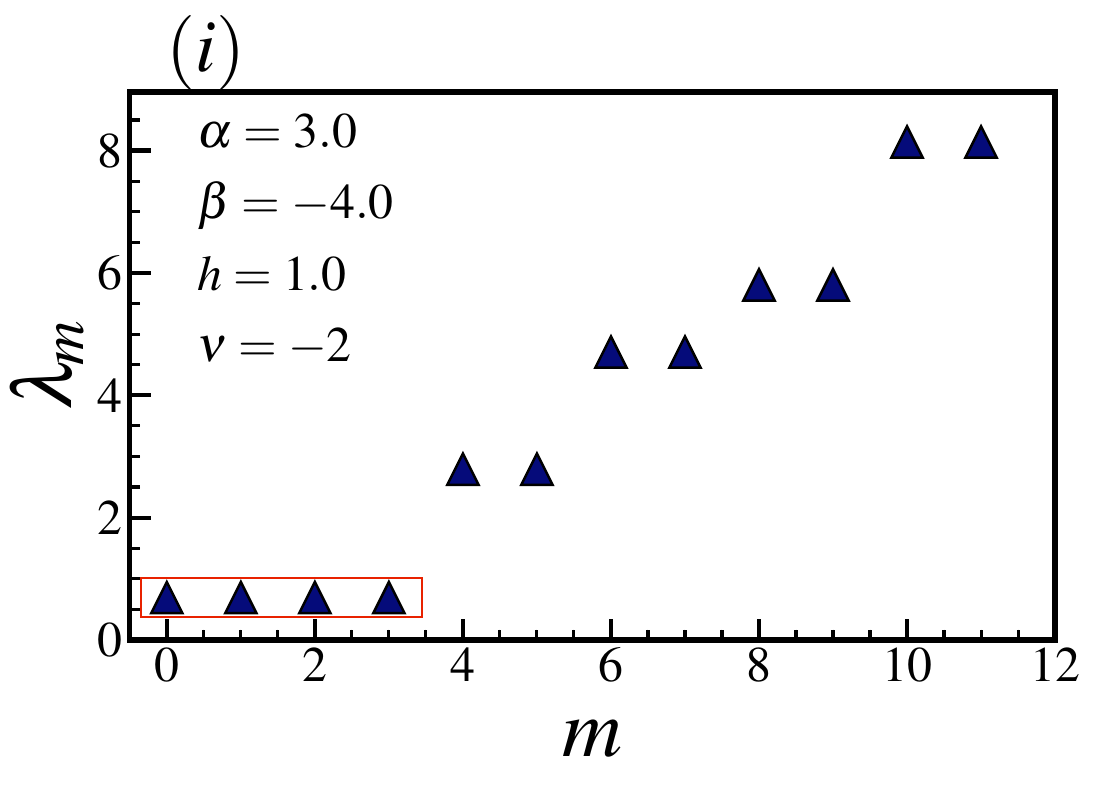}
	}

	\caption{. The bulk entanglement spectrum $\lambda_m$ is shown for different parameter sets: (a) $\beta=-8$, $h=1$; (b) $\beta=8$, $h=1$; (c) $\beta=0.6$, $h=0.05$; (d) $\beta=0.8$, $h=0.1$; (e) $\beta=-2$, $h=0.5$; (f) $\beta=2$, $h=1$; (g) $\beta=0$, $h=1$; and (i) $\beta=-4$, $h=1$. All results are obtained for a chain of size $N=1000$ and $\alpha=3$.
	}
	\label{Fig12}
	
\end{figure*}

To further characterize the nontrivial phases identified from the winding number, we examine the bulk entanglement spectrum (ES) \cite{li101entanglement,pollmann2010entanglement,yu2024universal}. Unlike the entanglement entropy, the entanglement spectrum retains information about the full structure of the reduced density matrix. In practice, the spin chain is divided into two equal subsystems, and the reduced density matrix of one subsystem, $\rho_A$, is used to define the entanglement Hamiltonian through
\begin{equation}
	\tilde{H}_A=-\log \rho_A .
\end{equation}
The entanglement spectrum ${\lambda_m}$ is then given by the eigenvalues of $\tilde{H}_A$, ordered in ascending order.

The entanglement spectrum is particularly useful for characterizing topological phases because it captures nonlocal information beyond conventional observables. In systems exhibiting nontrivial topology, the low-lying structure of the spectrum reflects the underlying bulk--boundary correspondence, with degeneracies in the low-lying entanglement levels serving as signatures of symmetry-protected edge modes. Consequently, distinct topological sectors can often be distinguished directly from the degeneracy structure of the ES. In the present model, the entanglement spectrum provides an independent characterization of the topological phases identified from the winding number and allows us to follow the evolution of the edge-state structure across the phase boundaries.

Figure~\ref{Fig12} presents the bulk entanglement spectrum for representative points at $\alpha=3.0$, where the phase diagram contains the richest topological structure. In this regime, the system supports five distinct nontrivial topological phases characterized by winding numbers $\nu=\pm3,\pm1,$ and $-2$. The different topological sectors are reflected in the degeneracy structure of the lowest entanglement levels.

The spectra shown in Figs.~\ref{Fig12}(a) and \ref{Fig12}(b), obtained for $(\beta,h)=(-8,1)$ and $(8,1)$, respectively, together with those in Figs.~\ref{Fig12}(c) and \ref{Fig12}(d) for $(\beta,h)=(0.6,0.05)$ and $(0.8,0.1)$, belong to phases with winding numbers $\nu=\pm3$. In all four cases, the lowest entanglement level is sixfold degenerate. 

By contrast, the spectra in Figs.~\ref{Fig12}(e) and \ref{Fig12}(f), associated with the $\nu=-1$ phase, as well as Fig.~\ref{Fig12}(g) for $\nu=+1$, show a twofold degeneracy of the lowest entanglement level. Finally, the spectrum shown in Fig.~\ref{Fig12}(i), corresponding to $\nu=-2$, exhibits a fourfold degeneracy.

The observed degeneracy pattern suggests a close correspondence between the winding number and the low-energy structure of the entanglement spectrum. In particular, the degeneracy of the lowest entanglement level follows the relation $2|\nu|$, consistent with the bulk--boundary correspondence expected in one-dimensional topological superconducting systems. Within the Kitaev framework, this degeneracy can be interpreted in terms of virtual Majorana edge modes localized near the bipartition boundaries. As a result, phases with larger $|\nu|$ exhibit enhanced degeneracy and a more intricate topological structure.  The degeneracy structure of the entanglement spectrum therefore follows the magnitude of the winding number, linking the bulk topological invariant to the low-lying entanglement structure.

These results provide an independent consistency check of the topological classification obtained from the winding number analysis. The entanglement spectrum not only distinguishes between different topological sectors, but also reflects the hierarchy of their edge-state structure through the degeneracy of the lowest-lying entanglement levels. The emergence of sixfold degeneracy in the $\nu=\pm3$ phases further highlights the role of extended interactions in stabilizing higher-winding topological phases beyond those typically realized in nearest-neighbor Kitaev chains. More broadly, the robustness of these features and their associated edge excitations may be relevant for engineered quantum materials and platforms aimed at topological quantum information processing and protected quantum transport.

\section{Conclusion} \label{sec:conclusion}

\begin{figure}[h]
	\centerline{\includegraphics[width=0.9\linewidth,height=0.7\linewidth]{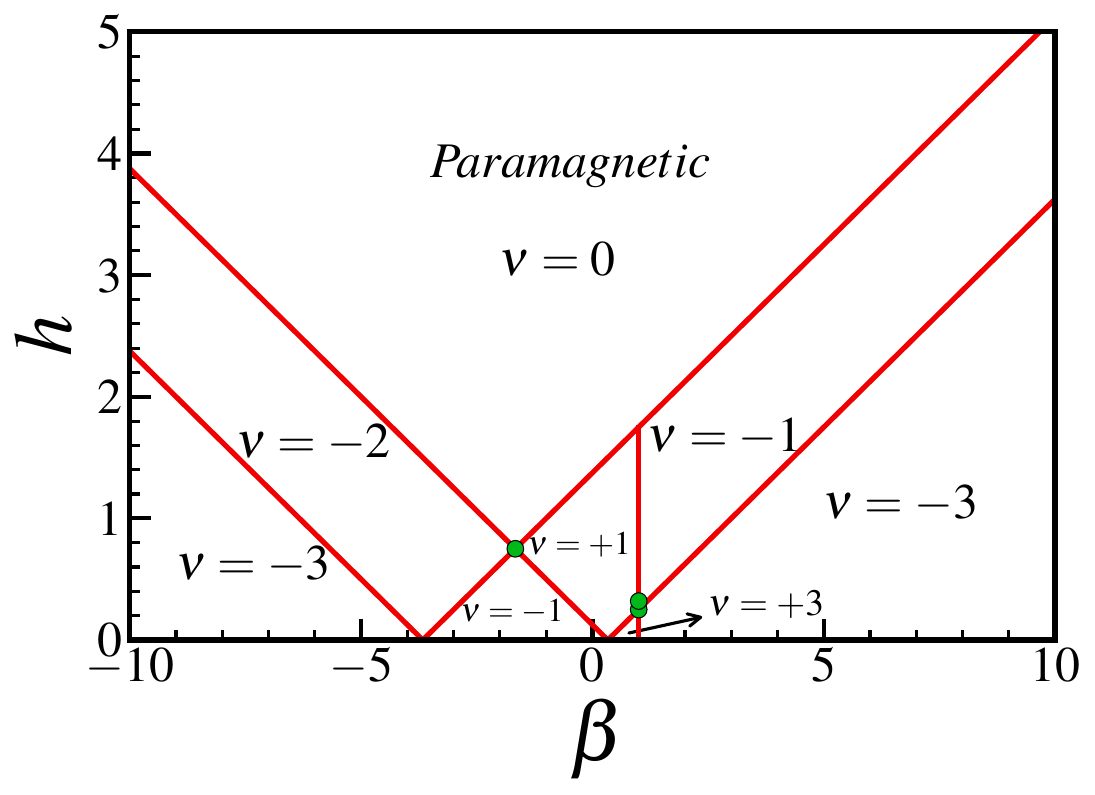}
	}

	\caption{The topological phase diagram of the spin-1/2 XX chain model with an anisotropic four-spin interaction in the presence of a transverse magnetic field.
	}
	\label{Fig13}
	
\end{figure}

In this work, we have established the ground-state phase structure of the spin-$1/2$ XX chain with anisotropic four-spin cluster interactions in a transverse magnetic field. The extended cluster coupling gives rise to a remarkably rich phase diagram in which conventional ordered regimes coexist with chiral and nematic phases, while the interplay between anisotropy, cluster interactions, and magnetic field produces a sequence of distinct topological sectors. The resulting phase diagram, summarized schematically in Fig.~\ref{Fig13}, identifies the critical lines and topological sectors through their winding numbers.

A central feature of the phase diagram is the emergence of multiple topologically inequivalent gapped phases characterized by quantized winding numbers
$\nu=0,\pm1,-2,\pm3$. In particular, the extended four-spin interaction stabilizes phases with winding numbers of magnitude larger than those typically accessible in short-range one-dimensional models. For $\alpha=3$, five distinct nontrivial topological sectors are realized, including the $\nu=\pm3$ phases. These higher-winding phases demonstrate that cluster interactions can substantially enlarge the topological structure of the phase diagram by supporting multiple topological sectors within a single one-dimensional spin model.

The topological sectors are separated by quantum critical lines at which the bulk excitation gap closes and the winding number changes. The resulting phase boundaries therefore provide a direct connection between the closing of the quasiparticle gap and changes in the topological invariant. Increasing the transverse magnetic field progressively reshapes this structure and ultimately drives the system toward the topologically trivial polarized phase. Thus, the magnetic field provides a direct control parameter for the creation, evolution, and destruction of the higher-winding topological phases.

The phase diagram also contains distinct interaction-dominated regimes with finite chiral and nematic correlations. These phases emerge from the extended cluster interaction and occupy finite regions of parameter space, demonstrating that the cluster coupling does not merely modify the topological sector of the model but also produces qualitatively different forms of quantum order. Their suppression with increasing transverse field accompanies the evolution toward the polarized regime, revealing a close interplay between conventional order, unconventional correlations, and topology.

The bulk entanglement spectrum provides an independent characterization of the topological structure. The lowest entanglement levels exhibit characteristic degeneracy patterns in the different winding sectors: twofold, fourfold, and sixfold structures are associated with $|\nu|=1$, $2$, and $3$, respectively. The resulting approximate scaling of the lowest-level degeneracy as $2|\nu|$ is consistent with the bulk--boundary correspondence of the fermionic representation, where higher winding numbers correspond to an increasing number of Majorana-like boundary modes. In particular, the sixfold structure observed in the $\nu=\pm3$ sectors provides a direct entanglement signature of the higher-winding topology generated by the four-spin interaction.

Taken together, the excitation spectrum, winding number, order-parameter structure, and bulk entanglement spectrum establish a coherent picture of the phase diagram. The model simultaneously supports conventional and unconventional ordered phases, chiral and nematic regimes, and a hierarchy of topological phases with winding numbers extending to $|\nu|=3$. 

More broadly, these results demonstrate that anisotropic four-spin interactions provide a controllable mechanism for realizing higher-winding topological phases in one-dimensional quantum systems. The coexistence of multiple topological sectors, chiral and nematic regimes, and their characteristic entanglement signatures shows how extended interactions can qualitatively enrich the topology and ordering structure of correlated quantum matter. Because the cluster interaction and transverse field provide independent control parameters, the model offers a useful setting for exploring and manipulating higher-winding phases and their associated boundary degrees of freedom in experimentally engineered one-dimensional platforms. The stabilization of higher-winding phases and their edge structures may provide useful insights for topological quantum information processing and protected transport in interacting low-dimensional systems.

\bibliography{ref}

\end{document}